\documentclass[a4paper,11pt]{article}
\usepackage{jheppub} % for details on the use of the package, please see the JINST-author-manual
\usepackage{lineno}
\usepackage{graphicx}% Include figure files
\usepackage{wrapfig}
\usepackage{cancel}
\usepackage{dcolumn}% Align table columns on decimal point
\usepackage{bigints}
\usepackage{bm}% bold math
\usepackage[usenames,dvipsnames]{color}
\usepackage{soul}
\usepackage{subcaption}
\usepackage{enumitem}
\usepackage{xspace}
\usepackage{booktabs}
\usepackage{lipsum}
\usepackage[per-mode=power]{siunitx} % Add support for units
\usepackage[capitalise]{cleveref}
\usepackage{ragged2e}
\usepackage{mathrsfs}
\usepackage{cleveref}
\usepackage{hyperref}
\usepackage[normalem]{ulem}
\usepackage{amsmath}
\usepackage{wrapfig}
\usepackage{tcolorbox}
\usepackage{stackengine}
\usepackage{physics}
\usepackage{aas_macros}
\usepackage{feynmp-auto}

\DeclareSIUnit\year{yr} 

\DeclareCaptionJustification{justified}{\justifying}
\newcommand{\bx}{{\bm x}}
\newcommand{\by}{{\bm y}}
\newcommand{\bk}{{\bm k}}

\newcommand{\bp}{{\bm p}}
\newcommand{\bq}{{\bm q}}
\newcommand{\bv}{{\bm v}}

\newcommand{\beq}{\begin{equation}\begin{aligned}}
\newcommand{\eeq}{\end{aligned}\end{equation}}
\newcommand{\Mpc}{\,{\rm Mpc}}
\newcommand{\aeq}{a_{\rm eq}}
\newcommand{\bD}{{\bm{\mathcal{D}}}}

\newcommand{\br}{{\bm{r}}}
\newcommand{\df}{{{\small{DF}}}}
\def\ddelta{{\mathchar '26\mkern -10mu\delta_D}}
\newcommand{\ma}[1]{{\textcolor{Magenta}{{#1}}}}

\definecolor{rp}{cmyk}{0.2, 1, 0.6, 0}
\definecolor{rp}{cmyk}{0.2, 1, 0.6, 0}
\definecolor{green2}{cmyk}{0.27, 0, 1, 0.52}

\usepackage{color}
\hypersetup{
    colorlinks=true,       % false: boxed links; true: colored links
    linkcolor=green2,          % color of internal links
    citecolor=green2,        % color of links to bibliography
    filecolor=magenta,      % color of file links
    urlcolor=green2           % color of external links
}
\usepackage{physics}
\usepackage{tensor}
\title{\large{Structure Formation with Warm White Noise:\\
\small{Effects of Finite Number Density and Velocity Dispersion in Particle and Wave Dark Matter}}}
\author[a]{\small{Mustafa A. Amin}}
\emailAdd{mustafa.a.amin@rice.edu}
\author[b]{, M. Sten Delos}
\emailAdd{mdelos@carnegiescience.edu}
\author[c]{, Mehrdad Mirbabayi}
\emailAdd{mirbabayi@ictp.it}

\affiliation[a]{Department of Physics and Astronomy, Rice University,
Houston, TX, 77005, U.S.A.}
\affiliation[b]{Carnegie Observatories, 813 Santa Barbara Street, Pasadena, CA 91101, USA}
\affiliation[c]{International Centre for Theoretical Physics, Trieste, Italy}
\abstract{We investigate the evolution of density perturbations in dark matter, including the new combined effects of finite number density and non-zero velocity dispersion. Using a truncated BBGKY hierarchy, we derive analytical expressions for the dark matter power spectrum during radiation and matter domination. A component of {\it warm white noise} emerges in our analysis, which arises due to the finite number density and undergoes scale-dependent evolution because of the velocity dispersion. Although free streaming erases adiabatic initial perturbations on small scales, warm white noise persists below the free-streaming length and grows during matter domination, with growth suppressed below the dark matter Jeans length. Our calculated power spectra agree with $N$-body simulations in the linear regime and accurately predict halo mass functions in the nonlinear regime. Effects of warm white noise can emerge on observable quasi-linear scales for ultralight dark matter produced after inflation with a subhorizon correlation length. Our formalism is applicable to these scenarios (with de Broglie-scale quasi-particles), to cases in which dark matter includes macroscopic structures (such as primordial black holes), and to traditional warm and cold dark matter scenarios.}

\begin{document}
\maketitle
\flushbottom

%%%%%%%%%%%%%%%%%%%%%%%%%%%%%%%%
\section{Introduction}
\label{sec:intro}
%%%%%%%%%%%%%%%%%%%%%%%%%%%%%%%%
There is significant evidence for the existence of dark matter from observations of cosmological structures on large scales ($\gtrsim \rm Mpc$ co-moving) \cite{Cirelli:2024ssz,Marsh:2024ury}. However, probing structure formation on smaller scales ($\lesssim \rm Mpc$ co-moving) offers a unique opportunity to uncover the microscopic nature of dark matter and its production mechanism. A variety of current and upcoming observations aim to probe the small-scale matter power spectrum \cite{Drlica-Wagner:2022lbd}. These include analyses of the Lyman-$\alpha$ forest, galaxy satellite populations, gravitational lensing, stellar streams, 21-cm line intensity mapping, dynamical heating of stars, and the abundance of high-redshift galaxies
\cite{Mondino:2020rkn,Sabti:2021unj,Gilman:2021gkj,Delos:2021ouc,Drlica-Wagner:2022lbd,Boylan-Kolchin:2022kae,Chung:2023syw,Irsic:2023equ,Delos:2023dwq,Esteban:2023xpk,Nadler:2024ims,Xiao:2024qay,Ji:2024ott,deKruijf:2024voc,Buckley:2025zgh}. To fully leverage these data, it is beneficial to develop theoretical models that connect the observed small-scale power spectrum to the microscopic properties and initial conditions of the dark matter.

On small scales, the linear matter power spectrum exhibits features that are sensitive to the nature of dark matter. For instance, in models with sufficient velocity dispersion in the early universe ($\gtrsim 10\,{\rm km}\,\,\rm{s}^{-1}$ at matter radiation equality), such as those involving ``warm" dark matter, the adiabatic power spectrum is suppressed below co-moving scales of $\sim \rm Mpc$. This ``free-streaming suppression" has been extensively studied in microscopic particle dark matter models \cite{Narayanan:2000tp,Hansen:2001zv,Lewis:2002nc,Green:2003un,Viel:2005qj,Lesgourgues:2006nd,Viel:2007mv,Boyarsky:2008xj,Erickcek:2011us,Lancaster:2017ksf,Irsic:2017ixq,wdm,Miller:2019pss,Erickcek:2021fsu,Ballesteros:2020adh,Sarkar:2021pqh,Garcia:2023qab}, and more recently in wave dark matter models \cite{Amin:2022nlh,Liu:2024pjg,Ling:2024qfv,Long:2024imw}.

Another universal feature emerges on sufficiently small scales where the Poissonian distribution of the underlying constituents of the dark matter becomes important. These constituents may be particles, quasi-particles in wave dark matter, primordial black holes \cite{Green:2020jor}, axion mini-clusters \cite{Ellis:2022grh}, solitons \cite{Zhou:2024mea} or other bound structures (e.g.~\cite{Kaplan:2024ydw}). This Poissonian behavior introduces a white-noise component in the power spectrum at small scales. If the co-moving number density of these constituents is $\bar{n}\lesssim 10^8 \, \rm Mpc^{-3}$, then the resulting white-noise spectrum could fall within the reach of observational probes of structure in the quasi-linear regime ($k\lesssim 10~\rm{Mpc}^{-1}$), such as Lyman-$\alpha$ and 21-cm maps and the galaxy distribution. For higher $\bar n$, the white-noise contribution would boost dark matter substructure deep in the nonlinear regime, but the influence of such structure could still be detected. For example, effects on the power spectrum could be detected up to $k\sim 10^2~\rm{Mpc}^{-1}$ with strong lensing \cite{Gilman:2021gkj} and dwarf galaxy structures \cite{Esteban:2023xpk}, up to $k\sim 10^3~\rm{Mpc}^{-1}$ with stellar kinematics \cite{Graham:2024hah}, and up to $k$ in excess of $10^6~\mathrm{Mpc}^{-1}$ with pulsar timing \cite{Ramani:2020hdo,Lee:2020wfn,Lee:2021zqw,Delos:2021rqs} and microlensing of high-redshift stars \cite{Dai:2019lud,Blinov:2021axd}.

\begin{figure}
    \centering
\includegraphics[width=1\linewidth]{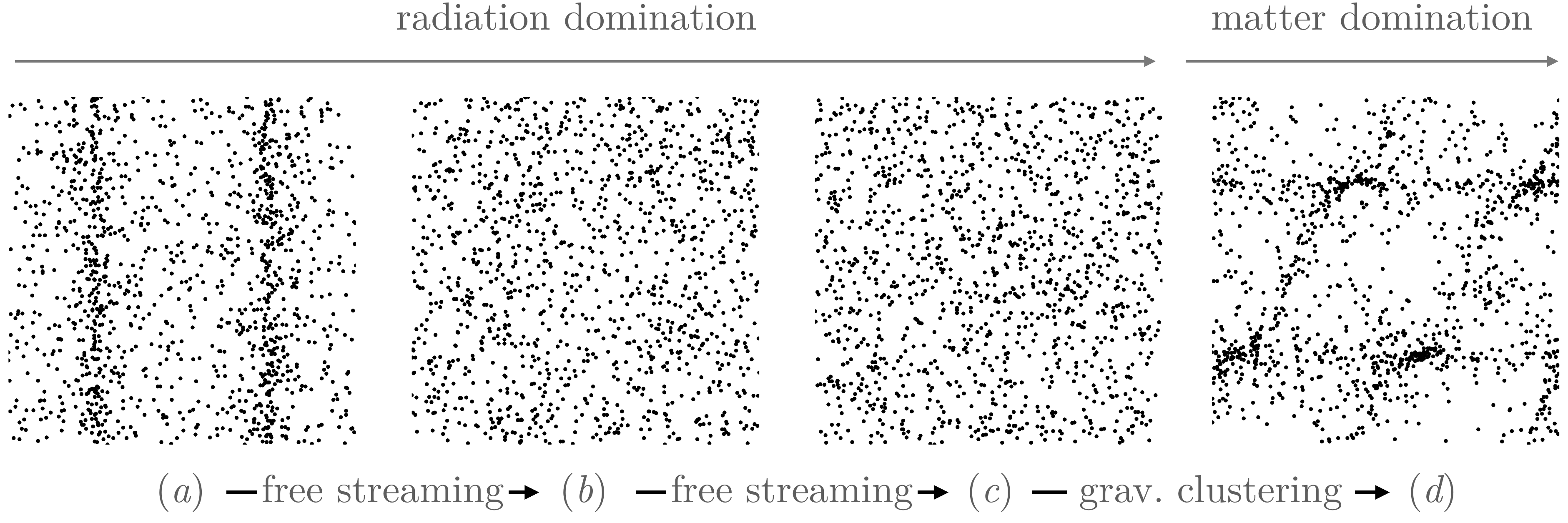}
    \caption{During radiation domination, free streaming erases existing correlations below the free-streaming length, revealing more of the underlying Poisson distribution of particles: $(a)\rightarrow (b)$. However, free streaming cannot change the underlying Poisson distribution: $(b)\rightarrow (c)$. During matter domination, gravitational clustering builds correlations above the Jeans length: $(c)\rightarrow (d)$. The points could represent de Broglie-scale quasi-particles in wave dark matter, composite/macroscopic dark matter (including primordial black holes or substructures) or ``usual" particle dark matter.}
    \label{fig:Points}
\end{figure}

In this work, we focus on the evolution of this white-noise contribution in the presence of a velocity dispersion. We analytically derive the time evolution of the matter power spectrum during the radiation- and matter-dominated eras. Our result exhibits several key physical effects, which we illustrate in Figure~\ref{fig:Points}. During the radiation era, free streaming erases initial (e.g., adiabatic) density perturbations, but it cannot erase the underlying Poisson distribution. During matter domination, that Poisson distribution can then give rise to gravitational clustering, even below the free-streaming length. The velocity dispersion instead suppresses perturbation growth below the Jeans length, which is smaller than the free-streaming length and shrinks further over time. We find excellent agreement between the analytically calculated power spectrum and $N$-body simulations. The analytic prediction also leads to accurate halo mass functions.

%~~~~~~~~~~~~~~~~~~~~~~~~~~%~~~~~~~~~~~~~~~~~~~~~
%~~~~~~~~~~~~~~~~~~~~~~~~~~

Our calculation of the power spectrum involves solving a truncated BBGKY (Bogoliubov–Born–Green–Kirkwood–Yvon) hierarchy. The setup, including the truncated BBGKY hierarchy, can be found in textbooks \cite{Nicholson:1983,Binney:1987}. The result for the free-streaming suppression for the adiabatic part is also well known. However, the form of the solution we are able to write down for the evolution of the white noise part is different from existing literature.  In this solution, the connection between the shape of the initial phase space distribution function ({\df}) and the shape of the final power spectrum is transparent for both the adiabatic and white-noise parts. 

Although our formalism rigorously applies to particle dynamics, we expect our results to approximately apply also to wave dark matter as long as we restrict ourselves to scales sufficiently larger than the de Broglie scale. Ultralight wave dark matter is of particular interest because it can simultaneously produce free-streaming suppression and white noise both at a level accessible to observations of quasi-linear structure \cite{Amin:2022nlh}. For instance, if excitations of the dark matter field are produced after inflation, as in axion models with post-inflation Peccei–Quinn symmetry breaking, their finite correlation length leads to a clumpiness of the density field with an initially white noise spectrum. The same clumpiness also implies a non-trivial velocity dispersion. A simplified model would then be to introduce quasi-particles which represent the coherent patches and whose velocity dispersion matches that of the underlying wave dark matter field.\footnote{Clumpiness at the de Broglie scale has been widely studied in the context of halos of ultralight dark matter \cite{Hu:2000ke,2019MNRAS.485.2861C,2019ApJ...871...28B,Dalal:2020mjw,2021ApJ...915...27B,Mocz:2017wlg,Ferreira:2020fam,Hui:2021tkt,May:2021wwp,Amin:2022pzv,Gosenca:2023yjc,Powell:2023jns,Eberhardt:2024ocm,Boddy:2025oxn,Eberhardt:2025tao,Luu:2024gnk}. In this case the velocity dispersion arises due to local gravitational clustering rather than early-universe initial conditions.} We expect that our formalism can be also extended to include additional effects present in wave dark matter at small scales.

Our present formalism is also specialized to non-relativistic gravitational dynamics of single-component dark matter, but it could be extended to include interactions beyond gravity, relativistic effects, and multi-component dark matter. Moreover, in the future, baryonic effects, which have been ignored in the present work, will need to be considered for detailed comparison with observations on small scales.

As an historical aside, the use of the Liouville equation and the (truncated) BBGKY hierarchy has a rich history in gravitational (and plasma) physics. 
 One can derive fluid equations based on the collisionless Boltzmann equation, also called the Jeans or Vlasov equation  \cite{Jeans:1915,Henon:1982}, which can then be solved with relative ease in many circumstances in cosmology (see for example, \cite{Thorne:2021}). In our work, we are going beyond this collisionless equation to include collision terms, but with a simple truncation scheme that works in the linear regime of density fluctuations. For related earlier work, see for example refs.~\cite{Gilbert:1971,Fall:1976a,Fall:1976b}.

%As an historical aside, the use of the Liouville equation and the (truncated) BBGKY hierarchy has a rich history in gravitational (and plasma) physics. One of the earliest and most impactful results for astrophysics is that from Jeans \cite{Jeans:1915}. The Jeans equation is referred to as the Vlasov equation in Plasma physics, although as argued by \cite{Henon:1982}, it should more appropriately be called the ``collisionless" Boltzmann equation in both cases. One can derive fluid equations based on the collisionless Boltzmann equation, which can then be solved with relative ease in many circumstances in cosmology (see for example, \cite{Thorne:2021}). In our work, we are going beyond this collisionless equation to include collision terms, but with a simple truncation scheme that works in the linear regime of density fluctuations. For related earlier work, see for example refs.~\cite{Gilbert:1971,Fall:1976a,Fall:1976b}.

%~~~~~~~~~~~~~~~~~~~~~~~~~~~~~~
\subsection{Notation \& Conventions}
\label{sec:notconv}
The background Friedmann–Lemaître–Robertson–Walker spacetime is given by
\beq
ds^2=-dt^2+a^2(t)d\bx\cdot d\bx\,,
\eeq
where $a(t)$ is the scale factor, $\bx$ are co-moving co-ordinates, and we set $c=1$. We assume a spatially flat background expansion history dominated by radiation and matter, with the Hubble expansion rate given by
\beq
H(y)=\frac{k_{\rm eq}}{\sqrt{2}a_{\rm eq}}y^{-2}\sqrt{1+y}\,,\quad \textrm{where}\quad y\equiv \frac{a}{a_{\rm eq}}.
\eeq
The scale factor at matter-radiation equality is $a_{\rm eq}\approx 1/3388$, and the co-moving wavenumber associated with the horizon size at that time is $k_{\rm eq}=a_{\rm eq}H(a_{\rm eq})\approx 0.01/\Mpc$ \cite{Planck:2018vyg}.

Our Fourier conventions for finite volumes $V$ and infinite volumes, respectively, are
\beq
f(\bx)&=V^{-1}\sum_\bk e^{i\bk \cdot \bx}f_\bk,   \qquad f_\bk=\int_V d\bx \,e^{-i\bk \cdot \bx}f(\bx), \quad \int_V d\bx e^{\pm i(\bk+\bk')\cdot \bx}=V\delta_{\bk,-\bk'}\,,\\
f(\bx)&=\int \frac{d\bk}{(2\pi)^3} e^{i\bk \cdot \bx}f(\bk),\enspace f(\bk)=\int d\bx\, e^{-i\bk \cdot \bx}f(\bx),\enspace \int d\bx e^{\pm i(\bk+\bk')\cdot \bx}=(2\pi)^3\delta_D(\bk+\bk')\,.
\eeq
Here $\delta_{\bk,-\bk'}$ is the Kronecker delta and $\delta_D$ is the Dirac delta function. 
Note that the finite- and infinite-volume cases are linked by:
\beq\label{eq:finite_infinite}
&\qquad\int_\bk \quad\longleftrightarrow\quad \int \frac{d\bk}{(2\pi)^3}\quad\longleftrightarrow\quad \frac{1}{V}\sum_\bk\,,\\
&\ddelta(\bk+\bk')\longleftrightarrow (2\pi)^3 \delta_D(\bk+\bk') \longleftrightarrow V\delta_{\bk,-\bk'}\,,
\eeq
and formally $V\leftrightarrow (2\pi)^3\delta_D({\bf 0})$.
The first entry on each line of equation~\eqref{eq:finite_infinite} represents the shorthand notation
\beq
f(\bx)&=\int_\bk e^{i\bk \cdot \bx}f_\bk,\qquad f_\bk=\int_\bx e^{-i\bk \cdot \bx}f(\bx),\quad \int_\bx e^{\pm i(\bk+\bk')\cdot \bx}=\ddelta(\bk+\bk'),
\eeq
where the integral without the explicit measure can stand in for finite- or infinite-volume cases.

For a statistically homogeneous field $f$, the power spectrum $P_f$ is defined:
\beq
\langle f(\bk)f(\bk')\rangle =(2\pi)^3 P_f(\bk)\delta_D(\bk+\bk')\,, \qquad \langle f_\bk f_{\bk'}\rangle =V P_f(\bk)\delta_{\bk,-\bk'}.
\eeq

%%%%%%%%%%%%%%%%%%%%%%%%%%%%%%%%
\subsection{Summary of Derivation and Results}
\label{sec:derivation-summary}

Section~\ref{sec:derivation} of this paper is devoted to deriving the power spectrum $P_\delta$ of the field of density contrasts $\delta(\bx)=[\rho(\bx)-\bar\rho]/\bar\rho$. Here is a schematic representation of our derivation:
\beq
&P_\delta(t,k)\\
&\quad\Uparrow \\
&\underbrace{f^{(2)}(t,\bx_1,\bx_2,\bp_1,\bp_2)}_{\substack{\text{$2$-particle phase space} \\ \text{distribution function}}}\\
&\quad\Uparrow \\&\underbrace{\mathcal{L}^{(1)}_tf^{(1)}(t,\bp_1)=S[f^{(2)}],\quad \mathcal{L}^{(2)}_tf^{(2)}(t,\bx_1,\bx_2,\bp_1,\bp_2)\approx S[f^{(1)}]}_{\textrm{(truncated) BBGKY hierarchy}}\,,
\eeq
where $\mathcal{L}_t^{(1,2)}$ symbolically represent time-evolution operators and $S[f^{(1,2)}]$ are source functions in the evolution equations. The input will be the initial 1-particle phase space distribution function ({\small{DF}}) $f^{(1)}(t_0,\bp)=f_0(p)$, and initial adiabatic perturbations if they exist.
%Within the perturbative regime, we will see that it is possible to neglect the time dependence of $f^{(1)}$ and the influence of 3-particle and higher DFs.
We assume statistical homogeneity and isotropy, so $f^{(1)}$ is independent of position and direction.
We restrict ourselves to non-relativistic particles, clustering under Newtonian gravity in an expanding universe.
%While not necessary in the early parts of the derivation, the final results  assume statistical homogeneity and isotropy. 

%~~~~~~~~~~~~~~~~~~~~~~~~~~%~~~~~~~~~~~~~~~~~~~~~
\begin{figure}
    \centering
\includegraphics[width=0.95\linewidth]{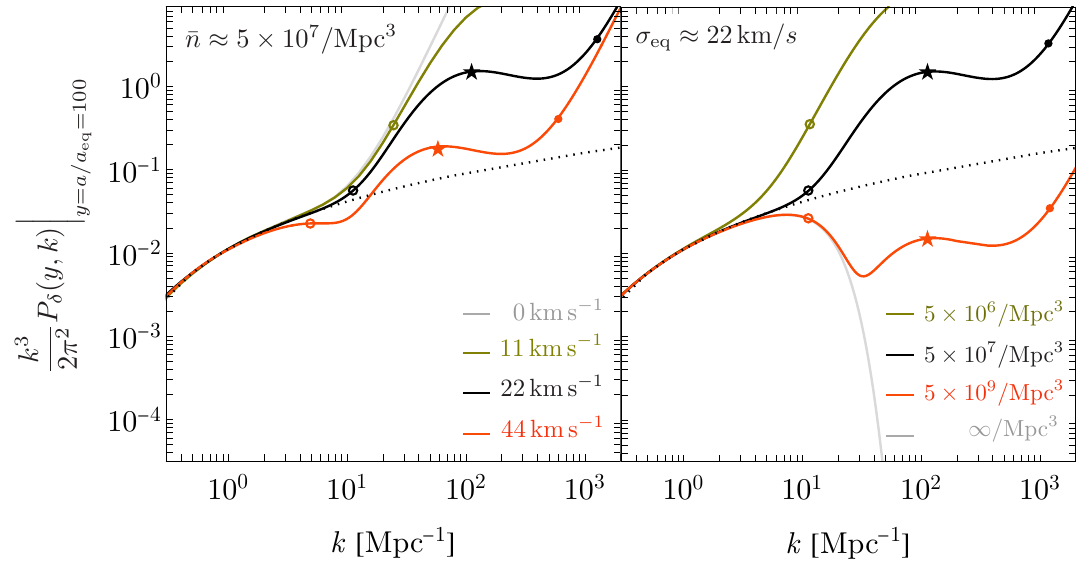}
    \caption{Small length-scale features in the dimensionless density power spectrum depend on the co-moving number $\bar{n}$ (setting the amplitude of the white noise contribution) and 1D velocity dispersion $\sigma_{\rm eq}$ (at matter-radiation equality, which determines the free-streaming and Jeans scales).
    %We adopt a Maxwellian velocity distribution.
    For the left panel, we keep $\bar{n}$ fixed, whereas on the right $\sigma_{\rm eq}$ is held fixed. We show linear-theory power spectra at $y=a/a_{\rm eq}=100$ evaluated using the results of this work. The filled dots indicate the Jeans scales $k_{\rm J}$ at this time, whereas the stars are the Jeans scales $k_{\rm J,eq}$ at $y=1$. The open circles indicate the free-streaming scale $k_{\rm fs}$. The dotted curve is the ``usual" cold dark matter (CDM) without any significant velocity dispersion or white noise $(\bar{n},\sigma_{\rm eq})\rightarrow(\infty,0)$. Existing observations in the quasi-linear regime only permit major deviations from the dotted curve for $k\gtrsim 10\,\Mpc^{-1}$.}
    \label{fig:PlotParamPS}
\end{figure}
%~~~~~~~~~~~~~~~~~~~~~~~

The final result can be written as follows:
\beq
\label{eq:MainResultPS}
P_{\delta}(y,k)&=\underbrace{P_{\delta_{\rm ad}}(y_0,k)\left[\mathcal{T}_k^{\rm ad}(y,y_0)\right]^2}_{\textrm{adiabatic IC + evolution}}+\underbrace{\frac{1}{\bar{n}}\left[1+3\int_{y_0}^y \frac{dy'}{\sqrt{1+y'}}\mathcal{T}^{(a)}_k(y,y')\mathcal{T}^{(b)}_k(y,y')\right]}_{\textrm{white noise IC + evolution}},
\eeq
where $y=a/a_{\rm eq}$ and $y_0$ is at an initial time when all wavenumber-$k$ modes of interest are subhorizon and the ``particles" of interest are non-relativistic; the initial conditions (IC) are specified at that time. Here, $\bar{n}$ is the co-moving number density of ``particles". The three different $\mathcal{T}_k^{\textrm{ad},(a),(b)}$ in the above expressions are growth functions, which describe how density perturbations evolve due to gravitational clustering and free streaming.
%capture the evolution of the power spectrum in the presence of gravitational clustering and free streaming, and 
They are given by:
\beq
\mathcal{T}^{\rm ad}_k(y,y_0)&=\mathcal{T}^{(a)}_k(y,y_0) + \frac{1}{2}\frac{d\ln P_{\delta_{\rm ad}}(y_0,k)}{d\ln y_0}\sqrt{1+y_0}\,\mathcal{T}^{(b)}_k(y,y_0),\\
\mathcal{T}^{(a)}_k(y,y')&=T_{\rm fs}(y,y',k)+\frac{3}{2}\int_{y'}^y \frac{dy''}{\sqrt{1+y''}}\mathcal{T}^{(b)}_k(y,y'')T_{\rm fs}(y'',y',k),\\
%\mathcal{T}^{(b)}_k(y,y')&=\mathcal{F}(y,y')T_{\rm fs}(y,y',k)+\frac{3}{2}\int_{y'}^y \frac{dy''}{\sqrt{1+y''}}\mathcal{T}^{(b)}_k(y'',y')\mathcal{F}(y,y'')T_{\rm fs}(y,y'',k).\\
\mathcal{T}^{(b)}_k(y,y')&=\mathcal{F}(y,y')T_{\rm fs}(y,y',k)+\frac{3}{2}\int_{y'}^y \frac{dy''}{\sqrt{1+y''}}\mathcal{T}^{(b)}_k(y,y'')\mathcal{F}(y'',y')T_{\rm fs}(y'',y',k).\\
\eeq
Here, $\mathcal{F}(y,y')=\ln\left[(y/y')(1+\sqrt{1+y'})^2/(1+\sqrt{1+y})^2\right]$ captures the functional dependence of the co-moving distance traveled by non-relativistic particles during the time interval between $y'$ and $y$. These equations also rely on the free-streaming ``transfer function", $T_{\rm fs}$, which can be calculated based on an (isotropic) initial phase space distribution function $f_0(p)$ as
\beq
T_{\rm fs}(y,y',k)&=4\pi\int dp\, p^2 f_0(p)\,\textrm{sinc}\left[\frac{p}{m a_{\rm eq}}\frac{k}{k_{\rm eq}}\sqrt{2}\mathcal{F}(y,y')\right],
\eeq
where $m$ is the particle mass, and $\textrm{sinc}(x)=\sin(x)/x$.

The only equation that is nontrivial to solve is the Volterra integral equation for $\mathcal{T}^{(b)}_k$. We provide an efficient numerical algorithm for evaluating the $\mathcal{T}_k^{\textrm{ad},(a),(b)}$ and the power spectrum $P_\delta$ at \url{https://github.com/delos/warm-structure-growth} \cite{delos_2025_17064722}.

Our result for the scale-dependent evolution of the power spectrum exhibits the following key features:
\begin{enumerate}
    \item A free-streaming cutoff in the adiabatic spectrum, arising due to erasure of existing correlations (such as initial adiabatic power) below the free-streaming length.
    \item Suppression of gravitational growth of the initial Poisson noise below the Jeans length.
\end{enumerate}
The free-streaming length is the distance traveled by dark matter particles up to a given time, whereas the Jeans length is essentially the distance they travel over a Hubble time. Figure~\ref{fig:PlotParamPS} shows examples of the total power spectrum that can emerge after the two aforementioned effects.
Depending on the co-moving number density $\bar n$ and the velocity dispersion of the dark matter, the dimensionless power spectrum can begin to dip at $k\gtrsim k_{\rm fs}$ due to the free-streaming suppression before rising again due to the white noise (an effect previously noticed by refs.~\cite{Amin:2022nlh,Ling:2024qfv}). Due to the Jeans suppression, the white-noise component of the dimensionless power spectrum shallows for $k\gtrsim k_{\rm J,eq}$ (the Jeans wavenumber at matter-radiation equality). This characteristic behavior is a consequence of delayed growth at small scales: modes with $k>k_{\rm J,eq}$ do not begin to grow around matter-radiation equality, and their growth only begins when the Jeans wavenumber $k_{\rm J}=k_{\rm J,eq}\sqrt{a/\aeq}$ eventually rises to be of order $k$.\footnote{These scalings imply that the power spectrum for $k_{\rm J,eq}\ll k\ll k_{\rm J}$ would approach $P_\delta(k)\propto k^{-4}$ or [$k^3/(2\pi^2)]P_\delta(k)\propto k^{-1}$. In practice, the matter-dominated epoch is not long enough for this asymptotic scaling to emerge cleanly.} 

Section~\ref{sec:result} further details the time evolution of the adiabatic and white-noise contributions to the power spectrum (see Figs. \ref{fig:PwEvolution} and~\ref{fig:PbothEvolutionLateApprox}).
We also show in section~\ref{sec:N-Compare} that the power spectrum we derive matches $N$-body simulations during radiation and matter domination as long fractional perturbations are small (see Figs.~\ref{fig:plotNumCompare} and \ref{fig:plotBothNumCompare}), and moreover, that when fractional perturbations are large, it provides a critical input to accurately calculate the abundance of dark matter halos (see Figs.~\ref{fig:plotMassFuncCompare} and \ref{fig:plotMassFuncCompareIsoAd}).

On a technical note, we begin with the full BBGKY hierarchy which results from the Liouville equation (see for example \cite{Binney:1987}). The tractability of our calculation relies on assuming that the connected part (in the sense of the Mayer Cluster expansion \cite{mayer1948statistical}) of the 3-particle {\df} is negligible, and that the connected 2-particle {\df} is small compared to the disconnected one. The former allows us to truncate the BBGKY hierarchy, and the latter reduces the problem to a single first-order (in time) partial differential equation for the 2-particle {\df}. The adiabatic part of the power spectrum evolves via the homogeneous solution of this equation in Fourier space. The inhomogeneous solution yields the evolution of white noise.

%%%%%%%%%%%%%%%%%%%%%%%%%%
\section{Derivation: from Liouville Equation to Power Spectrum Evolution}
\label{sec:derivation}
%%%%%%%%%%%%%%%%%%%%%%%%%%
For the sake of simplicity, we initially ignore cosmic expansion in the derivation. It is re-instated in the final results by appropriate transformation from physical to co-moving co-ordinates and a motivated choice of time variable.

Consider an ensemble of ``systems", each having $N$ identical particles\footnote{See ref.~\cite{Amin:2025ayf} for a derivation with non-identical particles.} confined to a volume $V$. Each particle has the mass $m$. The position and momentum of the $i$th particle are $\bx_i$ and $\bp_i=m\dot{\bx}_i$. The particles interact via Newtonian gravity, so the force acting on the $i$th particle is
\beq
\label{eq:Newton}
%\dot{\bx}_i&=\bp_i/m,\\
\dot{\bp}_i&=-m\sum_{j=1,j\ne i}^N\nabla_{\bx_i}\Psi(|\bx_i-\bx_j|)\,,\quad\textrm{where}\quad
%\Psi(t,\bx_i)=-Gm \sum_{j=1,j\ne i}^N \frac{1}{|\bx_i-\bx_j|}.
\Psi(r)=-\frac{Gm}{r}.
\eeq
For each system, the mass density at any given time is $\rho(t,\bx)=m\sum_{i=1}^N \delta_D(\bx-\bx_i)$, with a Fourier transform $\rho_\bk(t)=m\sum_{i=1}^N e^{-i\bk\cdot\bx_i}$. Our main goal is understanding the time-evolution of the two-point correlation function $\langle \rho_\bk(t)\rho_{\bk'}(t)\rangle$, where $\langle \hdots\rangle$ indicate an ensemble average. This correlation function can be written as a sum of a ``diagonal" and ``off-diagonal" piece:
\beq
\frac{1}{m^2}\langle \rho_\bk(t)\rho_{\bk'}(t)\rangle
&=\sum_{i=1}^N\langle  e^{-i(\bk+\bk')\cdot\bx_i}\rangle+\sum_{i=1}^N\sum_{j=1,j\ne i}^N\langle  e^{-i(\bk\cdot\bx_i+\bk'\cdot\bx_j)}\rangle.
\eeq
Here $\langle A \rangle \equiv \int (\prod_{i=1}^N d\bx_i d\bp_i)\,F(t,\bx_1,\hdots\bx_N,\bp_1,\hdots\bp_N) A(t,\bx_1,\hdots\bx_N,\bp_1,\hdots\bp_N)$. Here, 
\newline $F(t,\bx_1,\hdots\bx_N,\bp_1,\hdots\bp_N)\prod_{i=1}^N d\bx_i d\bp_i$ is the probability of a system being in a state so that particle $1$ is within the phase space volume $d\bx_1d\bp_1$ around the point $(\bx_1,\bp_1)$, particle $2$ is within the phase space volume $d\bx_2d\bp_2$ around the point $(\bx_2,\bp_2)$ and so on. Since no particle is special,
\beq
\label{eq:rhocorr_indistinguishable}
\frac{1}{m^2}\langle \rho_\bk(t)\rho_{\bk'}(t)\rangle&=N\langle  e^{-i(\bk+\bk')\cdot\bx_1}\rangle+{N(N-1)}\langle  e^{-i(\bk\cdot\bx_1+\bk'\cdot\bx_2)}\rangle.\\
%&=\frac{N}{V}\int d\bx_1 P(t,\bx_1)e^{-i(\bk+\bk')\cdot\bx_1}+\frac{N(N-1)}{V}\int d\bx_1 d\bx_2 P(t,\bx_1,\bx_2) e^{-i(\bk\cdot\bx_1+\bk'\cdot\bx_2)},\\
\eeq
So, to evaluate the expectation values, we only need the marginalized probability densities:
\beq
P(t,\bx_1)&=\!\!\int\!\! d\bp_1\!\!\left(\prod_{i=2}^N d\bx_i d\bp_i\!\right)\!F(t,\bx_1,\hdots\bx_N,\bp_1,\hdots\bp_N)\equiv\!\! \int\!\! d\bp_1 \frac{f^{(1)}(t,\bx_1,\bp_1)}{V},\\
P(t,\bx_1,\bx_2)&=\!\!\int\!\! d\bp_1d\bp_2\!\!\left(\prod_{i=3}^N d\bx_i d\bp_i\!\right)\!F(t,\bx_1,\hdots\bx_N,\bp_1,\hdots\bp_N)
\\&\equiv\!\! \int\!\! d\bp_1 d\bp_2 \frac{f^{(2)}(t,\bx_1,\bx_2,\bp_1,\bp_2)}{V^2},
\eeq
where we defined the 1 and 2 particle phase-space distribution functions $f^{(1)}$ and $f^{(2)}$. These functions (or more specifically, the spatial Fourier transform) are what we need to find the two point correlation function. While it might seem that we would need the solution of the $6N$ dimensional $F$ to calculate $f^{(1)}$ and $f^{(2)}$, it turns out that under certain approximations which hold for our system of interest, a closed set of coupled equations for $f^{(1)}$ and $f^{(2)}$ can be derived. In the following sections, we will begin with the equations for the $6N$-dimensional $F$. We will then arrive at a coupled set of equations for $f^{(1)}$ and $f^{(2)}$. Indeed, it turns out that we only need the time evolution of $f^{(2)}$. 

To see some of this simplification, note that under the assumption of (statistical) spatial homogeneity and isotropy, $f^{(1)}(t,\bx_1,\bp_1)\rightarrow f^{(1)}(t,\bp_1)$ and $f^{(2)}(t,\bx_1,\bx_2,\bp_1,\bp_2)\rightarrow f^{(2)}(t,|\bx_1-\bx_2|,\bp_1,\bp_2)$. In this case, evaluating the ensemble averages in eq.~\eqref{eq:rhocorr_indistinguishable} yields
\beq\label{eq:rhocorr_homogeneous}
\frac{1}{\bar{\rho}^2}\langle \rho_\bk(t)\rho_{\bk'}(t)\rangle&=\frac{1}{\bar{n}}\left[1+\bar{n} \int d\bp_1 d\bp_2 f^{(2)}_k(t,\bp_1,\bp_2)\right]V\delta_{\bk,-\bk'}
\eeq
for $k>0$, where $f^{(2)}_k(t,\bp_1,\bp_2)=\int_\by e^{-i\bk\cdot \by}f^{(2)}(t,|\by|,\bp_1,\bp_2)$, and we defined $\bar{n}=N/V\approx (N-1)/V$ and used that $\bar\rho=m\bar n$. The first term is the Poisson noise. The second term, which contains the time evolution, is entirely determined by $f^{(2)}_k$. So our goal is to solve for the time evolution of this momentum integral of $f^{(2)}_k$. We will also need an initial condition $f^{(2)}_k(t_0,\bp_1,\bp_2)$. We will consider two cases: zero initial correlation, corresponding to $f^{(2)}_k(t_0,\bp_1,\bp_2)=0$ (for $k>0$), and correlations set in accordance with the initial adiabatic density power spectrum, corresponding to $\int d\bp_1 d\bp_2f^{(2)}_k(t_0,\bp_1,\bp_2)=P_{\delta_{\rm ad}}(t_0,k)$.

\subsection{Evolution of the distribution function}

\subsubsection{Liouville Equation}

The probability density $F$ over the $6N$-dimensional phase space satisfies the Liouville equation \cite{Gibbs:1884}:
\beq
\partial_t F+\sum_{i=1}^N\dot{\bx}_i\cdot \nabla_{\bx_i} F+\sum_{i=1}^N\dot{\bp}_i\cdot \nabla_{\bp_i} F=0,
\eeq
where $\dot{\bx}_i=\bp_i/m$ and $\dot{\bp}_i$ is given by eq.~\eqref{eq:Newton}.
It is convenient to define the acceleration of the $i$th particle due to the force from the $j$th one as
\beq
\label{eq:acc_ij}
\bm{a}_{ij}\equiv -\nabla_{\bx_i}\Psi(|\bx_i-\bx_j|)=-Gm\frac{\bx_i-\bx_j}{|\bx_i-\bx_j|^3},
\eeq
if $i\neq j$, and $\bm{a}_{ii}=0$. Then the total acceleration of the $i$th particle is
\beq
\frac{\dot{\bp}_i}{m}=\bm{a}_i=\sum_{j=1}^N\bm{a}_{ij},
\eeq
and the Liouville equation becomes
\beq
\partial_t F+\frac{1}{m}\sum_{i=1}^N\bp_i\cdot \nabla_{\bx_i} F+m\sum_{i=1}^N\sum_{j=1}^N\bm{a}_{ij}\cdot \nabla_{\bp_i} F=0.
\eeq
%~~~~~~~~~~~~~~~~~~~~~~~
\subsubsection{BBGKY Hierarchy}
%~~~~~~~~~~~~~~~~~~~~~~~
The Liouville equation is equivalent to the BBGKY hierarchy of equations for the reduced $s$-particle distributions:
\beq
\frac{f^{(s)}(\bx_1,\hdots,\bx_s,\bp_1,\hdots,\bp_s)}{V^s}=\int \!\!\left(\prod_{i=s+1}^{N}d\bx_id\bp_i\!\right)\!F(\bx_1,\hdots,\bx_N,\bp_1,\hdots,\bp_N).
\eeq
The details of this derivation can be found in textbooks; we follow ref.~\cite{Nicholson:1983} except that we use momentum instead of velocity.
For $s<N$, the evolution of the $s$-particle distribution depends on the $(s+1)$-particle distribution:
\beq
&\partial_t f^{(s)}+\frac{1}{m}\sum_{i=1}^s\bp_i\cdot \nabla_{\bx_i} f^{(s)}+m\sum_{i=1}^s\sum_{j=1}^s\bm{a}_{ij}\cdot \nabla_{\bp_i} f^{(s)}\\
=&-m\frac{(N-s)}{V}\sum_{i=1}^s\int d\bx_{s+1}d\bp_{s+1}\,\bm{a}_{i,s+1}\cdot\nabla_{\bp_i}f^{(s+1)}.
\eeq
In particular, for $s=1,2$, we have the following equations for the reduced distributions $f^{(1)}(t,\bx_1,\bp_1)$ and $f^{(2)}(t,\bx_1,\bx_2,\bp_1,\bp_2)$:
\beq
&\partial_t f^{(1)}+\frac{1}{m}\bp_1\cdot \nabla_{\bx_1} f^{(1)}=-m\frac{(N-1)}{V}\int d\bx_{2}d\bp_{2}\,\bm{a}_{12}\cdot\nabla_{\bp_1}f^{(2)},\label{eq:BBGKY_f1}
\eeq\beq
&\partial_t f^{(2)}+\frac{1}{m}\left(\bp_1\cdot \nabla_{\bx_1}+\bp_2\cdot \nabla_{\bx_2}\right) f^{(2)}+m\left(\bm{a}_{12}\cdot\nabla_{\bp_1}+\bm{a}_{21}\cdot\nabla_{\bp_2}\right)f^{(2)}\\
&=-m\frac{(N-2)}{V}\int d\bx_{3}d\bp_{3}\,\left(\bm{a}_{13}\cdot\nabla_{\bp_1}+\bm{a}_{23}\cdot\nabla_{\bp_2}\right)f^{(3)}.\label{eq:BBGKY_f2}
\eeq
with $f^{(3)}=f^{(3)}(t,\bx_1,\bx_2,\bx_3,\bp_1,\bp_2,\bp_3)$. Note that this is not a closed system, since an equation for $f^{(3)}$ has not been specified in terms of $f^{(2)}$ and $f^{(1)}$.
%~~~~~~~~~~~~~~~~~~~~~~~~~~~~~
\subsubsection{Mayer Cluster Expansion}
\label{sec:Mayer}
%~~~~~~~~~~~~~~~~~~~~~~~~~~~~~
Without loss of generality, we express the reduced distributions $f^{(s)}$ using the Mayer cluster expansion \cite{Nicholson:1983}:
\beq
\label{eq:MayerExp}
f^{(1)}(1)&=f(1),\\
f^{(2)}(12)&=f(1)f(2)+g(12),\\
f^{(3)}(123)&=f(1)f(2)f(3)+f(1)g(23)+f(2)g(13)+f(3)g(12)+h(123),
\eeq
where we have adopted the shorthand $f(1)=f(t,\bx_1,\bp_1)$, $g(12)=g(t,\bx_1,\bx_2,\bp_1,\bp_2)$, and so on. The function $g$ is the connected two-particle correlation function, while $h$ is the three-particle counterpart.
We assume that these correlation functions are small: $g\ll ff$, $h\ll g f$. In particular, we will approximate $h=0$, so that we have a closed set of equations for $g$ and $f$. These assumptions are expected to hold as long as fractional density perturbations are $\ll 1$.\footnote{More precisely, we neglect the influence of the $(N+1)$-particle correlation function on how the $N$-particle correlation function evolves. We will show that this is valid for $N=1$, i.e., that we can neglect how $g$ drives time evolution of $f$, as long as the system remains in the perturbative regime. We assume that this also holds for $N=2$, so that we can neglect how $h$ drives evolution in $g$. As we will see, results from this assumption match numerical simulations in the perturbative regime.} To reduce clutter, since $N\gg 1$, we set $(N-1)/V\approx (N-2)/V\approx N/V\equiv \bar{n} =\bar{\rho}/m$.

With these assumptions, and adopting the notation $d2=d\bx_2d\bp_2$ and so on, eq.~\eqref{eq:BBGKY_f1} becomes
\beq\label{eq:Mayer_f}
&\partial_t f(1)+\frac{1}{m}\bp_1\cdot \nabla_{\bx_1} f(1)+\left[\bar{\rho}\int d2\,\bm{a}_{12}f(2)\right]\!\cdot\!\nabla_{\bp_1}f(1)=-\bar{\rho}\int d2\,\bm{a}_{12}\cdot\nabla_{\bp_1} g(12).
\eeq
Now, by substituting the cluster expansion into the $f^{(2)}$ evolution equation \eqref{eq:BBGKY_f2} and using eq.~\eqref{eq:Mayer_f} to eliminate time derivatives of $f(1)$, we obtain an evolution equation for $g$:
\beq\label{eq:Mayer_g}
&\partial_t g(12)+\frac{1}{m}\left(\bp_1\cdot \nabla_{\bx_1}+\bp_2\cdot \nabla_{\bx_2}\right) g(12)+m\left(\bm{a}_{12}\cdot\nabla_{\bp_1}+\bm{a}_{21}\cdot\nabla_{\bp_2}\right)g(12)\\
&+\bar{\rho}\int d3\,\left\{\bm{a}_{13}\cdot\nabla_{\bp_1}\left[f(1)g(23)+f(3)g(12)\right]+\bm{a}_{23}\cdot\nabla_{\bp_2}\left[f(2)g(13)+f(3)g(12)\right]\right\}\\
&=-m\left(\bm{a}_{12}\cdot\nabla_{\bp_1}+\bm{a}_{21}\cdot\nabla_{\bp_2}\right)f(1)f(2).
\eeq

%~~~~~~~~~~~~~~~~~
\subsubsection{Statistical Homogeneity}
\label{sec:homog}
%~~~~~~~~~~~~~~~~~
We assume that we are dealing with systems that are statistically homogeneous and isotropic in space. That is, we assume the following forms for $f$ and $g$:

\beq
f(1)=f(t,\bp_1),\qquad g(12)=g(t,|\bx_1-\bx_2|,\bp_1,\bp_2).
\eeq
Then eq.~(\ref{eq:Mayer_f}) for $f(1)$ simplifies to 
\beq
\label{eq:f1time}
\partial_t f(t,\bp_1)=-\bar{\rho}\int d2\,\bm{a}_{12}\cdot\nabla_{\bp_1} g(12),
\eeq
where we used
\beq
\label{eq:isot}
\int d2\,\bm{a}_{12}f(2)= 0,
\eeq
which follows from statistical homogeneity.

Turning to eq.~\eqref{eq:Mayer_g} for $g(12)$, we drop a term of order $g/ff$ (in comparison with the ``source" term on the right-hand side) and use eq.~\eqref{eq:isot} to get
\beq
\label{eq:g12time0}
&\partial_t g(12)\!+\!\frac{1}{m}\left(\bp_1\!\cdot\! \nabla_{\bx_1}+\bp_2\!\cdot\! \nabla_{\bx_2}\right) g(12)\!+\!\bar{\rho}\!\!\int \!\!d3\,\left[g(23)\bm{a}_{13}\!\cdot\!\nabla_{\bp_1}f(t,\bp_1)\!+\!g(13)\bm{a}_{23}\!\cdot\!\nabla_{\bp_2}f(t,\bp_2)\right]\\
&=-m\left(\bm{a}_{12}\!\cdot\!\nabla_{\bp_1}\!+\!\bm{a}_{21}\!\cdot\!\nabla_{\bp_2}\right)f(t,\bp_1)f(t,\bp_2).
\eeq
To calculate the time-evolution of the density power spectrum, we will be interested in $g$ only; see eq.~\eqref{eq:PwnEvol1}. Since the time variation of $f$ depends on $g$ which is supposed to be ``small", at leading order we can ignore the time variation of $f$ in the equation for $g$. We will also assume that this initial $f$ is isotropic in momentum. These considerations mean that we use $f(t,\bp)\approx f(t_0, p)\equiv f_0(p)$ in eq.~\eqref{eq:g12time0}, yielding:
\beq
\label{eq:g12time}
&\partial_t g(12)+\frac{1}{m}\left(\bp_1\!\cdot\! \nabla_{\bx_1}+\bp_2\!\cdot\! \nabla_{\bx_2}\right) g(12)+\bar{\rho}\!\int\! d3\,\left[g(23)\bm{a}_{13}\!\cdot\!\nabla_{\bp_1}f_0(p_1)+g(13)\bm{a}_{23}\!\cdot\!\nabla_{\bp_2}f_0(p_2)\right]\\
&=-m\left(\bm{a}_{12}\cdot\nabla_{\bp_1}+\bm{a}_{21}\cdot\nabla_{\bp_2}\right)f_0(p_1)f_0(p_2).
\eeq
We expect our assumption that the time variation in $f$ can be ignored to eventually break down when the velocity dispersion generated by gravitational clustering becomes comparable to the input velocity dispersion.
%There is a somewhat lengthy set of manipulations which transform eq.~\eqref{eq:f1time} into the Lenard-Balescu equation \cite{Lenard:1960,Balescu:1960}, which in turn can be written as a Fokker Planck equation \cite{Landau:1936dvu}. For a pedagogical treatment, see ref.~\cite{Nicholson:1983} in the context of statistically homogeneous plasmas. Using this Fokker-Planck equation, it is easy to check that the Maxwell-Boltzmann distribution ($f_0(p)\propto e^{-p^2/(2m^2 \sigma^2)}$) leads to $\partial_t f=0$. This lends some further justification to our approximation of negligible evolution in $f$, at least for a {\df} close to Maxwellian.}

Equations and some solutions for the evolution of $f(1)$ and $g(12)$ were written down in the 1970s to understand the growth of correlations from an initially uncorrelated, statistically homogeneous and isotropic collection of point particles (see for example, \cite{Fall:1976a,Fall:1976b}, and also \cite{Gilbert:1971}). However, the general form of the solutions we obtain for the evolution of the power spectrum in radiation and matter dominated eras were not provided in these works. 
%~~~~~~~~~~~~~~~~~
\subsection{Power Spectrum Evolution}
%~~~~~~~~~~~~~~~~~
Recall that we wish to understand the evolution of the power spectrum of density perturbations:
\beq
\langle\delta_{\bk_1}(t)\delta_{\bk_2}(t)\rangle &\equiv P_{\delta}(t,k_1)\ddelta(\bk_1+\bk_2),\\
\eeq
where $\delta_{\bk}(t)=\rho_{\bk}(t)/\bar{\rho}$ (for $\bk\neq 0$) and $\ddelta(\bk_1+\bk_2)=(2\pi)^3\delta_D(\bk_1+\bk_2)\leftrightarrow V\delta_{\bk_1,-\bk_2}$. The time evolution of the power spectrum can be expressed in terms of Fourier transforms of the connected 2-point correlation function $g(12)$ as
\beq
\label{eq:PwnEvol1}
P_{\delta}(t,k_1)\ddelta(\bk_1\!+\!\bk_2)&=\frac{1}{\bar{n}}\left[1+\bar{n}\int d\bp_1 d\bp_2 \tilde g_{k_1}(t,\bp_1,\bp_2)\right]\ddelta(\bk_1\!+\!\bk_2),\\
&=\underbrace{\frac{1}{\bar{n}}\ddelta(\bk_1\!+\!\bk_2)}_{\textrm{white noise}}+\underbrace{\int d\bp_1 d\bp_2 g_{\bk_1\bk_2}(t,\bp_1,\bp_2)}_{\textrm{time evolution of correlations}},\\
&=\underbrace{\frac{1}{\bar{n}}\ddelta(\bk_1\!+\!\bk_2)+\int\! d\bp_1 d\bp_2 g_{\bk_1\bk_2}^{S\ne 0}(t,\bp_1,\bp_2)}_{P_{\delta_{\rm wn}}(t,k_1)\ddelta(\bk_1+\bk_2)}+\underbrace{\int\! d\bp_1 d\bp_2 g_{\bk_1\bk_2}^{S=0}(t,\bp_1,\bp_2)}_{P_{\delta_{\rm ad}}(t,k_1)\ddelta(\bk_1+\bk_2)}.
\eeq
In the last line, we split $g_{\bk_1\bk_2}=g_{\bk_1\bk_2}^{S\ne 0}+g_{\bk_1\bk_2}^{S=0}$ into sourced and source-free solutions, a distinction that will become clear.
In the above equations, we have assumed a statistically homogeneous and isotropic ensemble (see \eqref{eq:rhocorr_homogeneous}), where $\tilde{g}_k(t,\bp_1,\bp_2)$ is the Fourier transform of $g(12)=g(t,|\bx_1-\bx_2|,\bp_1,\bp_2)$, whereas $g_{\bk_1\bk_2}(t,\bp_1,\bp_2)$ is the Fourier transform of $g(t,\bx_1,\bx_2,\bp_1,\bp_2)$. In going from $f^{(2)}_k$ in \eqref{eq:rhocorr_homogeneous} to $\tilde{g}_k$ above, we used the Mayer cluster expansion for $f^{(2)}$ (see \eqref{eq:MayerExp}). We also assumed $k_1>0$, so that $f^{(2)}_{k_1}(t,\bp_1,\bp_2)=\tilde g_{k_1}(t,\bp_1,\bp_2)$ due to spatial homogeneity of the 1-particle DFs. The reason for writing the evolution in terms of $g_{\bk_1\bk_2}(t,\bp_1,\bp_2)\equiv \tilde{g}_{k_1}(t,\bp_1,\bp_2)\ddelta(\bk_1+\bk_2)$ is that evolution equations are symmetrical in $\bk_1$ and $\bk_2$. As seen in the third line, the total power spectrum is a sum of two contributions
\beq
P_\delta(t,k)=P_{\delta_{\rm wn}}(t,k)+P_{\delta_{\rm ad}}(t,k),
\eeq
which will be related to the sourced and source-free solutions of $g_{\bk_1,\bk_2}$ respectively.

In what follows, we will derive a formal solution for $g_{\bk_1,\bk_2}(t,\bp_1,\bp_2)$ and its momentum integral. To this end, first note that the acceleration \eqref{eq:acc_ij} can be written as
\beq\label{eq:aij_Fourier}
\bm{a}_{ij}=4\pi G m \int_\bq \frac{i\bq}{q^2} e^{i\bq\cdot(\bx_i-\bx_j)},
\eeq
where $\int_\bq =\int d\bq/(2\pi)^3=V^{-1}\sum_\bq$. 
By Fourier transforming equation~\eqref{eq:g12time} (in both position variables) and using equation~\eqref{eq:aij_Fourier}, we obtain the evolution equation for $g_{\bk_1\bk_2}$,
\beq
\label{eq:g}
&\partial_t g_{\bk_1\bk_2}(t,\bp_1,\bp_2)+\frac{i}{m}\left(\bp_1\cdot {\bk_1}+\bp_2\cdot {\bk_2}\right) g_{\bk_1\bk_2}(t,\bp_1,\bp_2)\\
&+4\pi G m\bar{\rho}\int d\bp_3\,\left[g_{\bk_2,\bk_1}(t,\bp_2,\bp_3)\frac{i\bk_1}{k_1^2} \cdot\nabla_{\bp_1}f_0(p_1)+g_{\bk_1,\bk_2}(t,\bp_1,\bp_3) \frac{i\bk_2}{k_2^2}\cdot\nabla_{\bp_2}f_0(p_2)\right]\\
&=-4\pi Gm^2 \left(\frac{i\bk_1}{k_1^2}\cdot\nabla_{\bp_1}+\frac{i\bk_2}{k_2^2}\cdot\nabla_{\bp_2}\right)f_0(p_1)f_0(p_2)\ddelta(\bk_1+\bk_2),
\eeq
where it is useful to keep in mind that $g_{\bk_1\bk_2}\propto \ddelta(\bk_1+\bk_2)$. 
This equation has the form:\footnote{The labels $(s)$ that appear below do not correspond to those in the BBGKY hierarchy, i.e. they are unrelated to $(s)$ in $f^{(s)}$.}
\beq
\label{eq:DgS}
\mathcal{D}{g}_{\bk_1\bk_2}=\underbrace{4\pi Gm(S^{(1)}_{\bk_1}+S^{(2)}_{\bk_2})\ddelta(\bk_1+\bk_2)}_{\textrm{source $S$}},
\eeq
where $\mathcal{D}$ is the integro-differential operator on the left-hand side and
\beq
S^{(1)}_{\bk_1}=-m \frac{i\bk_1}{k_1^2}\cdot\nabla_{\bp_1}f_0(p_1)f_0(p_2),
\qquad
S^{(2)}_{\bk_2}=-f_0(p_1)m \frac{i\bk_2}{k_2^2}\cdot\nabla_{\bp_2}f_0(p_2).
\eeq
Its general solution is of the form 
\beq
g_{\bk_1\bk_2}(t,\bp_1,\bp_2)=g^{S=0}_{\bk_1\bk_2}(t,\bp_1,\bp_2)+g^{S\ne0}_{\bk_1\bk_2}(t,\bp_1,\bp_2),
\eeq
where the first term is the homogeneous solution, i.e., the solution of the source-free equation $\mathcal{D}{g}_{\bk_1\bk_2}=0$ with non-vanishing initial condition $g^{S=0}_{\bk_1\bk_2}(t_0,\bp_1,\bp_2)\ne 0$. The second term is the particular solution consistent with the non-zero source term and vanishing initial condition $g^{S\ne 0}_{\bk_1\bk_2}(t_0,\bp_1,\bp_2)=0$.

We first provide a formal expression for the homogeneous solution, which will then immediately allow us to construct the evolution of the power spectrum for adiabatic initial conditions (with $g^{S=0}_{\bk_1\bk_2}(t_0,\bp_1,\bp_2)\ne 0$). We will then construct $g_{\bk_1\bk_2}^{S\ne 0}$ via a Green's function. This will allow us to write down the power spectrum evolution for white noise initial conditions (with $g^{S\ne 0}_{\bk_1\bk_2}(t_0,\bp_1,\bp_2)=0$). Putting these together will yield the complete evolution of the density power spectrum.

The source-free equation, $\mathcal{D}{g}_{\bk_1\bk_2}=0$, has solutions of the form
\beq
g_{\bk_1\bk_2}(t,\bp_1,\bp_2)=\gamma^{(i)}_{\bk_1}(t,\bp_1)\gamma^{(j)}_{\bk_2}(t,\bp_2)\ddelta(\bk_1+\bk_2),
\eeq
where the $\gamma^{(i,j)}_\bk$ are functions satisfying\footnote{The general solution to $\mathcal{D}{g}_{\bk_1\bk_2}=0$ would be an infinite sum ${g}_{\bk_1\bk_2}=\sum_{ij}\gamma^{(i)}_{\bk_1}\gamma_{\bk_2}^{(j)}\ddelta(\bk_1+\bk_2)$.}
\beq
\label{eq:gammaEq}
&\partial_t \gamma_{\bk}+\frac{i}{m}(\bk\cdot \bq) \,\gamma_{\bk}+4\pi G\bar{\rho}m\frac{i\bk}{k^2}\cdot\nabla_{\bq}f_0(q)\int d\bp\,\gamma_{\bk}(\bp)=0.
\eeq
The formal solution to eq.~\eqref{eq:gammaEq} can be written as
\beq
\label{eq:gammasol}
&\gamma_{\bk}(t,\bq)\!=\!\gamma_{\bk}(t_0,\bq)e^{-i\bk\cdot\frac{\bq}{m}(t-t_0)}\!-\!4\pi G m\bar{\rho}\frac{i\bk}{k^2}\!\cdot\!\nabla_{\bq}f_0(q)\!\!\int_{t_0}^t \!dt'' e^{-i\bk\cdot\frac{\bq}{m}(t-t'')}\!\!\int \!d\bp\,\gamma_{\bk}(t'',\bp).
\eeq
% \beq
% \label{eq:gammasol}
% &\gamma_{\bk_s}(t,t_0,\bq)\!=\!\gamma_{\bk_s}(t_0,\bq)e^{-i\bk_s\cdot\frac{\bq}{m}(t-t_0)}\!+\!4\pi G m\bar{\rho}\frac{i\bk_s}{k_s^2}\!\cdot\!\nabla_{q}f_0(q)\!\!\int_{t_0}^t \!dt' e^{-i\bk_s\cdot\frac{\bq}{m}(t-t')}\!\!\int \!d\bp\,\gamma_{\bk_s}(t',t_0,\bp).
% \eeq
An initial condition $\gamma_{\bk}(t_0,\bq)$ is needed to proceed further. 

Before proceeding, we note that equation~\eqref{eq:gammaEq} could have also been derived as the evolution equation for the perturbation to the 1-particle \df, if we had allowed the 1-particle \df\ to depend on position, as in standard approaches to cosmological perturbation theory (e.g. ref.~\cite{Brandenberger:1987}).\footnote{We emphasize a conceptual difference between our approach and standard cosmological perturbation theory. We assume in this work that the universe is statistically homogeneous, and we analyze the evolution of the statistical ensemble of realizations of a perturbed universe. In the standard approach, one fixes an (arbitrary) specific realization of the cosmological perturbations and analyzes how that realization evolves.}
That is, if the 1-particle \df\ were $f(t,\bk,\bp)=f_0(p)+\gamma_\bk(t,\bp)$, then equation~\eqref{eq:gammaEq} would follow from the collisionless Boltzmann equation for $f$ at linear order in $\gamma_\bk$. Consequently, the momentum integrals of the $\gamma_\bk$ are precisely growth functions for warm dark matter that describe how particular realizations of the density field $\delta_\bk$ evolve over time.

%~~~~~~~~~~~~~~~~~~~~~~~~~~~~~~~~~~~~~
\subsubsection{Fundamental Solutions: the Growth Functions for Warm Dark Matter}
\label{sec:growthFunctions}
%~~~~~~~~~~~~~~~~~~~~~~~~~~~~~~~~~~~~~

Consider the following two initial conditions:
\beq 
\label{eq:gammaabIC}
&(a)\qquad\gamma^{(a)}_{\bk}(t_0,\bq)=f_0(q),\\
&(b)\qquad \gamma^{(b)}_{\bk}(t_0,\bq)=-m (i\bk/k^2)\cdot \nabla_{\bq}f_0(q).
\eeq
The $\bq$-dependence of these choices makes them pure bulk perturbations to the density and velocity,\footnote{Given an overall density shift $\delta(\bk)$ and momentum shift $\bar{\bq}(\bk)$, the 1-particle {\df} becomes $f(t_0,\bk,\bq)=f_0(\bq-\bar{\bq}(\bk))+\delta(\bk)f_0(\bq)\approx f_0(\bq)+\delta(\bk)f_0(\bq)-\bar{\bq}(\bk)\cdot\nabla_\bq f_0(\bq)$ at lowest order in $\bar{\bq}(\bk)$. The $\bq$-dependence in these perturbation terms matches those of $\gamma^{(a)}_{\bk}$ and $\gamma^{(b)}_{\bk}$ in equation~\eqref{eq:gammaabIC}.} while the ($\bk$-dependent) normalizations are chosen for future convenience. The product of these initial $\gamma$ also matches the form of the source terms in eq.~\eqref{eq:g}.
With each initial condition, the solution $\gamma_\bk^{(a,b)}(t,\bq)$ in \eqref{eq:gammasol} can be integrated over $\bq$ to get $T^{(a,b)}_\bk(t,t_0)\equiv \int d\bq \gamma_\bk^{(a,b)}(t,\bq)$ which satisfy:
\beq
\label{eq:TaTb}
T_{\bk}^{(a)}(t,t_0)&=T_{\rm fs}(k(t-t_0))+4\pi G \bar{\rho}\int_{t_0}^t dt'' (t-t'')T_{\rm fs}(k(t-t'')) T_{\bk}^{(a)}(t'',t_0),\\
T_{\bk}^{(b)}(t,t_0)&=(t-t_0)T_{\rm fs}(k(t-t_0))+4\pi G \bar{\rho}\int_{t_0}^t dt'' (t-t'')T_{\rm fs}(k(t-t'')) T_{\bk}^{(b)}(t'',t_0),
\eeq
with the free-streaming effects being represented via:
\beq
T_{\rm fs}(k(t-t''))
=\int d\bp f_0(p) e^{-i\bk\cdot \frac{\bp}{m}(t-t'')}
=4\pi \int dp\, p^2f_0(p)\,{\rm sinc}\!\left[\frac{p}{m}k(t\!-\!t'')\right],
\eeq
where ${\rm sinc}(x)=\sin(x)/x$.
We will see shortly that an initial density perturbation (with zero bulk velocity) evolves in proportion with $T_\bk^{(a)}$, while a pure initial velocity perturbation evolves in proportion with $T_\bk^{(b)}$. That is, the $T_\bk^{(a,b)}$ are growth functions for warm dark matter. They will form the building blocks for the power spectrum evolution.

Equations~\eqref{eq:TaTb} are Volterra integral equations, which must be solved numerically in general. We discuss their structure in Appendix~\ref{sec:Diagram}, and we show that these equations can be put into the form:
\beq\label{eq:TaTbAlt}
T^{(a)}_\bk(t,t_0)&=T_{\rm fs}(k(t-t_0))+4\pi G \bar{\rho}\int_{t_0}^t dt'' T^{(b)}_\bk(t,t'')T_{\rm fs}(k(t''-t_0)),
\\
T^{(b)}_\bk(t,t_0)&=(t-t_0)T_{\rm fs}(k(t-t_0))+4\pi G \bar{\rho}\int_{t_0}^t dt'' T^{(b)}_\bk(t,t'')(t''-t_0)T_{\rm fs}(k(t''-t_0)).
\eeq
That is, $T^{(a)}_k$ reduces to an integral over $T^{(b)}_k$, and we only need to solve one Volterra equation for $T^{(b)}_k$.

Conveniently, due to how the $\gamma^{(a,b)}_\bk$ were normalized, $T_{\bk}^{(a)}(t_0,t_0)=1$ and $T_{\bk}^{(b)}(t_0,t_0)=0$, and their initial time derivatives are $0$ and $1$, respectively. It is also useful to note that the growth functions are isotropic: $T_{\bk}^{(a,b)}(t,t_0)=T_{k}^{(a,b)}(t,t_0)$.

%~~~~~~~~~~~~~~~~~~~~~~~~~~~~~~~~~~~~~
\subsubsection{Evolution from Adiabatic Initial Conditions}
%~~~~~~~~~~~~~~~~~~~~~~~~~~~~~~~~~~~~~

We now consider the source-free evolution of initial correlations, with the aim of evaluating the time evolution of the adiabatic part of the power spectrum,
\beq
\label{eq:PdeltaS=0}
P_{\delta_{\rm ad}}(t,k_1)\ddelta(\bk_1+\bk_2)
&=\int d\bp_1d\bp_2g^{S=0}_{\bk_1\bk_2}(t,\bp_1,\bp_2).
\eeq
Although $g^{S=0}_{\bk_1\bk_2}$ can in general involve a sum over products of solutions to eq.~\eqref{eq:gammaEq}, its freedom is reduced if we assume that, at each $\bk$, there are no initial correlations in momentum space. Then the correlation function factorizes as
\beq
\label{eq:gS=0}
g^{S=0}_{\bk_1\bk_2}(t,\bp_1,\bp_2)
&=\gamma_{\bk_1}(t,\bp_1)\gamma_{\bk_2}(t,\bp_2)\ddelta(\bk_1+\bk_2)
\eeq
for some $\gamma_\bk$ satisfying eq.~\eqref{eq:gammaEq}. The particular $\gamma_\bk$ is determined when we specialize to adiabatic initial conditions, which represent overall bulk perturbations to the density and velocity.\footnote{However, the initially adiabatic perturbations are bulk density and velocity perturbations only to the extent that they have not yet been affected significantly by free streaming of particles. Thus, for the $\gamma_\bk$ in equation~\eqref{eq:gamma_ad} to be appropriate, in principle the initial time $t_0$ must be either the time that the perturbations are first sourced or the time that (non-relativistic) free streaming begins, whichever is later. See Section~\ref{sec:relativistic} for further discussion.} Then it must be a simple linear combination of the $\gamma_\bk^{(a,b)}$ given in eq.~\eqref{eq:gammaabIC}. Given an initial adiabatic power spectrum $P_{\delta_{\rm ad}}(t_0,k)$ and its time derivative $dP_{\delta_{\rm ad}}(t_0,k)/dt_0$ at the time $t_0$, the normalizations of the $\gamma_\bk^{(a,b)}$ require the particular combination
\beq
\label{eq:gamma_ad}
\gamma_{\bk}(t,\bp)
=\sqrt{P_{\delta_{\rm ad}}(t_0,k)}\gamma_\bk^{(a)}(t,\bp)
+\frac{d\sqrt{P_{\delta_{\rm ad}}(t_0,k)}}{dt_0}\gamma_\bk^{(b)}(t,\bp).
\eeq
Using this solution in equations \eqref{eq:gS=0} and~\eqref{eq:PdeltaS=0}, we obtain
\beq
\label{eq:Pad(t)}
P_{\delta_{\rm ad}}(t,k)&=P_{{\delta}_{\rm ad}}(t_0,k)\left[T^{\rm ad}_{k}(t,t_0)\right]^2,\\
T^{\rm ad}_k(t,t_0)&=T_k^{(a)}(t,t_0)+\frac{d\ln \sqrt{P_{\delta_{\rm ad}}(t_0,k)}}{dt_0}T_k^{(b)}(t,t_0),
\eeq
with the $T^{(a,b)}_\bk$ given by equations~\eqref{eq:TaTbAlt}.

There is nothing new here, and this is the usual solution expected for warm dark matter with some initial density and bulk velocity perturbations. At early times during radiation domination, one can ignore the gravitational terms proportional to $G$ in the equations for $T_k^{(a,b)}$ (see \eqref{eq:TaTb}) and hence also in $T_k^{\rm ad}$. Then $T_{\rm fs}$ leads to erasure of correlated fluctuations below of a free-streaming length via $P_{\delta_{\rm ad}}(t,k)\sim P_{\delta_{\rm ad}}(t_0,k)T_{\rm fs}^2(k(t-t_0))$. At late times during matter domination, the $G$ terms which correspond to gravitational clustering can dominate. In this case, $T_{\rm fs}$ prevents the power spectrum from growing significantly below the Jeans length.

%~~~~~~~~~~~~~~~~~~~~~~~~~~~~~~~
\subsubsection{Evolution of White Noise}
%~~~~~~~~~~~~~~~~~~~~~~~~~~~~~~~

We now focus on the evolution of the power spectrum under the ongoing influence of the sourcing terms in equation~\eqref{eq:g} when there are no initial correlations in position space, so the initial power spectrum is white noise. Then we can write 
\beq
P_{\delta_{\rm wn}}(t,k_1)\ddelta(\bk_1+\bk_2)
&=\frac{1}{\bar{n}}\ddelta(\bk_1+\bk_2)+\int d\bp_1 d\bp_2 g^{S\ne 0}_{\bk_1\bk_2}(t,\bp_1,\bp_2)\,,\\
&=\frac{1}{\bar{n}}\ddelta(\bk_1+\bk_2)+\int d\bp_1 d\bp_2 \left[g^{(1)}_{\bk_1\bk_2}(t,\bp_1,\bp_2)+g^{(2)}_{\bk_1\bk_2}(t,\bp_1,\bp_2)\right]\,,
\eeq
with the initial condition $g^{S\ne 0}_{\bk_1\bk_2}(t_0,\bp_1,\bp_2)=0$. Here, $\mathcal{D}g^{S\ne 0}_{\bk_1\bk_2}=\mathcal{D}\left[g^{(1)}_{\bk_1\bk_2}+g^{(2)}_{\bk_1\bk_2}\right]=4\pi Gm \left[S^{(1)}_{\bk_1}+S^{(2)}_{\bk_2}\right]\ddelta(\bk_1+\bk_2)$ (cf. eq.~\eqref{eq:DgS}).

We can solve for $g^{(1)}_{\bk_1\bk_2}$ from:
\beq
\mathcal{D}g^{(1)}_{\bk_1\bk_2}=4\pi Gm S^{(1)}_{\bk_1}\ddelta(\bk_1+\bk_2).
\eeq 
The solution can be written as
\beq
g^{(1)}_{\bk_1\bk_2}(t,\bp_1,\bp_2)=4\pi Gm\int_{t_0}^t dt' \mathcal{G}^{(1)}_{\bk_1,\bk_2}(t,t',\bp_1,\bp_2)\ddelta(\bk_1+\bk_2),%-\int_{t_0}^t dt' \mathcal{G}^{(2)}_k(t,t',\bp_1,\bp_2),
\eeq
where $\mathcal{G}^{(1)}_{\bk_1\bk_2}$ is a Green's function:
\beq
&\mathcal{G}^{(1)}_{\bk_1\bk_2}(t,t',\bp_1,\bp_2)=\gamma_{\bk_1}^{(i)}(t,t',\bp_1)\gamma_{\bk_2}^{(j)}(t,t',\bp_2)\Theta(t-t')
\eeq
for some $\gamma_{\bk_s}^{(i,j)}(t,t',\bp_s)$ given by eq.~\eqref{eq:gammasol} with $t_0\rightarrow t'$. The initial conditions for $\gamma_{\bk_s}^{(i,j)}$ needed to specify the Green's function can be determined from the defining equation for the Green's function:
\beq
&\mathcal{D}\mathcal{G}^{(1)}_{\bk_1\bk_2}(t,t',\bp_1,\bp_2)=\delta_D(t-t')S^{(1)}_{\bk_1}(\bp_1,\bp_2).
\eeq
Note that integrating this equation for $\mathcal{G}^{(1)}_{\bk_1\bk_2}$ across $t=t'$, we have $\mathcal{G}^{(1)}_{\bk}(t'+\epsilon,t',\bp_1,\bp_2)=S^{(1)}_{\bk_1}(\bp_1,\bp_2)$ for arbitrarily small $\epsilon>0$. This can be satisfied by choosing $\gamma_{\bk_1}^{(i)}(t'+\epsilon,\bp_1)=-m (i\bk_1/k_1^2)\cdot\nabla_{\bp_1}f_0(p_1)$ and $\gamma_{\bk_2}^{(j)}(t'+\epsilon,\bp_2)=f_0(p_2)$, which are the same as $\eqref{eq:gammaabIC}$ with $t_0\rightarrow t'+\epsilon$. So the general solution with these ``initial" conditions is
\beq
{g}^{(1)}_{\bk_1\bk_2}(t,t_0,\bp_1,\bp_2)=4\pi Gm \int_{t_0}^t dt'\gamma^{(b)}_{\bk_1}(t,t',\bp_1)\gamma^{(a)}_{\bk_2}(t,t',\bp_2)\ddelta(\bk_1+\bk_2),
\eeq
Integrating both sides over $\bp_1$ and $\bp_2$, we have
\beq
\int d\bp_1 d\bp_2\,{g}^{(1)}_{\bk_1\bk_2}(t,t_0,\bp_1,\bp_2)=4\pi Gm\int_{t_0}^t dt'T_{k_1}^{(b)}(t,t')T^{(a)}_{k_2}(t,t')\ddelta(\bk_1+\bk_2),
\eeq
where $T_{\bk_s}^{(a,b)}(t,t')$ are given by \eqref{eq:TaTb} or \eqref{eq:TaTbAlt} with $t_0\rightarrow t'$.
% \beq
% T_{\bk_1}^{(a)}(t,t')=\int d\bp\, \gamma^{(1)}_{\bk_1}(t,t',\bp),\qquad T_{\bk_2}^{(b)}(t,t')=\int d\bp\, \gamma^{(1)}_{\bk_2}(t,t',\bp),%=\int d\bp\, \gamma^{(1)}_{-\bk}(t,t',-\bp)
% \eeq
% and they satisfy:
% %\footnote{ If we had a Yukawa force, then the only change necessary would be(check!):
% % \beq
% % T_{\bk}^{(a)}(t,t_0)&=T_{\rm fs}(k(t-t_0))+4\pi G \bar{\rho}\frac{k^2}{k^2+\mu^2}\int_{t_0}^t dt' (t-t') T_{\bk}^{(a)}(t',t_0) T_{\rm fs}(k(t-t')),\\
% % T_{\bk}^{(b)}(t,t_0)&=(t-t_0)\frac{k^2}{k^2+\mu^2}T_{\rm fs}(k(t-t_0))+4\pi G \bar{\rho}\frac{k^2}{k^2+\mu^2}\int_{t_0}^t dt' (t-t') T_{\bk}^{(b)}(t',t_0)T_{\rm fs}(k(t-t')).
% % \eeq
% % }
% \beq
% T_{\bk}^{(a)}(t,t_0)&=T_{\rm fs}(k(t-t_0))+4\pi G \bar{\rho}\int_{t_0}^t dt' (t-t') T_{\bk}^{(a)}(t',t_0) T_{\rm fs}(k(t-t')),\\
% T_{\bk}^{(b)}(t,t_0)&=(t-t_0)T_{\rm fs}(k(t-t_0))+4\pi G \bar{\rho}\int_{t_0}^t dt' (t-t') T_{\bk}^{(b)}(t',t_0)T_{\rm fs}(k(t-t')).
% \eeq
% The free-streaming effects are captured by:
% \beq
% T_{\rm fs}(k(t-t'))=\int d\bp f_0(p) e^{-i\bk\cdot \frac{\bp}{m}(t-t')}.
% \eeq

We repeat this exercise for $\mathcal{D}g^{(2)}_{\bk_1 \bk_2}=4\pi Gm S^{(2)}_{\bk_2}\ddelta(\bk_1+\bk_2)$ and find that $\int d\bp_1 d\bp_2\,{g}^{(2)}_{\bk_1\bk_2}=\int d\bp_1 d\bp_2\,{g}^{(1)}_{\bk_1\bk_2}$ after using $\ddelta(\bk_1+\bk_2)$ to interchange $\bk_1$ and $\bk_2$. Adding up the contributions from $g^{(1,2)}_{\bk_2\bk_2}$, we have
\beq
\int d\bp_1 d\bp_2 \,g^{S\ne 0}_{\bk_1\bk_2}(t,\bp_1,\bp_2)
%&=\int d\bp_1 d\bp_2 \,g^{(1)}_{\bk_1\bk_2}(t,\bp_1,\bp_2)+\int d\bp_1 d\bp_2 \,g^{(2)}_{\bk_1\bk_2}(t,\bp_1,\bp_2),\\
% &=4\pi Gm\int_{t_0}^t dt' \left[T^{(a)}_{\bk_1}(t,t')T^{(b)}_{\bk_2}(t,t')+T^{(b)}_{\bk_1}(t,t')T^{(a)}_{\bk_2}(t,t')\right]\ddelta(\bk_1+\bk_2).
% \eeq
% Note that $T^{(a)}_\bk=T^{(a)}_k$, and similarly for $T^{(b)}_\bk=T^{(b)}_k$. Hence,
% \beq
% \int d\bp_1 d\bp_2 \,g_{\bk_1\bk_2}(t,\bp_1,\bp_2)
=8\pi Gm\int_{t_0}^t dt' T^{(a)}_{k_1}(t,t')T^{(b)}_{k_1}(t,t')\ddelta(\bk_1+\bk_2).
\eeq
Thus, the power spectrum when we start with no existing correlations is given by
% \beq
% \frac{1}{\bar{\rho}^2}\langle \rho_{\bk_1}(t) \rho_{\bk_2}(t)\rangle = \frac{1}{\bar{n}}\left[1+8\pi Gm\bar{n}\int_{t_0}^t dt' T^{(a)}_{k_1}(t,t')T^{(b)}_{k_1}(t,t')\right]\ddelta(\bk_1+\bk_2). 
% \eeq
% Or simply,
\beq
\label{eq:Pwn_t}
P_{\delta_{\rm wn}}(t,k)=\frac{1}{\bar{n}}\left[1+8\pi G\bar{\rho}\int_{t_0}^t dt' T^{(a)}_{k}(t,t')T^{(b)}_{k}(t,t')\right].
\eeq
This is one of the main new results of our work. The $1/\bar{n}$ is the initial white noise, while the term proportional to $G$ is the evolution of white noise. We show the diagrammatic interpretation of this expression in Appendix~\ref{sec:Diagram}.

Heuristically, at early times close to $t_0$, the evolution term is $G\bar{\rho}\int dt' (t-t')T_{\rm fs}^2(k(t-t'))$ capturing the initial growth of correlations. Unlike the white noise $1/\bar{n}$, this growth is subject to free streaming effects as is evident from $T_{\rm fs}^2$ inside the integral. At late times, this term will capture growth above, and lack thereof below a Jeans length.

On a technical note, notice that the above power spectrum $P_{\delta_{\rm wn}}$ requires an integral in time over the product of $T^{(a)}_kT^{(b)}_k$. In contrast, $P_{\delta_{\rm ad}}\propto (T^{(a)}_k+c T^{(b)}_k)^2$ with no additional time integral.
Conceptually, $P_{\delta_{\rm wn}}$ is built from density and velocity perturbations sourced at all times, whereas $P_{\delta_{\rm ad}}$ is built from such perturbations sourced only at an initial time.

%~~~~~~~~~~~~~~~~~~~~~~~~~~~~~~~~~~
\subsection{Power Spectrum Evolution in an Expanding Universe}
%~~~~~~~~~~~~~~~~~~~~~~~~~~~~~~~~~~

To convert the previous results to the case of an expanding universe, we express positions, momenta, wavenumbers and densities in terms of their co-moving counterparts $\{\bx,\bk,\bp\}\rightarrow \{a\bx, \bk/a,\bp/a\}$ and $\{\bar{n},\bar{\rho},f_0(q)\}\rightarrow \{a^{-3}\bar{n},  a^{-3}\bar{\rho},  a^3 f_0(q)\}$, where the variables after the arrows are co-moving quantities.  We make a further change to a new time variable $\eta$ which appears naturally when dealing with non-relativistic motion in an expanding universe:
\beq
d\eta=\frac{dt}{a^2(t)}.
\eeq
In particular, the co-moving distance traveled by a particle with peculiar (physical) velocity $\bv(t)=\bp/(ma(t))$ is given by
\beq
l_{\rm fs}(\bp,t)=\int_{t_0}^t \bv(t') \frac{dt'}{a(t')}=\frac{\bp}{m}\int_{t_0}^t \frac{dt'}{a^2(t')}=\frac{\bp}{m}(\eta-\eta_0).
\eeq
This is the scale that naturally enters the free-streaming transfer function, which becomes
\beq
%T_{\rm fs}(k(\eta-\eta'))= \int d\bp f_0(p)e^{-i\bk\cdot\frac{\bp}{m}(\eta-\eta')} = 4\pi \int dp\, p^2f_0(p)\,{\rm sinc}\left[\frac{p}{m}k(\eta-\eta')\right],
T_{\rm fs}(k(\eta-\eta'))= \int d\bp f_0(p)e^{-i\bk\cdot\frac{\bp}{m}(\eta-\eta')} = 4\pi \int dp\, p^2f_0(p)\,{\rm sinc}\!\left[\frac{p}{m}k(\eta\!-\!\eta')\right],
\eeq
where ${\rm sinc}(x)=\sin(x)/x$. As another example, eq.~\eqref{eq:gammaEq} becomes
\beq
\label{eq:gammaEq_eta}
%a^{-2}\partial_\eta \gamma_{\bk}+ a^{-2}\frac{i}{m}(\bk\cdot \bq) \,\gamma_{\bk}-4\pi G[\bar{\rho} a^{-3}]m [a^2\frac{i\bk}{k^2}\cdot\nabla_{q}][a^3f_0(q)]\int [a^{-3} d\bp]\,\gamma_{\bk}(\bp)=0\\
\partial_\eta \gamma_{\bk}+\frac{i}{m}(\bk\cdot \bq) \,\gamma_{\bk}-4\pi G\bar{\rho}a(\eta)m\frac{i\bk}{k^2}\cdot\nabla_{\bq}f_0(q)\int d\bp\,\gamma_{\bk}(\bp)=0.
\eeq
This suggests a quick replacement in the equations~\eqref{eq:TaTbAlt} for the growth functions $T^{(a,b)}$: 
\beq
t\rightarrow \eta',\qquad 4\pi G\bar{\rho}\,dt\rightarrow 4\pi G\bar{\rho}a(\eta)\,d\eta= (3/4)H^2(a_{\rm eq})\aeq^3 a(\eta)d\eta.
\eeq
Using this replacement we can find the corresponding expressions for the power spectrum evolution (eqs. \eqref{eq:Pad(t)} and \eqref{eq:Pwn_t}) in an expanding universe in terms of comoving quantities.
%\beq
%P_\delta (\eta,k)&=P_{{\delta}_{\rm ad}}(\eta_0,k)\left[T^{\rm ad}_{k}(\eta,\eta_0)\right]^2+\frac{1}{\bar{n}}\left[1+\frac{3}{2}H_{\rm eq}^2\aeq^3\int_{\eta_0}^\eta d\eta' a(\eta')T^{(a)}_{k}(\eta,\eta')T^{(b)}_{k}(\eta,\eta')\right],\\
%T^{\rm ad}_k(\eta,\eta_0)&=T_k^{(a)}(\eta,\eta_0)+\frac{1}{2}\frac{d}{d\eta_0}\left[\ln P_{\delta_{\rm ad}}(\eta_0,k)\right]\,T^{(b)}_k(\eta,\eta_0),\\
%T_{k}^{(a)}(\eta,\eta')&=T_{\rm fs}(k(\eta-\eta'))+\frac{3}{4}H_{\rm eq}^2\aeq^3\int_{\eta'}^\eta d\eta'' a(\eta'')  T_{k}^{(b)}(\eta,\eta'')T_{\rm fs}(k(\eta''-\eta')),\\
%T_{k}^{(b)}(\eta,\eta')&=(\eta-\eta')T_{\rm fs}(k(\eta-\eta'))+\frac{3}{4}H_{\rm eq}^2\aeq^3\int_{\eta'}^\eta d\eta''  a(\eta'') T_{k}^{(b)}(\eta,\eta'')(\eta''-\eta')T_{\rm fs}(k(\eta''-\eta')).
%\eeq

%~~~~~~~~~~~~~~~~~~~~~~~~~~~~~~~~~~
\subsubsection{The Main Result}
\label{sec:result}
%~~~~~~~~~~~~~~~~~~~~~~~~~~~~~~~~~~

It is convenient to write our results in terms of dimensionless variables, 
\beq
y=a/a_{\rm eq},
\qquad \eta-\eta'=\frac{\sqrt{2}}{a_{\rm eq}k_{\rm eq}}\mathcal{F}(y,y')\quad\textrm{with}\quad \mathcal{F}(y,y')=\ln\left[\frac{y}{y'}\left(\frac{1+\sqrt{1+y'}}{1+\sqrt{1+y}}\right)^2\right],
\eeq
where we plan to use $y=a/a_{\rm eq}$ as the time variable. Then the white-noise part becomes
\beq
\label{eq:PwEvolution}
P_{\delta_{\rm wn}}(y,k)=
\frac{1}{\bar{n}}\left[1+3\int_{y_0}^y \frac{dy'}{\sqrt{1+y'}}\mathcal{T}^{(a)}_k(y,y')\mathcal{T}^{(b)}_k(y,y')\right],
\eeq
and including adiabatic perturbations:
\beq\label{eq:PEvolution}
\boxed
{
P_{\delta}(y,k)=P_{\delta_{\rm ad}}(y_0,k)\left[\mathcal{T}_k^{\rm ad}(y,y_0)\right]^2+\frac{1}{\bar{n}}\left[1+3\int_{y_0}^y \frac{dy'}{\sqrt{1+y'}}\mathcal{T}^{(a)}_k(y,y')\mathcal{T}^{(b)}_k(y,y')\right],
}
\eeq
where
\beq\label{eq:PT}
\mathcal{T}^{\rm ad}_k(y,y_0)&=\mathcal{T}^{(a)}_k(y,y_0) + \frac{1}{2}\frac{d\ln P_{\delta_{\rm ad}}(y_0,k)}{d\ln y_0}\sqrt{1+y_0}\,\mathcal{T}^{(b)}_k(y,y_0),\\
% \mathcal{T}^{\rm ad}_k(y,y_0)&=\left[1+\frac{d\ln\delta_k}{d\ln y}\Big|_{y=y_0}\sqrt{1+y_0}\mathcal{F}(y,y_0)\right]T_{\rm fs}(y,y_0,k),\\
%&+\frac{3}{2}\int_{y_0}^y \frac{dy'}{\sqrt{1+y'}} \mathcal{F}(y,y')T_{\rm fs}(y,y',k)\mathcal{T}^{\rm ad}_{k}(y',y_0) ,\\
%\mathcal{T}^{(a)}_k(y,y')&=T_{\rm fs}(y,y',k)+\frac{3}{2}\int_{y'}^y \frac{dy''}{\sqrt{1+y''}}\mathcal{F}(y,y'')T_{\rm fs}(y,y'',k)\mathcal{T}^{(a)}_k(y'',y'),\\
\mathcal{T}^{(a)}_k(y,y')&=T_{\rm fs}(y,y',k)+\frac{3}{2}\int_{y'}^y \frac{dy''}{\sqrt{1+y''}}\mathcal{T}^{(b)}_k(y,y'')T_{\rm fs}(y'',y',k),\\
%\mathcal{T}^{(b)}_k(y,y')&=\mathcal{F}(y,y')T_{\rm fs}(y,y',k)+\frac{3}{2}\int_{y'}^y \frac{dy''}{\sqrt{1+y''}}\mathcal{T}^{(b)}_k(y'',y')\mathcal{F}(y,y'')T_{\rm fs}(y,y'',k).\\
\mathcal{T}^{(b)}_k(y,y')&=\mathcal{F}(y,y')T_{\rm fs}(y,y',k)+\frac{3}{2}\int_{y'}^y \frac{dy''}{\sqrt{1+y''}}\mathcal{T}^{(b)}_k(y,y'')\mathcal{F}(y'',y')T_{\rm fs}(y'',y',k).\\
\eeq
Here, the free-streaming transfer function is
\beq
T_{\rm fs}(y,y',k)&=4\pi\int dp\, p^2 f_0(p)\,\textrm{sinc}\left[\frac{p}{m a_{\rm eq}}\frac{k}{k_{\rm eq}}\sqrt{2}\mathcal{F}(y,y')\right].
\eeq
If the distribution function has a characteristic momentum scale $q_*$, so that $f_0(q)=A(q_*) B(q/q_*)$, then a convenient dimensionless parameter 
\beq
\alpha_k=\sqrt{2}\frac{k}{k_{\rm eq}}\frac{q_*}{a_{\rm eq} m},
\eeq
naturally appears, and the free streaming function $T_{\rm fs}$ becomes a function of $\alpha_k \mathcal{F}$. As a result, both $\mathcal{T}^{(a,b)}$ are functions of $\alpha_k$ with no further dependence on $k$. The behavior of these functions changes at $\alpha_k\sim 1$. 

\paragraph{Numerical Calculation} We provide efficient and fast numerical algorithms to solve the Volterra-type equations for $\mathcal{T}^{(a,b)}$ in equations~\eqref{eq:PT} for a given $T_{\rm fs}$ in Appendix \ref{sec:Eval}. An implementation can be found at \url{https://github.com/delos/warm-structure-growth} \cite{delos_2025_17064722}.
In Appendix~\ref{sec:Approx}, we also provide simpler, approximate formulae that work well for the Maxwell-Boltzmann momentum distribution, although they can be less accurate approximations in other cases.

\begin{figure}
    \centering
\includegraphics[width=1\linewidth]{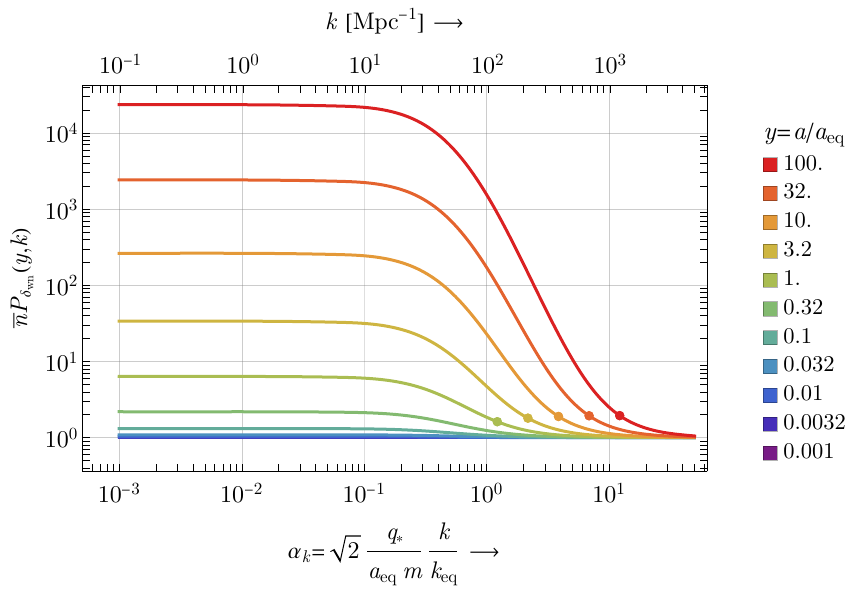}
    \caption{Evolution of the white-noise power spectrum as a function of wavenumber horizontal axis) and scale factor (colors). The colored lines are based on evaluation of our expression \eqref{eq:PwEvolution} for the evolution of the power spectrum. The dots indicate the co-moving Jeans scale $k_{\rm J}(a)$. The left edge of the plot evolves upwards as $(1+\frac{3}{2}y)^2$ in accordance with standard expectations, with growth at higher $k$ being Jeans-suppressed. The above calculations are done for a Maxwellian initial momentum distribution $f_0(q)=A e^{-q^2/2q_*^2}$, and the 1D velocity dispersion at equality is $\sigma_{\rm eq}=q_*/a_{\rm eq}m\approx 22\,{\rm km}\, {\rm s}^{-1}$. The dimensionless scale ($\alpha_k$) on the bottom axis is converted to $k$ in $\Mpc^{-1}$ (top axis) using this $\sigma_{\rm eq}$. Note that in terms of $\alpha_k$, the shape of $\bar{n}P_{\delta_{\rm wn}}(y,k)$ does not depend on $\sigma_{\rm eq}$.}
    \label{fig:PwEvolution}
\end{figure}

\begin{figure}
    \centering
\includegraphics[width=0.9\linewidth]{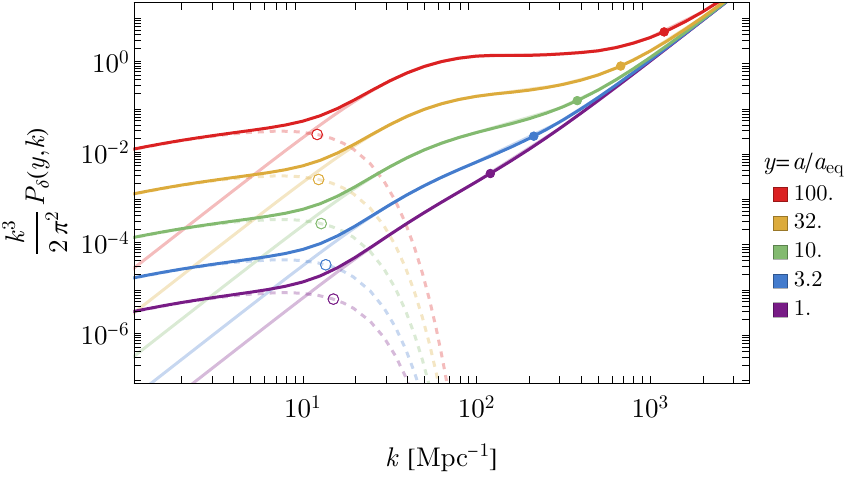}
    \caption{Evolution of the total dimensionless power spectrum (solid lines). The lighter curves are the separate adiabatic (dashed) and white-noise (solid) contributions. The open circles on the adiabatic part indicate the free-streaming wavenumber $k_{\rm fs}(y)$, while the closed circles on the white-noise part are the Jeans wavenumber $k_{\rm J}(y)$. We adopt a Maxwellian initial velocity distribution, with $\sigma_{\rm eq}\approx 22\,\textrm{km} \,\textrm{s}^{-1}$, and $\bar{n}\approx 5\times 10^7/\Mpc^3$. The adiabatic part has an amplitude consistent with Planck (2018) observations. The chosen parameters are motivated by ref.~\cite{Amin:2022nlh}, with $k_{\rm fs}\sim 10/\Mpc$ and $k_{\rm wn}=(2\pi^2\bar{n})^{1/3}\sim 10^3/\Mpc$ -- at the boundary of being consistent with observations in the quasi-linear regime.}
    \label{fig:PbothEvolutionLateApprox}
\end{figure}

\paragraph{Examples} 
As concrete examples, we consider an initially Maxwellian distribution
and a uniform-sphere (or ``top-hat'') distribution, which yield, respectively:
\beq
f_0(q)&=\frac{e^{-q^2/(2q_*^2)}}{(2\pi)^{3/2}q_*^{3}}\,,\quad T_{\rm fs}(y,y',k)=e^{-\alpha_k^2\mathcal{F}^2/2},\quad \langle q^2\rangle = 3q_*^2\\
f_0(q)&=\frac{\Theta(q_*-q)}{4\pi q_*^3/3}\,,\quad T_{\rm fs}(y,y',k)=\frac{3}{\alpha_k^3\mathcal{F}^3}\left(\sin [\alpha_k\mathcal{F}]-\alpha_k\mathcal{F}\cos [\alpha_k\mathcal{F}]\right),\quad \langle q^2\rangle = \frac{3}{5}q_*^2.
\eeq
where $\mathcal{F}=\mathcal{F}(y,y')$. For the Maxwell-Boltzmann case, Fig.~\ref{fig:PwEvolution} shows the growth of the white-noise part of the power spectrum, while Fig.~\ref{fig:PbothEvolutionLateApprox} shows the evolution of the dimensionless matter power spectrum including both the white noise and the initially adiabatic perturbations. In Appendix~\ref{sec:Flat}, we show similar results for the uniform-sphere case.

%~~~~~~~~~~~~~~~~~~~~~~~~~~%~~~~~~~~~~~~~~~~~~~~~
%~~~~~~~~~~~~~~~~~~~~~~~~~~

%~~~~~~~~~~~~~~~~~~~
%~~~~~~~~~~~~~~~~~

%~~~~~~~~~~~~~~~~~~~~~~~~~~~~~~~~
\subsubsection{Relevant Scales and Range of Validity}
%~~~~~~~~~~~~~~~~~~~~~~~~~~~~~~~~
\paragraph{Jeans and Free-streaming Scales:} In general, there are two relevant scales for understanding the cutoff in the power spectrum at late times. These are the Jeans scale (relevant for the white-noise part) and free-streaming scale (relevant for the adiabatic part). They are defined as
\beq
k_{\rm J}(y)&\equiv a(t)\sqrt{\frac{4\pi G \bar{\rho}(t)}{\sigma^2}}=\frac{\sqrt{3y}}{2}\frac{k_{\rm eq}}{\sigma_{\rm eq}}\,,\\
k_{\rm fs}(y)&\equiv \left[\int_{t_0}^t \sigma\frac{dt'}{a(t')}\right]^{-1}=\frac{1}{\sqrt{2}\mathcal{F}(y,y_0)}\frac{k_{\rm eq}}{\sigma_{\rm eq}},
\eeq
where $\sigma\equiv\langle \bv^2_{\rm p}(t)/3\rangle^{1/2}$ is the 1D peculiar velocity dispersion, and $\sigma_{\rm eq}\equiv\langle \bv_{\rm eq}^2/3\rangle^{1/2}$ is the 1D peculiar velocity dispersion at matter radiation equality. Here the expectation values are taken using $f_0(p)$. Note that $k_{\rm J}$ increases as $\sqrt{y}$ as the universe expands, whereas $k_{\rm fs}$ decreases as the universe expands. After equality, where $\mathcal{F}(1,y_0)\approx \ln(0.7/y_0)$, $\mathcal{F}$ increases very slowly with $y$, so $k_{\rm fs}$ shifts only very little over time. The cutoff in the white-noise part of the power spectrum typically resembles a power law in $k$, with suppression below $k_{\rm J}^{\rm eq} = k_{\rm J}(y=1)$. The suppression in the adiabatic part is sharper, with suppression beginning at $k_{\rm fs}^{\rm eq}=k_{\rm fs}(y=1)$. It is also worth noting that at equality, we expect $k^{\rm eq}_{\rm fs}\approx k^{\rm eq}_{\rm J}/\ln (y_0^{-1})<k_{\rm J}^{\rm eq}$. 

To provide a sense of the scales relevant for observations of the linear power spectrum (such as the Lyman-$\alpha$ forest), we note that:
\beq\label{eq:scales}
k_{\rm fs}(y)&\approx 15\,\Mpc^{-1}\times \frac{\mathcal{F}(1,10^{-3})}{\mathcal{F}(y,y_0)}\left(\frac{22\, {\rm km}\, {\rm s}^{-1}}{\sigma_{\rm eq}}\right),\\
k_{\rm J}(y)&\approx 120\,\Mpc^{-1}\times \sqrt{y} \left(\frac{22\, {\rm km}\, {\rm s}^{-1}}{\sigma_{\rm eq}}\right).
\eeq
Free streaming is associated with a longer length scale (smaller $k$) and so is usually expected to provide more stringent observational constraints. 

Note that for the examples in the present work we chose $y_0=10^{-3}$, determining the instant where non-relativistic free streaming can begin. The free-streaming scale is logarithmically sensitive to this choice. As we discuss in section~\ref{sec:relativistic}, non-relativistic streaming may be approximated to begin either when the dark matter becomes non-relativistic or when the initial perturbations are sourced (e.g., horizon entry), whichever is later, but a relativistic version of our calculation would be necessary to capture these considerations precisely.

\paragraph{Validity of Approximations:} Our derivation for the power spectrum rested on two assumptions that warrant further discussion: that the two-particle correlation function is small ($g\ll ff$) and that the one-particle {\df} does not evolve significantly in time ($\partial_t f\approx 0$). We first discuss the $g\ll ff$ assumption. Integrating this relation over momenta yields a condition on the correlation function in position space,
\beq
\label{eq:nonlinear}
1\gg\xi(r)
&=\int d\ln k \frac{k^3}{2\pi^2}\left[P_\delta(y,k)-\frac{1}{\bar{n}}\right]\frac{\sin(kr)}{kr}.
\eeq
%In the UV, the integral does not get significant contributions beyond $k_{\rm J}$. In the IR, we can limit the integral to the horizon scale.
This condition provides a domain of validity as a function of the length scale $r$ and time $y$. It is essentially a statement that fractional density perturbations must be small, meaning that our calculation is perturbative in the same sense as standard, linear-order cosmological perturbation theory.
Due to the log-measure, heuristically, requiring $(k^3/2\pi^2)[P_\delta(y,k)-\bar{n}^{-1}]\ll 1$ should typically be sufficient. 

The time evolution of the 1-particle {\df} is a more subtle matter. For certain power spectra $P_\delta$, velocities associated with gravitational growth of perturbations on the largest scales can easily contribute to the overall velocity variance $\langle v^2\rangle$ at a level exceeding the input velocity dispersion $\sigma$.\footnote{In the regime of small perturbations and approximating that $P_\delta(k)\propto a^2$ for $k<k_{\rm J}$, the ratio between the velocity variance due to gravitational growth and that due to the input dispersion can be shown to be
$\langle v^2\rangle/(3\sigma^2)
\approx \frac{2}{9}\int_0^{k_{\rm J}} d\ln k\, (k_{\rm J}/k)^2[k^3/(2\pi^2)]P_\delta(k)$. If the dimensionless power spectrum $[k^3/(2\pi^2)]P_\delta(k)$ is sufficiently shallow, this integral can be dominated by low-$k$ contributions, giving rise to $\langle v^2\rangle\gg 3\sigma^2$ long before any nonlinear clustering occurs.}
However, these velocities are dominated by large-scale bulk flows, which cannot affect dynamics on much smaller scales (as they can be removed with a change of reference frame). A velocity \textit{dispersion} only develops in nonlinearly evolved systems after stream crossing. Therefore, we expect that our neglect of growth of the velocity dispersion is always valid within the small-perturbation regime corresponding to eq.~\eqref{eq:nonlinear}.
Beyond the velocity dispersion, another potential concern is that an arbitrary initial {\df} would relax toward Maxwellian due to 2-particle interactions. However, one can show using standard arguments (e.g.~\cite{Binney:1987}) that the relaxation time scale is of order $t_\mathrm{relax}\sim(\bar n/k_{\rm J}^3)H^{-1}$,\footnote{For a given particle, an encounter with another particle with impact parameter $b$ and relative velocity $v$ induces a velocity kick $\Delta v=2Gm/(bv)$. There are $dN_{\rm enc}\sim \bar{n}_p v t_{\rm relax} 2\pi bdb$ encounters per $b$ interval over time $t_{\rm relax}$, where $\bar{n}_p$ is the physical number density. Relaxation corresponds to $\int dN_{\rm enc}(\Delta v)^2\sim v^2$, which yields $t_{\rm relax} \sim v^3/[8\pi (\ln\Lambda) (Gm)^2\bar{n}_p]$, where $\ln\Lambda=\int db/b$ is the Coulomb logarithm. Taking $\ln\Lambda\sim 10$ and using the definition of $k_{\rm J}$ (with $v\sim\sigma$), we can obtain $t_\mathrm{relax}\sim(\bar n/k_{\rm J}^3)H^{-1}$.} implying that relaxation is unimportant as long as $k_{\rm J}\ll \bar{n}^{1/3}$. As this is always true when fractional perturbations longer than the Jeans scale are small, we conclude that the $\partial_t f=0$ assumption does not add any new restriction beyond that.

In the next section, we will confirm these considerations by showing in $N$-body simulations that $(k^3/2\pi^2)P_\delta(y,k)\lesssim 0.1$ is sufficient for excellent agreement with analytically computed power spectra. However, we will also show that the abundance of collapsed dark matter halos can be accurately predicted by these analytically computed spectra even when $(k^3/2\pi^2)P_\delta(y,k)\gtrsim 1$, just as with standard linear-order cosmological perturbation theory.

%%%%%%%%%%%%%%%%%%%%%%%%%%%%%%%%%%%%%
\section{Comparison with $N$-body Simulations}
\label{sec:N-Compare}
%%%%%%%%%%%%%%%%%%%%%%%%%%%%%%%%%%%%%

We now test our analytic results with cosmological $N$-body simulations.
We will show that the analytic power spectrum \eqref{eq:PEvolution} is accurate in the linear regime where perturbations are small. We will also show that, like the results of standard cosmological perturbation theory, our calculation is useful in the nonlinear regime because it yields accurate results when used as input to halo models.

The simulations use $N=512^3$ particles, each with mass $780\,M_{\odot}$, in a periodic volume of co-moving size $L_\mathrm{box}^3=(1.38 \Mpc)^3$.
The total mass in the simulation is about $10^{11}\,M_\odot$.
At the starting scale factor of $a_i=10^{-3}a_{\rm eq}$, particle positions are drawn independently from a uniform distribution over the full simulation box. The independent random positions are what lead to Poissonian white noise. Meanwhile, particle velocities are drawn from a Maxwell-Boltzmann velocity distribution with a 1D velocity dispersion of $\sigma\equiv\langle \bv^2/3\rangle^{1/2}=0.0724 c=21700~{\rm km}\,\,\rm{s}^{-1}$.

We evolve the simulation volumes using a version of \textsc{gadget-4} \cite{Springel:2020plp} that we modified to include a homogeneous radiation component (as in ref.~\cite{Delos:2018ueo}). For simplicity, we take the white-noise particles to be all of the matter. Otherwise, we adopt a cosmology in line with the Planck measurements \cite{Planck:2018vyg}. Most relevantly, we have $a_{\rm eq}=2.94\times 10^{-4}$ (where $a=1$ today) and $k_{\rm eq}=1.04\times 10^{-2}~\Mpc^{-1}$.

The simulation particles represent physical dark matter particles, so it would be appropriate to run a ``collisional'' simulation, which fully resolves short-range gravitational interactions between the simulation particles (e.g.~\cite{2020MNRAS.497..536W,Rantala:2022pqs}). This approach was used in a recent cosmological simulation with primordial black holes (PBHs) \cite{Delos:2024poq}. However, for the sake of computational efficiency, we will neglect short-range interactions by including standard \textsc{gadget-4} force softening. The co-moving softening length is taken to be $0.03\bar n^{-1/3}$, where $\bar n$ is the co-moving number density of simulation particles. Although short-range interactions are important in simulations of cold PBHs \cite{Delos:2024poq}, this is largely because structure formation in that scenario begins at the scale of just a few PBHs. For scenarios with warm white noise, structure formation can begin at larger scales, since it is Jeans-suppressed at few-particle scales.

%~~~~~~~~~~~~~~~~~~~~~~~~~~~~~~
\subsection{White Noise in the Linear Regime}
\label{sec:SimWN}
%~~~~~~~~~~~~~~~~~~~~~~~~~~~~~~

%~~~~~~~~~~~~~~~~~~~
\begin{figure}
    \centering
\includegraphics[width=\linewidth]{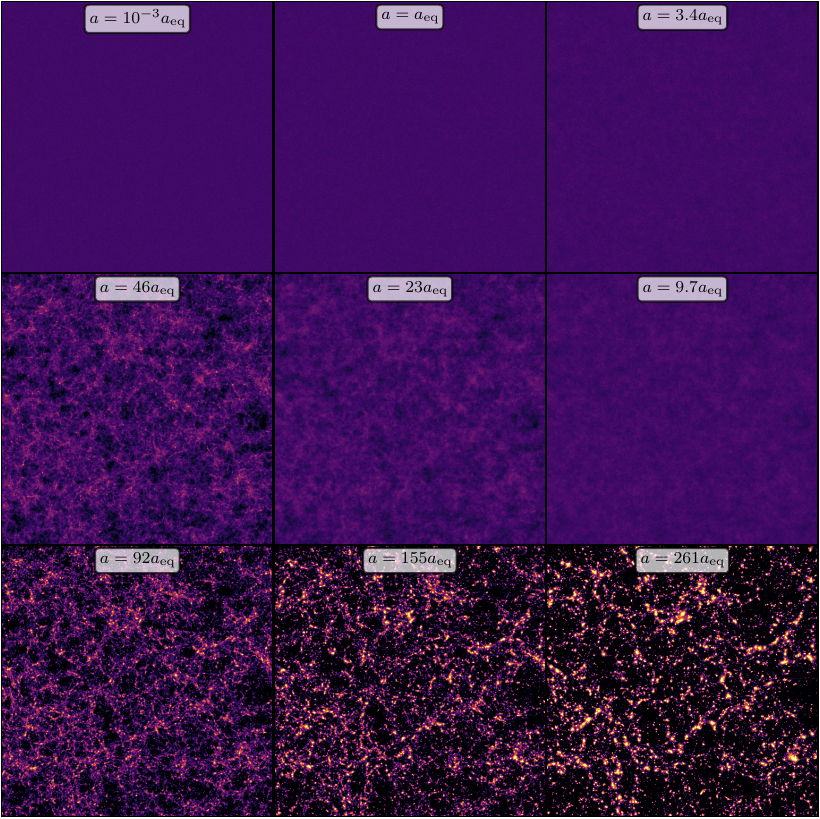}
    \caption{Gravitational clustering of a warm, Poisson distributed collection of particles (without initial adiabatic perturbations). We show, with a logarithmic color scale, the projected density across the $1.38\,\Mpc$ (co-moving) simulation volume. Note the lack of clustering during radiation domination ($a<\aeq$) and significant clustering during matter domination. For the scenario shown, the co-moving Jeans length during matter domination is $2\pi/k_{\rm J}\approx 0.05\,\Mpc \times \sqrt{a_{\rm eq}/a}$. Clustering is suppressed below this length scale.}
    \label{fig:fields}
\end{figure}
%~~~~~~~~~~~~~~~~~~~

First, we consider a scenario with no initial adiabatic power. In this case, the simulation particles are initially uniformly distributed in space with zero mean velocity (but still have the random positions and velocities discussed above). Figure~\ref{fig:fields} shows the evolution of the simulation volume from these initial conditions.
The Jeans wavelength at matter-radiation equality is about $2\pi/k_{\rm J}\simeq 0.05~\text{Mpc}$ (about 1/27 the width of the simulation volume), and this provides a characteristic scale at which structure formation begins. Collapsed halos become abundant by about $a=92\aeq$ (redshift $z=36$).

\begin{figure}
    \centering
\includegraphics[width=0.9\linewidth]{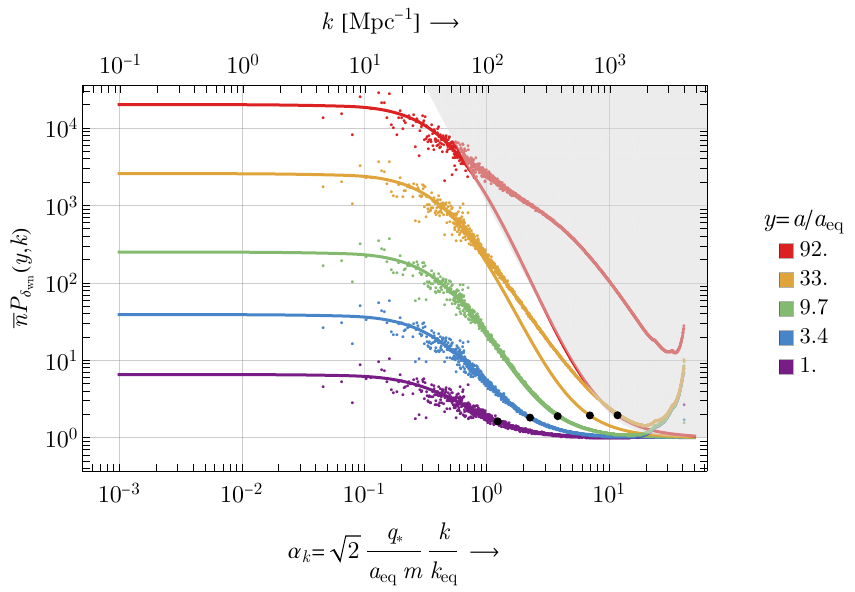}
    \caption{Comparing the analytic growth of the white-noise power spectrum with results from an $N$-body simulation. The solid curves are the analytic predictions, while the colored smaller dots are $N$-body results. The gray region marks the nonlinear regime, $(k^3/2\pi^2)[P_{\delta_{\rm wn}}(y,k)-\bar{n}^{-1}]>1$. The agreement between analytic and numerical results is excellent before the onset of nonlinearity. The model, initial conditions, and parameters are same as Fig.~\ref{fig:PwEvolution}. The large black dots indicate the Jeans wavenumber $k_{\rm J}(y)$.}
    \label{fig:plotNumCompare}
\end{figure}

Figure~\ref{fig:plotNumCompare} compares the matter power spectrum in the simulation to our analytic calculation. At early times, when $a\lesssim 10\aeq$, we find good agreement (except at very low $k$ due to cosmic variance in the simulation volume). At these times, density perturbations larger than the Jeans scale are still in the linear regime. Deviations from the analytic prediction become significant at later times as the growing density perturbations approach the nonlinear regime (indicated with gray shading).

%~~~~~~~~~~~~~~~~~~~
%~~~~~~~~~~~~~~~~~

%~~~~~~~~~~~~~~~~~~~~~~~~~~~~~~
\subsection{White Noise Beyond the Linear Regime}
\label{sec:SimWNnl}
%~~~~~~~~~~~~~~~~~~~~~~~~~~~~~~

For standard structure formation scenarios, the predictions of linear perturbation theory remain useful even deep in the nonlinear regime because they are the basis for halo models, such as Press-Schechter theory \cite{Press:1973iz} and variants (e.g.~\cite{Sheth:2001dp,Tinker:2008ff,Diemer:2020rgd,Ondaro-Mallea:2021yfv}). We now show that this remains true for the analytical white noise power spectrum \eqref{eq:PEvolution}, and we can use it to accurately predict the halo mass function.
The simplest halo models connect the halo mass function to the linear matter power spectrum $P_\delta(k)$ through the rms density variance 
\beq
\label{eq:sigmaM}
\sigma^2_M&=\int_0^{k_\mathrm{max}} d\ln k \frac{k^3}{2\pi^2}P_\delta(k)W^2(k R_M),
\eeq
in spheres of mass $M$, where $R_M\equiv[{3M}/{(4\pi\bar{\rho})}]^{1/3}$ is the radius of such a sphere and $W(z)\equiv {3}\left(\sin z-z\cos z\right)/z^3$ is the spherical top-hat window function in Fourier space. Ordinarily, the upper limit of the spectral integral is taken to be $k_\mathrm{max}\to\infty$.

\begin{figure}
    \centering
\includegraphics[width=0.9\linewidth]{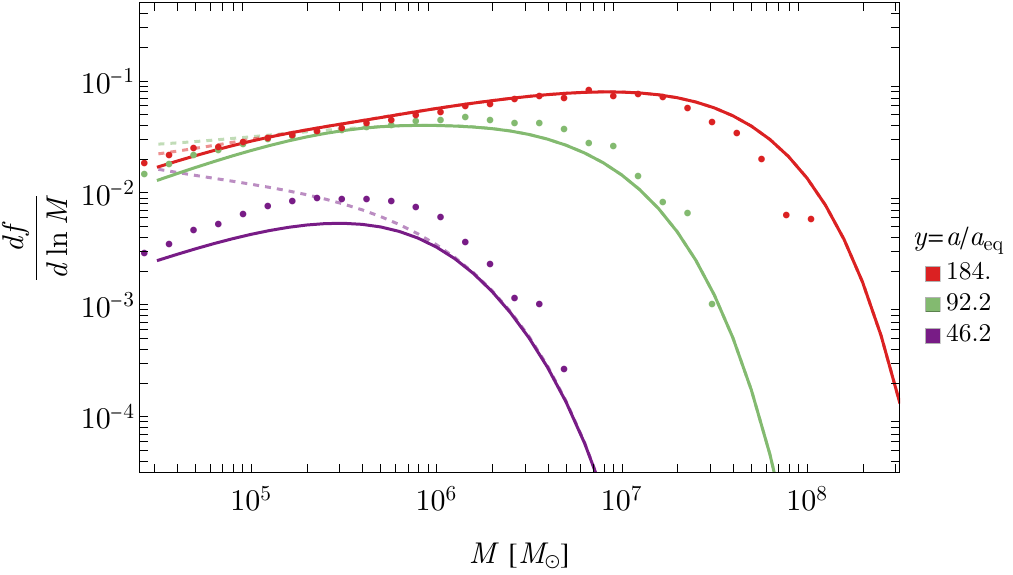}
    \caption{Comparing the halo mass function obtained using our analytically predicted power spectrum with results from an $N$-body simulation, for the case of white-noise initial conditions. We plot $df/d\ln M$, the differential fraction of mass in halos of mass $M$ per logarithmic interval in $M$.
    Solid lines show analytic predictions using the halo mass function from ref.~\cite{Delos:2023eve} with power spectrum cut off above $k_{\rm J}/4$, reflecting that smaller-scale perturbations cannot contribute to gravitational clustering.
    The prediction closely matches results from an $N$-body simulation (colored dots). The lighter dashed lines show analytic predictions when no cutoff is imposed in the linear matter power spectrum.}
    \label{fig:plotMassFuncCompare}
\end{figure}

Figure~\ref{fig:plotMassFuncCompare} shows the halo mass function in the simulation. We use the $M_{200}$ mass definition, which is the mass of the sphere whose enclosed density is 200 times the cosmological average.\footnote{\label{foot:halofinding} We center this sphere on the minimum of the gravitational potential. Halos are first identified using a friends-of-friends algorithm \cite{Davis:1985rj} with a co-moving linking length of $0.2\bar n^{-1/3}$ and a minimum of 32 particles.} We express the halo mass function as $df/d\ln M$, the differential \emph{fraction of dark matter} in halos of mass $M$ (per logarithmic mass interval). Note that $df/d\ln M$ is related straightforwardly to the differential number density of halos via $dn/d\ln M=(\bar{\rho}/M)df/d\ln M$, but the differential mass fraction is often a more useful quantity because it covers a much smaller dynamic range. Figure~\ref{fig:plotMassFuncCompare} shows that over time, an increasing proportion of the dark matter comes to reside in massive halos, although the mass fraction in low-mass halos does not appear to decrease.

%~~~~~~~~~~~~~~~~~~~
%~~~~~~~~~~~~~~~~~

We compare the halo mass functions in the simulation to the predictions of the formulation of excursion set theory presented by ref.~\cite{Delos:2023eve}, which is similar to the classic Press-Schechter approach \cite{Press:1973iz} but more closely follows its theoretical foundations in excursion set theory \cite{Bond:1990iw}. For convenience, we will employ the fitting function provided by ref.~\cite{Delos:2023eve},
\beq
\label{eq:dfdlnM}
\frac{df}{d\ln M}=0.658\sigma_M^{-0.582}e^{-1.056/\sigma_M^2}\Big|\frac{d\ln \sigma_M}{d\ln M}\Big|,
\eeq
which is a fit to the predictions of the excursion set theory (not a direct fit to simulations). Taking $k_\mathrm{max}\to\infty$ in equation~\eqref{eq:sigmaM}, the dashed lines in Fig.~\ref{fig:plotMassFuncCompare} show that this model accurately predicts the halo mass function in most regimes. We have also separately verified that it is more accurate than the Press-Schechter formula.\footnote{The standard Press-Schechter formula (with $\delta_{\rm c}\simeq 1.686$) is
\beq
\label{eq:dfdlnM_PS}
\frac{df}{d\ln M}=\sqrt{\frac{2}{\pi}}\frac{\delta_{\rm c}}{\sigma_M}e^{-\delta_{\rm c}^2/(2\sigma_M^2)}\Big|\frac{d\ln \sigma_M}{d\ln M}\Big|
\eeq
and is derived assuming density perturbations averaged on different mass scales are uncorrelated. The approach of ref.~\cite{Delos:2023eve} accounts for the correlations, and its predictions are well fit by equation~\eqref{eq:dfdlnM} instead.}

However, this calculation overpredicts the halo abundance at the low-mass end.
This discrepancy arises because not all of the density variation associated with the white-noise power spectrum is associated with gravitational clustering. As Fig.~\ref{fig:plotNumCompare} shows, the white-noise power spectrum always extends into the nonlinear regime at high enough $k$, but such small-scale perturbations do not necessarily contribute to gravitational clustering. Perturbations on such short length scales (below the Jeans length) are continually rearranging themselves due to random particle motion, so that the density perturbations average to zero over the long time scales relevant to gravitational clustering.

To address this problem, we make the simple ansatz that perturbation modes with $k>k_{\rm J}/4$ do not contribute to gravitational clustering, where $k_{\rm J}\equiv a\sqrt{4\pi G\bar\rho}/\sigma$ is the (co-moving) Jeans wavenumber given in equation~\eqref{eq:scales} (and $\sigma$ is the 1D peculiar velocity dispersion). That is, we cut off the spectral integral in equation~\eqref{eq:sigmaM} at $k_\mathrm{max}=k_{\rm J}/4$. The numerical factor of $1/4$ is not theoretically motivated, but it works well empirically. Figure~\ref{fig:plotMassFuncCompare} shows that with this ansatz, the mass functions predicted by equation~\eqref{eq:dfdlnM} match the simulation results reasonably well at all masses and redshifts.

In Appendix~\ref{sec:Flat}, we also consider the same simulation setup with a uniform-sphere velocity distribution instead of a Maxwell-Boltzmann velocity distribution. The analytically predicted power spectrum continues to be accurate in the linear regime for this scenario. Predicted halo mass functions are also reasonably accurate, but for the uniform-sphere scenario, the $k_\mathrm{max}=k_{\rm J}/4$ ansatz somewhat overestimates the abundance of low-mass halos at early times.

\subsection{White Noise with Initial Adiabatic Perturbations}
\label{sec:sim_ad}

We now modify the simulation initial conditions to include initial adiabatic density perturbations. We use the \textsc{CLASS} code \cite{Blas:2011rf} to evaluate the cold dark matter power spectrum at redshift $z=31$ given Planck cosmological parameters \cite{Planck:2018vyg}, and then we extrapolate the power spectrum backward to the simulation start time ($a=10^{-3}\aeq$) using a growth function from ref.~\cite{Hu:1995en} that is consistent with the non-relativistic, dark-matter-only physics that are present in the simulation. The purpose of this procedure is that the forward simulation evolution would then reproduce approximately the correct adiabatic modes at late times, even though the early evolution in the simulation (at $z\gtrsim 100$) is not exactly correct due to the absence of baryons.

However, even with $512^3$ particles, the simulation volume has few enough particles that the Poissonian particle noise would completely dominate over the adiabatic power almost up to the scale of the whole box.\footnote{Ordinary cosmological simulations evade this problem by initializing simulation particles on a grid or glass with no random thermal motion. This procedure suppresses the Poissonian particle noise, which is only a discreteness artifact if simulation particles are supposed to be tracers of a continuous dark matter distribution. In our scenario, the simulation particles are physical dark matter particles, so the Poissonian particle noise is physically real.} In order to ensure that we can resolve enough adiabatic power to test our prediction in equation~\eqref{eq:PEvolution}, we amplify the initial adiabatic power spectrum by a factor of 100 (so density perturbations are scaled by a factor of 10).

To incorporate these adiabatic perturbations, we start with the warm white noise initial conditions discussed above, in which particles have independently uniformly distributed positions and a Maxwell-Boltzmann velocity distribution with 1D dispersion $\sigma=0.0724 c$. Then we use the power spectrum to sample a random field of particle displacements and velocities (in accordance with ref.~\cite{Delos:2018ueo}). Finally, we interpolate over these fields to add bulk displacements and velocities to the white-noise particle load.

%~~~~~~~~~~~~~~~~~~~
\begin{figure}
    \centering
\includegraphics[width=0.9\linewidth]{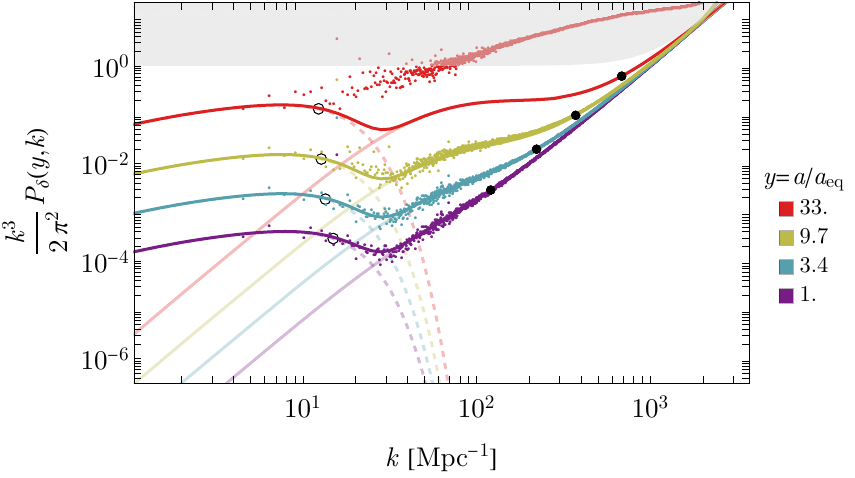}
    \caption{Comparing the analytic evolution of the dimensionless power spectrum with results from an $N$-body simulation for a scenario with both white noise and initial adiabatic perturbations. The solid curves are the analytic predictions, which are the sum of adiabatic (faint dashed) and white-noise (faint solid) contributions, while the colored dots are $N$-body results. The agreement between analytic and numerical results is excellent before the onset of nonlinearity (shaded area). Similarly to previous figures, we adopt $\sigma_{\rm eq}=q_*/a_{\rm eq}m\approx 22 \,{\rm km}\, s^{-1}$ and $\bar{n}\approx 5\times 10^7\,/\Mpc^3$, but here the initial adiabatic spectrum has been boosted by a factor of $100$ compared to Planck (2018) parameters in order to make it relevant within the $(1.38~\mathrm{Mpc})^3$ (co-moving) simulation volume. The open and filled black dots are the free-streaming and Jeans wavenumbers, respectively.}
    \label{fig:plotBothNumCompare}
\end{figure}
%~~~~~~~~~~~~~~~~~

%~~~~~~~~~~~~~~~~~~~
\begin{figure}
    \centering
\includegraphics[width=0.9\linewidth]{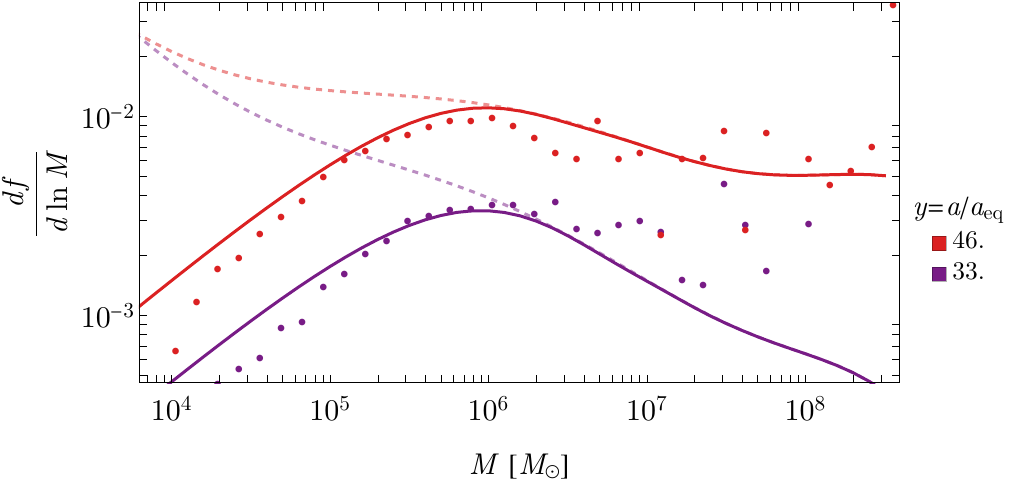}
    \caption{Halo mass functions for a scenario with both white noise and adiabatic initial perturbations. We show $df/d\ln M$, the differential fraction of mass in halos of mass $M$ per logarithmic interval in $M$. The colored points are from an $N$-body simulation, while the solid lines are the mass function from ref.~\cite{Delos:2023eve} evaluated using the analytic power spectrum cut off above $k_{\rm J}/4$ (as in Fig.~\ref{fig:plotMassFuncCompare}), reflecting inability of smaller-scale modes to contribute to gravitational clustering. This analytic prediction generally matches the $N$-body results well. The lighter dashed curves are analytic predictions with no cutoff imposed in the power spectrum.}
    \label{fig:plotMassFuncCompareIsoAd}
\end{figure}
%~~~~~~~~~~~~~~~~~

Figure~\ref{fig:plotBothNumCompare} compares the power spectrum in this simulation to the prediction \eqref{eq:PEvolution}. Similarly to the white-noise-only case, here the prediction matches the simulation well at early times, $y<10$, when the growing density perturbations are still in the linear regime. For this scenario, there is a dip in the matter power spectrum owing to the suppression of the adiabatic power due to free streaming, and this dip is followed by a rise at higher $k$ due to the white noise.

Figure~\ref{fig:plotMassFuncCompareIsoAd} shows the $M_{200}$ halo mass function in this simulation. Compared to the pure white-noise case, the adiabatic power causes halos at the high-mass end to become abundant much earlier, and low-mass halos become correspondingly less abundant. With the $k_\mathrm{max}=k_{\rm J}/4$ ansatz, equation~\eqref{eq:dfdlnM} continues to predict these halo mass functions with reasonable accuracy, although the abundance of low-mass halos is slightly overpredicted. Here, when evaluating $\sigma_M$, we modify equation~\eqref{eq:sigmaM} to account for the fact that the full simulation volume is restricted to be of average density.\footnote{The mean squared density contrast on a mass scale $M$ is
\beq
\sigma^2_M&=
\int_0^{k_\mathrm{max}} d\ln k \frac{k^3}{2\pi^2}P_\delta(k)W^2(k R_M)
-\frac{
\left[\int_0^{k_\mathrm{max}} d\ln k \frac{k^3}{2\pi^2}P_\delta(k) W(kr)\overline{W_\mathrm{box}(\bk)}\right]^2
}{
\int_0^{k_\mathrm{max}} d\ln k \frac{k^3}{2\pi^2}P_\delta(k) \overline{W^2_\mathrm{box}(\bk)}
}
\eeq
when conditioned on the density contrast being zero on the scale of the simulation volume (e.g.~\cite{Delos:2023eve}), where $W_\mathrm{box}(\bk)=\prod_{i=1}^3\textrm{sinc}(k_i L_\mathrm{box}/2)$ is the uniform-cube window function of length $L_\mathrm{box}$ and the overline represents the spherical average. Here $k_i$ are Cartesian components of $\bk$ and $L_\mathrm{box}$ is the co-moving length of the simulation volume. We use the close approximation $[\overline{W_\mathrm{box}(\bk)}]^2\simeq\overline{W^2_\mathrm{box}(\bk)}\simeq W^2(\sqrt{5/12}k L_\mathrm{box})$.}
We also do not accurately predict the very high-mass end of the halo mass function, possibly due to low-number statistics for halos of such masses. 

%~~~~~~~~~~~~~~~~~~~
\begin{figure}
    \centering
\includegraphics[width=\linewidth]{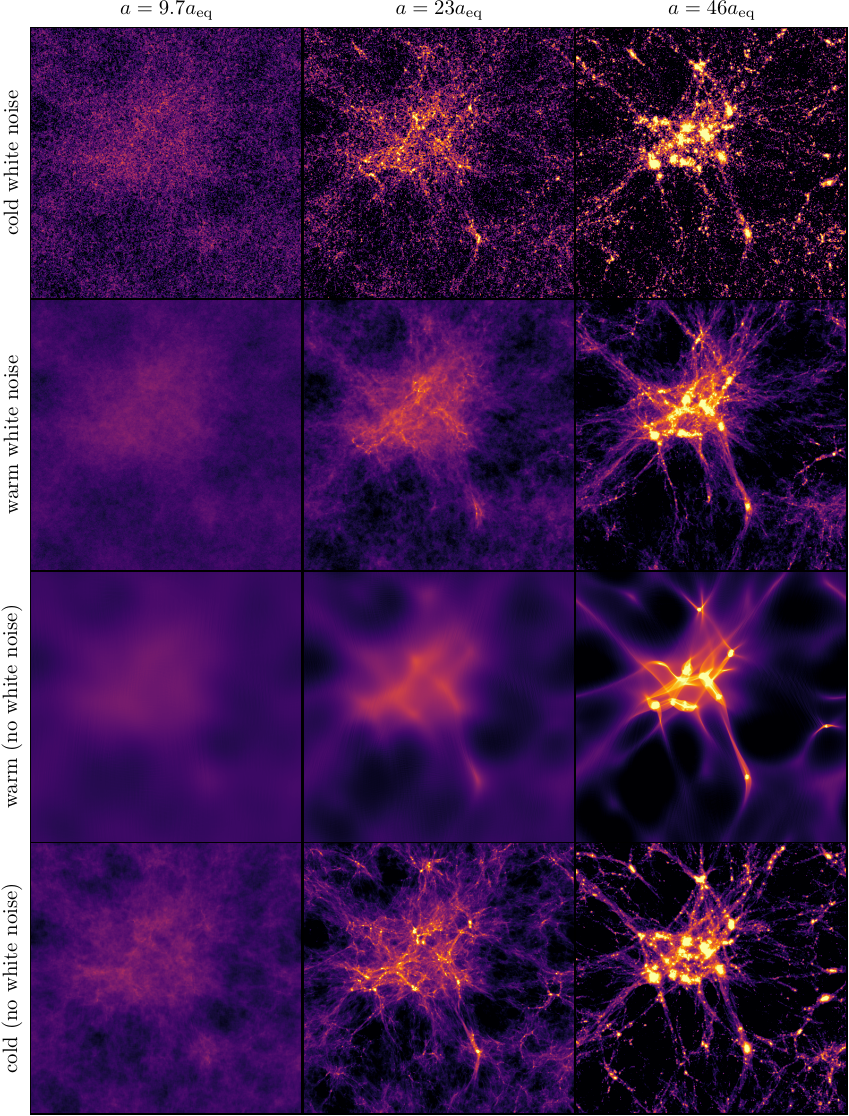}
    \caption{Comparing structure formation with warm white noise (second row from top) to other dark matter models; see the text for details. In all cases the adiabatic power is boosted by a factor of $100$ compared to Planck cosmological parameters \cite{Planck:2018vyg}. We show, with a logarithmic color scale, the density projected across the $(1.38\,\Mpc)^3$ co-moving box.
    }
    \label{fig:fieldsCompare}
\end{figure}
%~~~~~~~~~~~~~~~~~~~

We now compare how structure formation with warm white noise compares to other dark matter models. In all cases, we use the same initial velocity and displacement fields sampled above, corresponding to a power spectrum boosted by a factor of 100 compared to Planck cosmological parameters \cite{Planck:2018vyg}. We consider four models of dark matter, which are represented in the simulation as follows.
\begin{itemize}
    \item \textbf{Cold white noise}: Particles are Poisson distributed in the initial conditions (at $y=a/\aeq=10^{-3}$), but with no velocity dispersion, before being displaced in accordance with the initial velocity and displacement fields. For example, this could represent a typical primordial black hole cosmology \cite{Inman:2019wvr,Delos:2024poq}.
    \item \textbf{Warm white noise}: As described earlier, we start with Poisson-distributed particles with a Maxwell-Boltzmann initial velocity distribution (with 1D dispersion $\sigma=0.0724 c$) and include adiabatic perturbations.
    \item \textbf{Warm (no white noise)}: Here the simulation particles are each supposed to represent a large number of physical dark matter particles. To suppress discreteness noise, simulation particles are initially placed on a grid before being subjected to the initial velocity and displacement fields. Additionally, we do not include a velocity dispersion, since it would induce discreteness noise. Instead, we use equation~(\ref{eq:PEvolution}) to calculate the suppression of adiabatic perturbations at $y=10$ due to the initial Maxwell-Boltzmann velocity distribution, and we apply that suppression to the initial velocity and displacement fields. This corresponds to the standard approach for simulating warm particle dark matter (e.g.~\cite{Bode:2000gq}).
    \item \textbf{Cold (no white noise)}: Same as the warm case without white noise, but here we omit the suppression to initial perturbations.
\end{itemize}

Figure~\ref{fig:fieldsCompare} shows how structure formation proceeds for these different models. As is well known (e.g.~\cite{Afshordi:2003zb}), the cold white noise scenario produces an abundance of very small, dense structures. These structures turn out to be absent in the warm white noise case due to the Jeans scale. Note however that our cold white noise simulation does not properly include short-range gravitational interactions, which tend to suppress the abundance and internal density of the smallest structures \cite{Delos:2024poq}.

Meanwhile, the warm dark matter scenario without white noise produces no structure at all at very small scales. Compared to this, the particle noise in the warm white noise simulation introduces structure at smaller scales.
Indeed, the warm white noise scenario turns out to make a pattern of structures that is similar to that produced by cold dark matter without white noise. However, this is largely a coincidence owing to the similarity between the analytically predicted power spectrum in figure~\ref{fig:plotBothNumCompare} and a cold dark matter power spectrum. For example, if the adiabatic power were not boosted by a factor of 100, then the cold dark matter scenario without white noise would produce nonlinear structures significantly later than would the warm white noise.

\subsection{Halo Internal Structures}

%~~~~~~~~~~~~~~~~~~~
\begin{figure}
    \centering
    \includegraphics[width=0.837\linewidth]{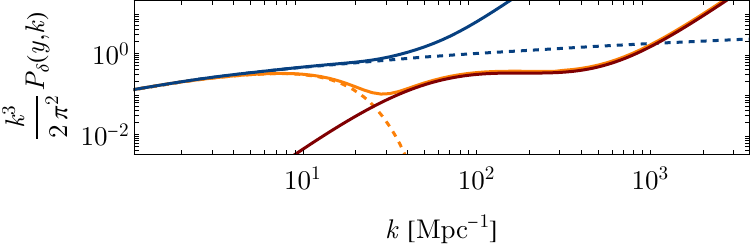}
    \newline\\
    \includegraphics[width=0.8\linewidth]{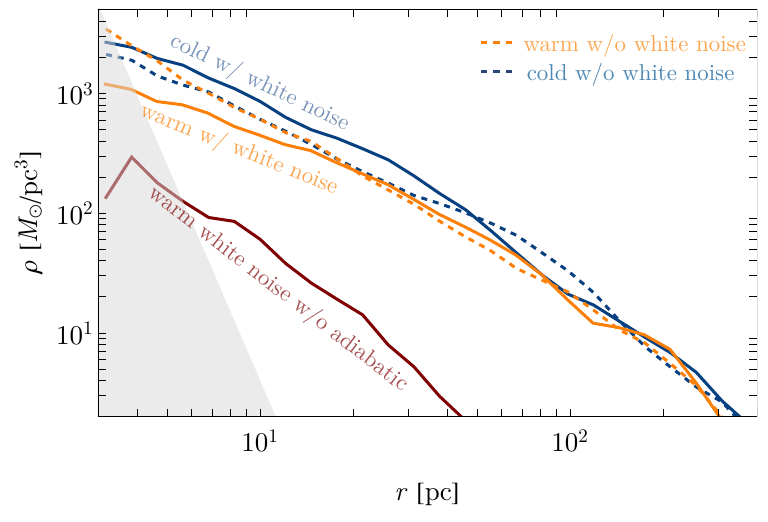}
    \caption{The bottom panel compares density profiles of a halo of mass $M\approx 10^9M_{\odot}$ at $y=a/a_{\rm eq}=46$ in the different sets of simulations for cold and warm dark matter (blue and orange curves) shown in Fig.~\ref{fig:fieldsCompare}. Their initial conditions include $100\times$ boosted adiabatic power with and without white noise. The maroon curve is for a halo of $M\approx 5\times 10^6M_{\odot}$ at $y=46$ from the warm white noise simulation in Fig.~\ref{fig:fields}, which does not include initial adiabatic perturbations. This halo is approximately representative of what would be expected to arise in a realistic warm white noise cosmology without the boost to adiabatic power. The shaded region marks the boundary of maximum density based on phase space considerations for the warm dark matter cases -- increased resolution would reveal a shallow central core.
    For reference, the upper panel shows the linear-theory dimensionless power spectra for these scenarios (evaluated using the formalism in this work) at the same time, $y=46$.}  
    \label{fig:profilesCompare}
\end{figure}
%~~~~~~~~~~~~~~~~~~~

Finally, we provide a brief discussion of the internal structures of dark matter halos in the different scenarios considered in this section. Figure~\ref{fig:profilesCompare} compares the density profiles\footnote{The density profile is evaluated about the minimum of the gravitational potential; see footnote~\ref{foot:halofinding}.} of the largest halo from each simulation at the time $y=46$. For all of the simulations with adiabatic power, this halo has a mass of about $M_{200}\approx 10^9\,M_\odot$ (1\% of the total mass in the simulation).
This halo has a similar internal structure in all cases, owing to the dominance of the $100\times$ boosted adiabatic power spectrum at this mass scale, but there are still noteworthy differences. As seen in Fig.~\ref{fig:profilesCompare},
warm white noise (solid orange) produces a less internally dense structure than does warm dark matter (without white noise, dashed orange) with the same velocity distribution. This outcome is somewhat surprising given that the warm white noise power spectrum matches or exceeds at all $k$ the power spectrum of warm dark matter without white noise.
The usual expectation is that more power leads to earlier halo formation and thus more internally dense halos \cite{Dalal:2010hy,Ludlow:2013bd,Delos:2019mxl,Delos:2022yhn}.
However, this difference in outcome is likely related to prompt cusp formation in the warm case without white noise \cite{White:2022yoc,Delos:2022bhp,Delos:2023exh,DelPopolo:2023tcp}. This prompt cusp formation has been shown in simulations to produce a compact $\rho\propto r^{-1.5}$ density cusp on the scale at which the matter power spectrum cuts off due to free streaming \cite{Ishiyama:2010es,Anderhalden:2013wd,Ishiyama:2014uoa,Polisensky:2015eya,Ogiya:2016hyo,Angulo:2016qof,Delos:2017thv,Delos:2018ueo,Delos:2019mxl,Ishiyama:2019hmh,Colombi:2020xbv,Delos:2022yhn,Ondaro-Mallea:2023qat}.
The halo in our warm simulation without white noise indeed possesses a precisely $\rho\propto r^{-1.5}$ density profile (dashed orange).
For warm white noise (solid orange), on the other hand, the white-noise contribution to the matter power spectrum appears to prevent prompt cusp formation from occurring.

The warm white noise halo is also less internally dense than the halos that form in both of the cold dark matter simulations (with and without white noise, solid and dashed blue). This outcome is unsurprising because the power spectra of cold dark matter and cold white noise lie above the power spectrum of warm white noise at all $k$ (see the upper panel of Fig.~\ref{fig:profilesCompare}), leading to earlier halo formation. Note however that, as we discussed above, the cold white noise simulation does not resolve short-range interactions that can be important in this scenario \cite{Delos:2024poq}.
Also, the high halo density in the simulations without white noise is an artificial consequence of our $100\times$ boosted adiabatic power spectrum. Without that boost, we expect warm white noise to produce denser halos than would dark matter without white noise, as we show next.

The maroon line in Fig.~\ref{fig:profilesCompare} shows the largest halo at the same time, $y=46$, in the warm white noise simulation without adiabatic power (sections \ref{sec:SimWN} and~\ref{sec:SimWNnl}). This halo has mass $M_{200}\approx 5 \times 10^{6}$ $M_\odot$. Note that $y=46$ corresponds to the redshift $z\approx 72$, which is largely before halos would form from standard adiabatic fluctuations. Consequently, this halo is fairly representative of what would actually arise in a warm white noise cosmology. It reaches an internal density of $\sim 100$ $M_\odot{\rm pc}^{-3}$, a significantly higher value than would arise comparably deep inside standard cold or warm dark matter halos.
Without the boost to adiabatic power, our simulations without white noise would not have even produced a halo by $y=46$.

Note that the density within the warm white noise halo is not expected to increase further at (unresolved) smaller radii, because it already saturates the phase-space limit (shaded region) discussed by ref.~\cite{Delos:2022yhn}. The initial velocity dispersion of the dark matter sets a maximum phase-space density at early times, and this density cannot be exceeded inside collapsed systems later. A halo's cold central density cusp should shallow into a finite-density core once it crosses into the shaded area. For example, this phenomenon may explain why the density profile of the halo forming without initial adiabatic power (maroon curve in Fig.~\ref{fig:profilesCompare}) does not continue rising towards the lowest radii.\footnote{We caution, however, that the discreteness noise at these radii is considerable, with the innermost displayed radial bin for this halo (maroon curve) containing only 13 particles (and enclosing about 50). In particular, the apparent decrease in density in this bin is almost certainly due to transient particle noise, because the ensemble-averaged density of an isotropic system typically cannot increase with radius (owing to how particle pericenters are distributed; see ref.~\cite{Dalal:2010hy}).}

Finally, we note that although our simulations do not resolve short-range gravitational interactions, we expect that these interactions would eventually give rise to gravo-thermal core collapse \cite{Lynden-Bell:1968eqn} inside warm white noise halos, greatly boosting the central density.
Core collapse is often considered in the context of dark matter with non-gravitational interactions \cite{Balberg:2002ue,Koda:2011yb}, but here it would arise entirely due to energy transfer in 2-body gravitational ``flyby'' collisions (similarly to core collapse in globular star clusters \cite{Makino:1996vi,Kremer:2019rid}).
Core collapse has been suggested to arise in halos of primordial black holes (cold white noise) \cite{Vaskonen:2019jpv,Stasenko:2023zmf,Stasenko:2024pzd}, but this process may be forestalled by the abundance of tightly bound binary systems that arise in this scenario prior to halo formation \cite{Delos:2024poq}. For warm white noise, however, the velocity dispersion should largely prevent the formation of binary systems outside of halos, so they should not present an obstacle to core collapse.

%%%%%%%%%%%%%%%%%%%%%%%%%%%%%%%%%%%
\section{Generalizations}
\label{sec:Generalizations}
%%%%%%%%%%%%%%%%%%%%%%%%%%%%%%%%%%%
%~~~~~~~~~~~~~~~~~~~~~~
\subsection{Warm Wave Dark Matter}
\label{sec:WDM}
%~~~~~~~~~~~~~~~~~~~~~~
We can translate our results to wave dark matter as follows. Suppose that we begin with a field $\phi$ with a mass $m_\phi$, with its initial power spectrum $q^3P_\phi(t_0,q)$ peaked at a $q=k_*$ such that $a(t_0)H(t_0)<k_*<a(t_0) m_\phi$.\footnote{Such peaked spectra without a significant zero mode in the dark matter field can arise from post-inflationary/causal production mechanisms. For example, see  refs.~\cite{Agrawal:2018vin,Irsic:2019iff,Gorghetto:2020qws,Buschmann:2021sdq,OHare:2021zrq,Adshead:2023qiw,Cyncynates:2023zwj,Saikawa:2024bta,Petrossian-Byrne:2025mto}. For inflationary production, such spectra arise for vector dark matter (e.g~\cite{Graham,Kolb:2020fwh}), and for some cases in scalars, e.g.~\cite{Redi:2022llj,Garcia:2023qab}. }  Then, $\sigma_{\rm eq}\sim k_*/m_{\phi}a_{\rm eq}$ and $\bar{n}\sim k_{\rm wn}^3\sim k_*^3$. Here $k_*^{-1}$ is the (comoving) de Broglie length scale which characterizes the comoving size and separation of the quasi-particles as well as the velocity dispersion. Note that $m$ used in the present work would be the mass of the quasi-particle, i.e., the mass within a de Broglie volume: $m\sim \bar{\rho}_\mathrm{com}/k_*^{3}\gg m_\phi$, where $\bar{\rho}_\mathrm{com}$ is the comoving dark matter density, and the momentum would be $q_*\sim (m/m_{\phi})k_*\gg k_*$. For fiducial parameters used here and in ref.~\cite{Amin:2022nlh}, we have $(m_\phi,k_*)\sim (10^{-19}\,,10^{-26})\,{\rm eV}$.\footnote{Note that $k_*/{\rm eV}\approx 6.4\times 10^{-30}k_*/\Mpc^{-1}$.} The free-streaming and Jeans scales \eqref{eq:scales} in this scenario are determined by $k_*$. These scales are different from the scale of the cutoff ($k\sim a\sqrt{m H}$ \cite{Hu:2000ke}) in the power spectrum for cold wave dark matter (dominated by a zero mode of the field). Also see the discussions in refs.~\cite{Amin:2022nlh,Liu:2024pjg,Ling:2024qfv}. We expect our framework to be applicable to wave dark matter as long as we restrict ourselves to length scales larger than $k_*^{-1}$.
%For wave dark matter, the key change is in the expressions for $\mathcal{T}_k^{(a,b)}$. Explicitly, an ansatz for the expected change is:
% \ma{
% \beq
% % \mathcal{T}^{\rm ad}_k(y,y_0)&=\mathcal{T}_k^{(a)}(y,y_0)+\frac{d\ln\sqrt{P_{\delta_{\rm ad}}(y_0,k)}}{d\ln y_0}\sqrt{1+y_0}\mathcal{T}_k^{(b)}(y,y_0)\\ \\
% % \mathcal{T}^{(a)}_k(y,y')&=\cos\left[\gamma_k\mathcal{F}(y,y')\right]T_{\rm fs}(y,y',k)\\
% % &\quad+\frac{3}{2}\int_{y'}^y \frac{dy''}{\sqrt{1+y''}}\frac{1}{\gamma_k}\sin\left[\gamma_k\mathcal{F}(y,y'')\right]T_{\rm fs}(y,y'',k)\mathcal{T}^{(a)}_k(y'',y'),\\
% % \mathcal{T}^{(b)}_k(y,y')&=\frac{1}{\gamma_k}\sin\left[\gamma_k\mathcal{F}(y,y')\right]T_{\rm fs}(y,y',k)\\
% % &\quad+\frac{3}{2}\int_{y'}^y \frac{dy''}{\sqrt{1+y''}}\frac{1}{\gamma_k}\sin\left[\gamma_k\mathcal{F}(y,y'')\right]T_{\rm fs}(y,y'',k)\mathcal{T}^{(b)}_k(y'',y').\\
% \mathcal{T}^{(a)}_k(y,y')&=\cos[\gamma_k\mathcal{F}(y,y')]T_{\rm fs}(y,y',k)\\
% &\quad+\frac{3}{2}\int_{y'}^y \frac{dy''}{\sqrt{1+y''}}\mathcal{T}^{(b)}_k(y,y'')\cos[\gamma_k\mathcal{F}(y'',y')]T_{\rm fs}(y'',y',k),\\
% \mathcal{T}^{(b)}_k(y,y')&=\frac{1}{\gamma_k}\sin[\gamma_k\mathcal{F}(y,y')]T_{\rm fs}(y,y',k)\\
% &\quad+\frac{3}{2}\int_{y'}^y \frac{dy''}{\sqrt{1+y''}}\mathcal{T}^{(b)}_k(y,y'')\frac{1}{\gamma_k}\sin[\gamma_k\mathcal{F}(y'',y')]T_{\rm fs}(y'',y',k).\\
% \eeq
% where $\gamma_k=k^2/(\sqrt{2}k_{\rm eq}a_{\rm eq}m)$. Note that solving for $k$ in $\gamma_k\mathcal{F}(y,y_0)\sim 1$ would yield the wave-dynamical Jeans scale \cite{Hu:2000ke} (also see the Appendix in \cite{Amin:2022nlh}).} 

In a follow-up work \cite{Amin:2025sla}, cosmological Schr\"{o}dinger-Poisson simulations and corresponding analytic calculations for structure formation in warm wave dark matter are provided. %(see also ref.~\cite{Liu:2025lts}).
Results consistent with the present work are seen in those simulations (above the de Broglie scale). In particular, the power spectrum shows: (i) a scale-dependent growth of the white-noise-like part during matter domination along with a Jeans suppression, and (ii) free-streaming suppression of the adiabatic part. The initial conditions relevant for these simulations, and evolution during radiation domination, are discussed in detail in ref.~\cite{Ling:2024qfv}.

A detailed derivation of a single- and multi-component Boltzmann-like equation appropriate for non-relativistic wave dynamics inside a halo is provided in ref.~\cite{Jain:2023ojg}. One of the key results there was the time-scale of soliton formation \cite{Levkov:2018kau,Jain:2023ojg}. It is not possible to get this time-scale directly in the current work. It would be interesting to merge the two approaches, which we leave for future work. Furthermore, the impact of non-standard expansion histories \cite{Long:2024imw} and non-minimal couplings \cite{Chen:2024pyr} can also be considered. While our focus here is on dark matter, we note at the end of inflation, peaked field spectra in the inflaton field are naturally generated via self-resonance \cite{Amin:2011hj}. It would be possible to apply our formalism to gravitational clustering in those very-early-universe scenarios once the field becomes non-relativistic, e.g. \cite{Amin:2019ums,Eggemeier:2021smj}, and/or if the field fragments into solitons (e.g.~\cite{Amin:2019ums,Lozanov:2022yoy,Lozanov:2023aez}). 

%~~~~~~~~~~~~~~~~~~~~~~~~~~~~~~~~~~~~~~~~~~~~~~
\subsection{Non-gravitational Interactions}
\label{sec:Non-gravInt}
%~~~~~~~~~~~~~~~~~~~~~~~~~~~~~~~~~~~~~~~~~~~~~~
Inclusion of non-gravitational interactions is possible within our framework. This requires us to consider acceleration due to forces other than gravity. That is, the acceleration of particle $i$ due to particle $j$ now becomes 
\beq
\bm{a}_{ij}=\bm{a}^{\rm grav}_{ij}+\bm{a}^{\rm int.}_{ij}.
\eeq
If $\bm{a}^{\rm int.}_{ij}=a^{-1}(t)\nabla_{\bx_i}V^{\rm int}_{ij}(a(t)|\bx_i-\bx_j|)$, then the derivation proceeds almost identically to before. For example, the additional interaction potential could be of Yukawa type, 
\beq
V^{\rm int.}(r)=-\alpha_\chi\frac{e^{-\mu r}}{r}.
\eeq
Instead of equation~\eqref{eq:aij_Fourier}, the acceleration due to this potential would be
\beq
\bm{a}^{\rm int.}_{ij}=4\pi G m \alpha_\chi \int_\bq \frac{i\bq}{q^2+\mu^2} e^{i\bq\cdot(\bx_i-\bx_j)}.
\eeq
%The Fourier transform of the acceleration due to this interaction $\propto \alpha_\chi i\bq/(q^2+\mu^2)$, allow for a easy inclusion with our framework.
This interaction is expected to affect the power spectrum for $k>\mu a(t)$. This type of additional interaction has been the focus of intense study to understand the small-scale structure in dark matter \cite{Tulin:2017ara,Nadler:2025jwh}, and warrants careful exploration. As special cases, this interaction also includes point or hard-sphere scattering type interactions \cite{Spergel:1999mh}. 

Generalizing further, if the differential cross section is available, we can relate this interaction potential to it via $d\sigma_{ij}/d\Omega\propto |\tilde{V}^{\rm int.}_{ij}(\bq)|^2$ (under the Born Approximation in the non-relativistic regime). Here, $\tilde{V}_{ij}^{\rm int.}(\bq)$ is the Fourier transform of the interaction potential. Hence, such interactions can also be included in the framework we have. Such non-gravitational interactions can also be included for wave dark matter discussed earlier \cite{Amin:2019ums, Jain:2023tsr,Mocz:2023adf,Painter:2024rnc,Kirkpatrick:2021wwz,Chen:2020cef}.

As has been appreciated for a long time, the equations for Newtonian gravitational dynamics of dark matter (or stellar systems) are similar to those for plasmas where electric fields are included but magnetic fields are ignored. If dark gauge fields are included, and the dark electric and magnetic fields are also relevant, one can imagine using the plasma physics literature to read off some results for the dark sector. %A related connection might also be possible using gravitomagnetic effects in general relativity.

\subsection{Subdominant Fractions}

Gravitational microlensing and stellar dynamics place strong constraints on scenarios in which all of the dark matter is in macroscopically massive ``particles'' \cite{Carr:2020gox,Mroz:2024mse}. 
Therefore, it would be natural to extend our treatment to scenarios in which these particles comprise a small fraction of the dark matter. In general, for particles with differential number density $dn/dm$ per mass interval and co-moving spatial volume, the initial white-noise power spectrum is
\beq
P_{\delta_{\rm wn}}(y_0,k) = \frac{1}{\bar\rho^2}\int dm\,m^2\frac{dn}{dm}.
\eeq
If a fraction $f$ of the dark matter mass is in massive ``particles'' of a single mass $m$, with the remainder being microscopic particle dark matter, then
\beq
P_{\delta_{\rm wn}}(y_0,k) = \frac{f m}{\bar\rho} = \frac{f^2}{\bar n},
\eeq
where $\bar n=f\bar\rho/m$ is the co-moving number density of the massive particles. In these expressions we take $\bar\rho$ to be the co-moving matter density.

This means that, for sufficiently high particle masses, a small fraction of the dark matter can still give rise to a significant white-noise contribution. However, the growth of structure with fractional warm white noise will differ from what we have calculated. The calculation for mixed dark matter scenarios appears in a follow-up work \cite{Amin:2025ayf} (see also ref.~\cite{Amin:2025nxm} for mixed wave dark matter cases).
For example, the massive ``particles" under consideration could be oscillons/Q-balls/solitons/boson stars (e.g.~\cite{Copeland:1995fq,Kusenko:1997si,Amin:2019ums,Gorghetto:2024vnp,Adshead:2021kvl,Jain:2021pnk,Chavanis_2011,Eby:2019ntd,Visinelli:2017ooc}), miniclusters (e.g.~\cite{Zurek:2006sy,Ellis:2022grh,Xiao:2021nkb,Visinelli:2018wza}), primordial black holes (e.g~\cite{Carr:2020gox,Liu:2022okz,Carr:2023tpt,Musco:2023dak,Zhang:2024ytf,Mroz:2024mse,Zhang:2025asq}), or substructure arising from enhanced primordial power at small scales (e.g.~\cite{Gosenca:2017ybi,Delos:2017thv,Delos:2018ueo,StenDelos:2022jld,Delos:2023fpm,Fakhry:2023prw,Graham:2024hah,deKruijf:2024voc,Dekker:2024nkb,Kumar:2025gon}). Hence such dynamics would find quite general applicability.

% For comparison purposes, we will fix the magnitude of  ${f_2^2}/{\bar{n}_2}$ to be equal to $1/{\bar{n}}$ that we have been using in most of the paper.
% We are interested in the evolution of this power spectrum. To proceed further, we need to initial 1 particle {\df} for each species. We will assume that they are both Maxwellian, with velocity dispersions $\langle v_{\rm eq}^2\rangle_{(s)}^{1/2}$. It is useful to consider three cases,
% \beq
% \langle v_{\rm eq}^2\rangle_{(1)}^{1/2}
% &= \langle v_{\rm eq}^2\rangle_{(2)}^{1/2}=\langle v_{\rm eq}^2\rangle^{1/2},\\
% \langle v_{\rm eq}^2\rangle_{(2)}^{1/2}
% &\ll \langle v_{\rm eq}^2\rangle_{(1)}^{1/2}=\langle v_{\rm eq}^2\rangle^{1/2},\\
% \langle v_{\rm eq}^2\rangle_{(1)}^{1/2}&
% \ll \langle v_{\rm eq}^2\rangle_{(2)}^{1/2}=\langle v_{\rm eq}^2\rangle^{1/2}.
% \eeq
% Let us heuristically compare the expected evolution of the total power spectrum to the case where we have single species with number density $\bar{n}$ and velocity dispersion $\langle v_{\rm eq}^2\rangle^{1/2}$.
% In the first case, we expect no difference in the evolution of the total power spectrum compared to the single species case. In the first and second case, the free-streaming suppression of the adiabatic piece appears due to the velocity dispersion is the first species. In the third species, no such suppression should be present. 

\subsection{Relativistic Physics}
\label{sec:relativistic}

We have focused on the regime of non-relativistic motion and Newtonian gravity. The relativistic regime complicates an analytical treatment in two main ways:
\begin{enumerate}
    \item Relativistic motion couples kinematics in different directions, because the distance covered along some axis depends not only on the velocity component along that axis but also on the velocity magnitude.
    \item Relativistic gravity depends not only on position but also on velocity.
\end{enumerate}

Relativistic physics are important for properly describing the early evolution of the adiabatic perturbations.
The non-relativistic growth functions $\mathcal{T}^{(a,b)}_\bk(y,y')$ that we have derived describe perturbation growth that begins at some preset initial time $y'$, and although the precise value of $y'$ is not important, the general order of magnitude must be picked carefully. In particular, the $y'\to 0$ limit is not meaningful. Roughly speaking, Newtonian free streaming at some scale $k$ begins either at horizon entry ($k\sim aH$), which is when the perturbation is sourced, or when the streaming motion becomes non-relativistic -- whichever is later.
But a relativistic version of our derivation would be needed to account for these effects precisely.
A relativistic description would also be needed to describe a fraction of the dark matter that is still relativistic at the time of matter-radiation equality. In the context of wave dark matter, some of these relativistic effects are included in the Klein-Gordon simulations of ref.~\cite{Ling:2024qfv}.

%%%%%%%%%%%%%%%%%%%%%%%%%%%%%%%%%%%%%
\section{Summary \& Conclusions}
\label{sec:Summary}
%%%%%%%%%%%%%%%%%%%%%%%%%%%%%%%%%%%%%

We have investigated the evolution of density perturbations in warm dark matter including the effect of a finite number density, which gives rise to an (initially) white-noise contribution to the power spectrum of density perturbations. Observationally accessible examples of such white-noise spectra (at $k\gtrsim 10/\rm Mpc$) with significant velocity dispersion can arise when light wave dark matter is produced after inflation. White-noise-related effects are also present for heavy primordial black holes, solitons, etc. composing (a fraction of) the total dark matter. The power spectrum enhancement on small scales due to white noise can have important consequences, such as accelerating early galaxy formation (e.g.~\cite{Hutsi:2022fzw,Parashari:2023cui,Iocco:2024rez}). To understand the growth of structure on small scales in these scenarios, we derive a formalism, based on solving the truncated BBGKY hierarchy, that provides the shape and time-evolution of the power spectrum.

We provide efficient code to numerically evaluate the results of our formalism, which are expressed as time integrals of solutions to a linear Volterra equation. These results include several key features. An initially white-noise power spectrum remains approximately constant in time during the radiation era, because it cannot be erased by free streaming of particles. During the matter-dominated era, self-gravity causes that power spectrum to grow below the Jeans wavenumber $k_{\rm J}$, but growth remains suppressed above $k_{\rm J}$. The formalism also naturally includes the contribution from initial adiabatic perturbations, for which we recover the standard growth and free-streaming suppression.

Our calculations of the matter power spectrum agree precisely with the results of numerical simulations as long as fractional density perturbations remain small. In the regime of large perturbations, we also used the predicted power spectra to estimate halo mass functions. In most regimes, the results agree well with the halo distributions that arise in numerical simulations of the same cosmological scenarios. At the smallest mass scales, the accuracy of our predicted halo mass functions is improved by making the ansatz that perturbations on wavenumbers $k>k_{\rm J}/4$ do not contribute to halo formation.
 
Beyond observations, we expect that our results can be useful for interpreting cosmological $N$-body simulations, in which warm white noise emerges as a discreteness artifact. Ordinarily, simulations are initialized to minimize white noise in the initial conditions, but it would still emerge during nonlinear evolution. In some cases, artificial warm white noise can also be relevant in the linear regime, for example if the simulation were initialized with a velocity dispersion (e.g.~\cite{Maccio:2012qf}) or if a velocity dispersion were to arise dynamically (e.g.~\cite{Ganjoo:2023fgg}).
 
Moreover, our framework includes traditional warm and cold dark matter models. In most cases, microscopic particle dark matter is far too warm for the Poissonian particle noise to give rise to any gravitational clustering. However, for some  models, such as that proposed by refs.~\cite{Boyle:2018tzc,Boyle:2018rgh,Turok:2023amx}, the dark matter is sufficiently cold and heavy that gravitational clustering can be relevant on scales on which the power spectrum is boosted by white noise. Moreover, even if the particle noise is not relevant, our results supply an alternative way to evaluate the effect of free streaming on the matter power spectrum, which is faster and simpler than the usual approach of integrating the Boltzmann hierarchy (e.g.~\cite{2011JCAP...09..032L}). This aspect of our calculation is similar to the approach suggested by ref.~\cite{Ji:2022iji} but is much simpler because we focus on the non-relativistic regime.

The formalism can be generalized in multiple ways. In follow-up papers, we have extended the framework to accommodate fractional amounts of dark matter in a warm and/or white-noise component \cite{Amin:2025ayf,Amin:2025nxm}, and we have modified it to apply to wave dark matter \cite{Amin:2025sla,Amin:2025nxm}. Another potential generalization would be to accommodate non-gravitational interactions between the dark matter particles.
%We expect that the formalism can be generalized in multiple ways, which we discussed but did not work out in detail. (1) It is possible to include fractional amounts of dark matter in a warm and/or white-noise component. (2) We can accommodate non-gravitational interactions between the dark matter particles. (3) We are working on a generalization to wave dark matter, where we expect the present results to hold above the de Broglie length. (4)
Finally, it would be useful to incorporate relativistic physics and the effects of baryons. Alternatively, it would be worthwhile to consider inclusion of the effects discussed here in existing numerical codes (e.g.~\cite{Blas:2011rf,2022ascl.soft03026G,Liu:2024yne}), which already account for relativistic effects and baryonic physics.

\newpage
%%%%%%%%%%%%%%%%%%%%%%%%%%%%%%%%%%%%%
\section*{Acknowledgements}
MA is supported by a DOE award DE-SC0021619. Part of this research was supported by grant NSF PHY-2309135 to the Kavli Institute for Theoretical Physics (KITP). Simulations for this work were carried out on the OBS HPC computing cluster at the Observatories of the Carnegie Institution for Science. We thank Fabian Schmidt for important insights regarding the analytic results. We also gratefully acknowledge access to simulation results of warm wave dark matter carried out by Simon May as well as conversations with him. We thank Adrienne Erickcek, Mudit Jain, Andrew Long, Siyang Ling, Moira Venegas, and Kaixin Yang for helpful discussions.

%%%%%%%%%%%%%%%%%%%%%%%%%%%%%%%%%%%%%

%\newpage
\bibliographystyle{JHEP}
\bibliography{main}

%~~~~~~~~~~~~~~~
\newpage
%%%%%%%%%%%%%%%%%%%%%%%%%%%%%%%%%%%%%%%%%%%%%%%%%%%%%%%%%%%%%%%%%%%%%%%%%%
\appendix
%%%%%%%%%%%%%%%%%%%%%%%%%%%%%%%%%%%%%%%%%%%%%%%%%%%%%%%%%%%%%%%%%%%%%%%%%%

%~~~~~~~~~~~~~~~~~~~~~~~~~~~~~~~~~~~~~~~~~
\section{Diagrammatic Interpretation of the Growth Functions}
\label{sec:Diagram}
%~~~~~~~~~~~~~~~~~~~~~~~~~~~~~~~~~~~~~~~~~

The warm dark matter growth functions are expressed as solutions to Volterra integral equations \eqref{eq:TaTb}, which may be written as
\beq
\label{eq:TaTb2}
T_{\bk}^{(a)}(t,t_0)
&=T_{\bk}^{(A)}(t,t_0)
+\int_{t_0}^t 4\pi G \bar{\rho}\,dt'\,T_{\bk}^{(B)}(t,t')\,T_{\bk}^{(a)}(t',t_0) ,\\
T_{\bk}^{(b)}(t,t_0)
&=T_{\bk}^{(B)}(t,t_0)
+\int_{t_0}^t 4\pi G \bar{\rho}\,dt'\,T_{\bk}^{(B)}(t,t')\,T_{\bk}^{(b)}(t',t_0),
\eeq
where we define
\beq
T_{\bk}^{(A)}(t,t_0)&=T_{\rm fs}(k(t-t_0)),\\
T_{\bk}^{(B)}(t,t_0)&=(t-t_0)T_{\rm fs}(k(t-t_0)).
\eeq
Similar expressions may be written for an expanding universe.
The physical meaning of these functions is that $T_{\bk}^{(A)}(t,t_0)$ and $T_{\bk}^{(B)}(t,t_0)$ evolve initial density or velocity perturbations, respectively, from $t_0$ to $t$ in the absence of self-gravity. Meanwhile, $T_{\bk}^{(a)}(t,t_0)$ and $T_{\bk}^{(b)}(t,t_0)$ do the same while accounting for self-gravity. The intuition for the integral in each of equations~\eqref{eq:TaTb2} is that, via the $T_{\bk}^{(B)}(t,t')$ factor, it sums over the velocity perturbations induced at all times $t'$ by the self-gravity of the perturbation $T_{\bk}^{(a,b)}(t',t_0)$.

The mathematical structure of these equations is clearest in a diagrammatic form. For some fixed $\bk$, let us define
\beq
\\
T_{\bk}^{(A)}(t,t_0) \equiv
\begin{fmffile}{symTa0}
\parbox{35pt}{
\begin{fmfgraph*}(30,30)
\fmfbottom{t0}
\fmflabel{$t_0$}{t0}
\fmftop{t}
\fmflabel{$t$}{t}
\fmf{wiggly,label=$A$}{t0,t}
\end{fmfgraph*}
}
\end{fmffile},
\hspace{2em}&\hspace{2em}
T_{\bk}^{(B)}(t,t_0) \equiv
\begin{fmffile}{symTb0}
\parbox{35pt}{
\begin{fmfgraph*}(30,30)
\fmfbottom{t0}
\fmflabel{$t_0$}{t0}
\fmftop{t}
\fmflabel{$t$}{t}
\fmf{plain,label=$B$}{t0,t}
\end{fmfgraph*}
}
\end{fmffile},
\\\\\\
T_{\bk}^{(a)}(t,t_0) \equiv
\begin{fmffile}{symTa}
\parbox{35pt}{
\begin{fmfgraph*}(30,30)
\fmfbottom{t0}
\fmflabel{$t_0$}{t0}
\fmftop{t}
\fmflabel{$t$}{t}
\fmf{dbl_wiggly,label=$a$}{t0,t}
\end{fmfgraph*}
}
\end{fmffile},
\hspace{2em}&\hspace{2em}\,\,
T_{\bk}^{(b)}(t,t_0) \equiv
\begin{fmffile}{symTb}
\parbox{35pt}{
\begin{fmfgraph*}(30,30)
\fmfbottom{t0}
\fmflabel{$t_0$}{t0}
\fmftop{t}
\fmflabel{$t$}{t}
\fmf{dbl_plain,label=$b$}{t0,t}
\end{fmfgraph*}
}
\end{fmffile}.
\\\\
\eeq
We may now write equations~\eqref{eq:TaTb2} as
\beq
\\
%T_{\bk}^{(a)}(t,t_0) \equiv
\begin{fmffile}{defTa}
\parbox{35pt}{
\begin{fmfgraph*}(30,60)
\fmfbottom{t0}
\fmflabel{$t_0$}{t0}
\fmftop{t}
\fmflabel{$t$}{t}
\fmf{dbl_wiggly,label=$a$}{t0,t}
\end{fmfgraph*}
}
=
\parbox{35pt}{
\begin{fmfgraph*}(30,60)
\fmfbottom{t0}
\fmflabel{$t_0$}{t0}
\fmftop{t}
\fmflabel{$t$}{t}
\fmf{wiggly,label=$A$}{t0,t}
\end{fmfgraph*}
}
+
\parbox{35pt}{
\begin{fmfgraph*}(30,60)
\fmfbottom{t0}
\fmflabel{$t_0$}{t0}
\fmftop{t}
\fmflabel{$t$}{t}
\fmf{plain,label=$B$}{t1,t}
\fmf{dbl_wiggly,label=$a$}{t0,t1}
\fmfblob{6pt}{t1}
\fmflabel{$t'$}{t1}
\end{fmfgraph*}
}
\end{fmffile},
\hspace{2em}&\hspace{2em}
%T_{\bk}^{(b)}(t,t_0) \equiv
\begin{fmffile}{defTb}
\parbox{35pt}{
\begin{fmfgraph*}(30,60)
\fmfbottom{t0}
\fmflabel{$t_0$}{t0}
\fmftop{t}
\fmflabel{$t$}{t}
\fmf{dbl_plain,label=$b$}{t0,t}
\end{fmfgraph*}
}
=
\parbox{35pt}{
\begin{fmfgraph*}(30,60)
\fmfbottom{t0}
\fmflabel{$t_0$}{t0}
\fmftop{t}
\fmflabel{$t$}{t}
\fmf{plain,label=$B$}{t0,t}
\end{fmfgraph*}
}
+
\parbox{35pt}{
\begin{fmfgraph*}(30,60)
\fmfbottom{t0}
\fmflabel{$t_0$}{t0}
\fmftop{t}
\fmflabel{$t$}{t}
\fmf{plain,label=$B$}{t1,t}
\fmf{dbl_plain,label=$b$}{t0,t1}
\fmfblob{6pt}{t1}
\fmflabel{$t'$}{t1}
\end{fmfgraph*}
}
\end{fmffile}.
\\\\
\eeq
Here we introduce another convention: the circle corresponds to an ``internal'' time vertex $t'$, which is to be integrated out between the time vertices immediately above and below it (so that $t_0<t'<t$ in this case) with the measure $4\pi G \bar{\rho}\,dt'$.

To reduce clutter, we will suppress the time vertex labels. By successive substitutions, we may now write out $T_{\bk}^{(a,b)}(t,t_0)$ explicitly as the infinite sums
\beq\label{eq:fullTaTb}
%T_{\bk}^{(a)}(t,t_0) \equiv
\begin{fmffile}{fullTa}
\parbox{35pt}{
\begin{fmfgraph*}(30,100)
\fmfbottom{t0}
\fmftop{t}
\fmf{dbl_wiggly,label=$a$}{t0,t}
\end{fmfgraph*}
}
=
\parbox{35pt}{
\begin{fmfgraph*}(30,100)
\fmfbottom{t0}
\fmftop{t}
\fmf{wiggly,label=$A$}{t0,t}
\end{fmfgraph*}
}
+
\parbox{35pt}{
\begin{fmfgraph*}(30,100)
\fmfbottom{t0}
\fmftop{t}
\fmf{plain,label=$B$}{t1,t}
\fmf{wiggly,label=$A$}{t0,t1}
\fmfblob{6pt}{t1}
\end{fmfgraph*}
}
+
\parbox{35pt}{
\begin{fmfgraph*}(30,100)
\fmfbottom{t0}
\fmftop{t}
\fmf{plain,label=$B$}{t1,t}
\fmf{plain,label=$B$}{t2,t1}
\fmf{wiggly,label=$A$}{t0,t2}
\fmfblob{6pt}{t1,t2}
\end{fmfgraph*}
}
+
\parbox{35pt}{
\begin{fmfgraph*}(30,100)
\fmfbottom{t0}
\fmftop{t}
\fmf{plain,label=$B$}{t1,t}
\fmf{plain,label=$B$}{t2,t1}
\fmf{plain,label=$B$}{t3,t2}
\fmf{wiggly,label=$A$}{t0,t3}
\fmfblob{6pt}{t1,t2,t3}
\end{fmfgraph*}
}
+
\parbox{35pt}{
\begin{fmfgraph*}(30,100)
\fmfbottom{t0}
\fmftop{t}
\fmf{plain,label=$B$}{t1,t}
\fmf{plain,label=$B$}{t2,t1}
\fmf{plain,label=$B$}{t3,t2}
\fmf{plain,label=$B$}{t4,t3}
\fmf{wiggly,label=$A$}{t0,t4}
\fmfblob{6pt}{t1,t2,t3,t4}
\end{fmfgraph*}
}
+
\ldots\,,
\end{fmffile}\\\\
%T_{\bk}^{(a)}(t,t_0) \equiv
\begin{fmffile}{fullTb}
\parbox{35pt}{
\begin{fmfgraph*}(30,100)
\fmfbottom{t0}
\fmftop{t}
\fmf{dbl_plain,label=$b$}{t0,t}
\end{fmfgraph*}
}
=
\parbox{35pt}{
\begin{fmfgraph*}(30,100)
\fmfbottom{t0}
\fmftop{t}
\fmf{plain,label=$B$}{t0,t}
\end{fmfgraph*}
}
+
\parbox{35pt}{
\begin{fmfgraph*}(30,100)
\fmfbottom{t0}
\fmftop{t}
\fmf{plain,label=$B$}{t1,t}
\fmf{plain,label=$B$}{t0,t1}
\fmfblob{6pt}{t1}
\end{fmfgraph*}
}
+
\parbox{35pt}{
\begin{fmfgraph*}(30,100)
\fmfbottom{t0}
\fmftop{t}
\fmf{plain,label=$B$}{t1,t}
\fmf{plain,label=$B$}{t2,t1}
\fmf{plain,label=$B$}{t0,t2}
\fmfblob{6pt}{t1,t2}
\end{fmfgraph*}
}
+
\parbox{35pt}{
\begin{fmfgraph*}(30,100)
\fmfbottom{t0}
\fmftop{t}
\fmf{plain,label=$B$}{t1,t}
\fmf{plain,label=$B$}{t2,t1}
\fmf{plain,label=$B$}{t3,t2}
\fmf{plain,label=$B$}{t0,t3}
\fmfblob{6pt}{t1,t2,t3}
\end{fmfgraph*}
}
+
\parbox{35pt}{
\begin{fmfgraph*}(30,100)
\fmfbottom{t0}
\fmftop{t}
\fmf{plain,label=$B$}{t1,t}
\fmf{plain,label=$B$}{t2,t1}
\fmf{plain,label=$B$}{t3,t2}
\fmf{plain,label=$B$}{t4,t3}
\fmf{plain,label=$B$}{t0,t4}
\fmfblob{6pt}{t1,t2,t3,t4}
\end{fmfgraph*}
}
+
\ldots\,.
\end{fmffile}
\eeq
Note that each ``internal'' time vertex (indicated with a circle) contributes a factor of $G\bar\rho$, so the $i$-th term on the right-hand side of each of these equations is proportional to $(G\bar\rho)^{i-1}$. The interpretation is that the first term in each of these equations does not account for gravity (as we noted above), and each successive term adds a correction due to the gravitational influence of the previous term. Symbolically, these series expansions correspond to
\beq
\label{eq:TaTb_expand}
T_{\bk}^{(a)}(t,t_0)
&=T_{\bk}^{(A)}(t,t_0)
+\sum_{s=1}^\infty(4\pi G \bar{\rho})^s
\!\int\!\!\left(\prod_{i=1}^s dt_i\right)T_{\bk}^{(B)}(t,t_s)\left[\prod_{i=2}^s T_{\bk}^{(B)}(t_i,t_{i-1})\right]T_{\bk}^{(A)}(t_1,t_0) ,\\
T_{\bk}^{(b)}(t,t_0)
&=T_{\bk}^{(B)}(t,t_0)
+\sum_{s=1}^\infty(4\pi G \bar{\rho})^s
\!\int\!\!\left(\prod_{i=1}^s dt_i\right)T_{\bk}^{(B)}(t,t_s)\left[\prod_{i=1}^s T_{\bk}^{(B)}(t_i,t_{i-1})\right],
\eeq
where we enforce the appropriate time ordering by defining $T_{\bk}^{(A)}(t,t')=T_{\bk}^{(B)}(t,t')=0$ if $t<t'$ (or equivalently, we multiply each integrand by a function that is 1 if $t>t_s>t_{s-1}>...>t_0$ and 0 otherwise).

Now from equations \eqref{eq:fullTaTb}, it is clear that
\beq
%T_{\bk}^{(a)}(t,t_0) \equiv
\begin{fmffile}{newTa}
\parbox{35pt}{
\begin{fmfgraph*}(30,40)
\fmfbottom{t0}
\fmftop{t}
\fmf{dbl_wiggly,label=$a$}{t0,t}
\end{fmfgraph*}
}
=
\parbox{35pt}{
\begin{fmfgraph*}(30,40)
\fmfbottom{t0}
\fmftop{t}
\fmf{wiggly,label=$A$}{t0,t}
\end{fmfgraph*}
}
+
\parbox{35pt}{
\begin{fmfgraph*}(30,40)
\fmfbottom{t0}
\fmftop{t}
\fmf{dbl_plain,label=$b$}{t1,t}
\fmf{wiggly,label=$A$}{t0,t1}
\fmfblob{6pt}{t1}
\end{fmfgraph*}
}
\end{fmffile},
\hspace{2em}&\hspace{2em}
\begin{fmffile}{newTb}
\parbox{35pt}{
\begin{fmfgraph*}(30,40)
\fmfbottom{t0}
\fmftop{t}
\fmf{dbl_plain,label=$b$}{t0,t}
\end{fmfgraph*}
}
=
\parbox{35pt}{
\begin{fmfgraph*}(30,40)
\fmfbottom{t0}
\fmftop{t}
\fmf{plain,label=$B$}{t0,t}
\end{fmfgraph*}
}
+
\parbox{35pt}{
\begin{fmfgraph*}(30,40)
\fmfbottom{t0}
\fmftop{t}
\fmf{dbl_plain,label=$b$}{t1,t}
\fmf{plain,label=$B$}{t0,t1}
\fmfblob{6pt}{t1}
\end{fmfgraph*}
}
\end{fmffile},
\eeq
or symbolically,
\beq
\label{eq:TaTbNew}
T_{\bk}^{(a)}(t,t_0)
&=T_{\bk}^{(A)}(t,t_0)
+\int_{t_0}^t 4\pi G \bar{\rho}\,dt'\,T_{\bk}^{(b)}(t,t')\,T_{\bk}^{(A)}(t',t_0) ,\\
T_{\bk}^{(b)}(t,t_0)
&=T_{\bk}^{(B)}(t,t_0)
+\int_{t_0}^t 4\pi G \bar{\rho}\,dt'\,T_{\bk}^{(b)}(t,t')\,T_{\bk}^{(B)}(t',t_0),
\eeq
proving equations~\eqref{eq:TaTbAlt} and implying that there is no need to solve a separate Volterra equation for $T_{\bk}^{(a)}$. 
The expression for $T_{\bk}^{(b)}$ here is still a Volterra equation, but we include it because it nicely mirrors the expression for $T_{\bk}^{(a)}$.
Intuition for equations~\eqref{eq:TaTbNew} is that perturbation growth with self-gravity ($T^{(a,b)}$) can be represented as the evolution of the same perturbation without self-gravity ($T^{(A,B)}$) plus the sum over all of the new perturbations $T^{(b)}$ that are sourced by the gravity of the original perturbation, where the new perturbations are themselves evolved subject to their own self-gravity.

Finally, we note that the white noise power spectrum in equation~\eqref{eq:Pwn_t} can be written in this notation as well, if we define the power spectrum of pure Poisson noise to be
\beq
\frac{1}{\bar n} \equiv
\begin{fmffile}{sym1}
\parbox{35pt}{
\begin{fmfgraph*}(30,40)
\fmfbottom{t0}
\fmftop{t}
\fmf{dashes}{t0,t}
\end{fmfgraph*}
}
\end{fmffile}.
\eeq
Now the white noise power spectrum $P_{\delta_{\rm wn}}(t,k)$ may be written
\beq
P_{\delta_{\rm wn}}(t,k)
=
\begin{fmffile}{PwnDiag}
\parbox{35pt}{
\begin{fmfgraph*}(30,60)
\fmfbottom{t0}
\fmftop{t}
\fmf{dashes}{t0,t}
\end{fmfgraph*}
}
+ 2
\parbox{70pt}{\hspace{12pt}
\begin{fmfgraph*}(40,60)
\fmfbottom{t0}
\fmftop{t,tt}
\fmf{dbl_plain,label=$b$,right=.39}{t1,tt}
\fmf{dbl_wiggly,label=$a$,left=.39}{t1,t}
\fmf{dashes,tension=2}{t0,t1}
\fmfblob{6pt}{t1}
\end{fmfgraph*}
}
\end{fmffile}.
%\right).
\eeq
As before, upper ``external'' vertices correspond to the final time $t$.
Intuitively, the interpretation is that the $1/\bar n$ white noise ``floor'' of the power spectrum continuously sources perturbations of both ``initial displacement'' and ``initial velocity'' type, and their product contributes to the power spectrum.

%~~~~~~~~~~~~~~~~~~~~~~~~~~~~~~~~~~~~~~~~~
\section{Numerically Evaluating the Growth Functions}
\label{sec:Eval}
%~~~~~~~~~~~~~~~~~~~~~~~~~~~~~~~~~~~~~~~~~
The time evolution of the power spectrum relies on the ``initial velocity'' growth function $\mathcal{T}^{(b)}(y,y',k)$ which satisfies a Volterra equation \eqref{eq:PT}. This growth function may be efficiently evaluated using an iterative scheme that is closely related to the series expansion in equation~\eqref{eq:fullTaTb} or \eqref{eq:TaTb_expand}. For $i\geq 0$, we define $\mathcal{T}^{(b,i)}_k(y,y')$ as
\beq\label{eq:Tbi}
\mathcal{T}^{(b,0)}_k(y,y')&=\mathcal{F}(y,y')T_{\rm fs}(y,y',k),
\\
\mathcal{T}^{(b,i+1)}_k(y,y')&=\mathcal{T}^{(b,0)}_k(y,y')+\frac{3}{2}\int_{y'}^y \frac{dy''}{\sqrt{1+y''}}\mathcal{T}^{(b,i)}_k(y,y'')\mathcal{T}^{(b,0)}_k(y'',y').
\eeq
Then $\mathcal{T}^{(b)}_k(y,y')=\lim_{i\to\infty}\mathcal{T}^{(b,i)}_k(y,y')$. Note that $\mathcal{T}^{(b,i)}_k$ corresponds to the series expansion \eqref{eq:fullTaTb} or \eqref{eq:TaTb_expand} truncated after the $(i+1)$th term. In practice, we find that extremely good convergence is achieved up to the time $y=10^3$ with $i\approx 10$ iterations. Fewer iterations are needed to integrate over shorter periods of matter domination. To carry out the integral in equation~\eqref{eq:Tbi}, we find that integration time steps as coarse as $\Delta\ln y\approx 0.2$ are sufficient. We provide an implementation of this iterative calculation in the \textsc{Python} language at \url{https://github.com/delos/warm-structure-growth} \cite{delos_2025_17064722}. This implementation can accurately evaluate the white-noise power spectrum in just a few seconds on a personal computer.

Beyond warm white noise, we also point out that this computation and code may be used more generally as a fast alternative to a full Boltzmann solver such as \textsc{CLASS} \cite{2011JCAP...09..032L} to study the effect of non-cold relics on the matter power spectrum, at least in the non-relativistic regime. Our iterative approach is similar to that suggested by ref.~\cite{Ji:2022iji}.

\section{Fitting Functions and Approximate Evaluation of Power Spectrum}
\label{sec:Approx}
%~~~~~~~~~~~~~~~~~~~~~~~~~~~~~~~~~~~~~~~~~

\begin{figure}
    \centering
\includegraphics[width=1\linewidth]{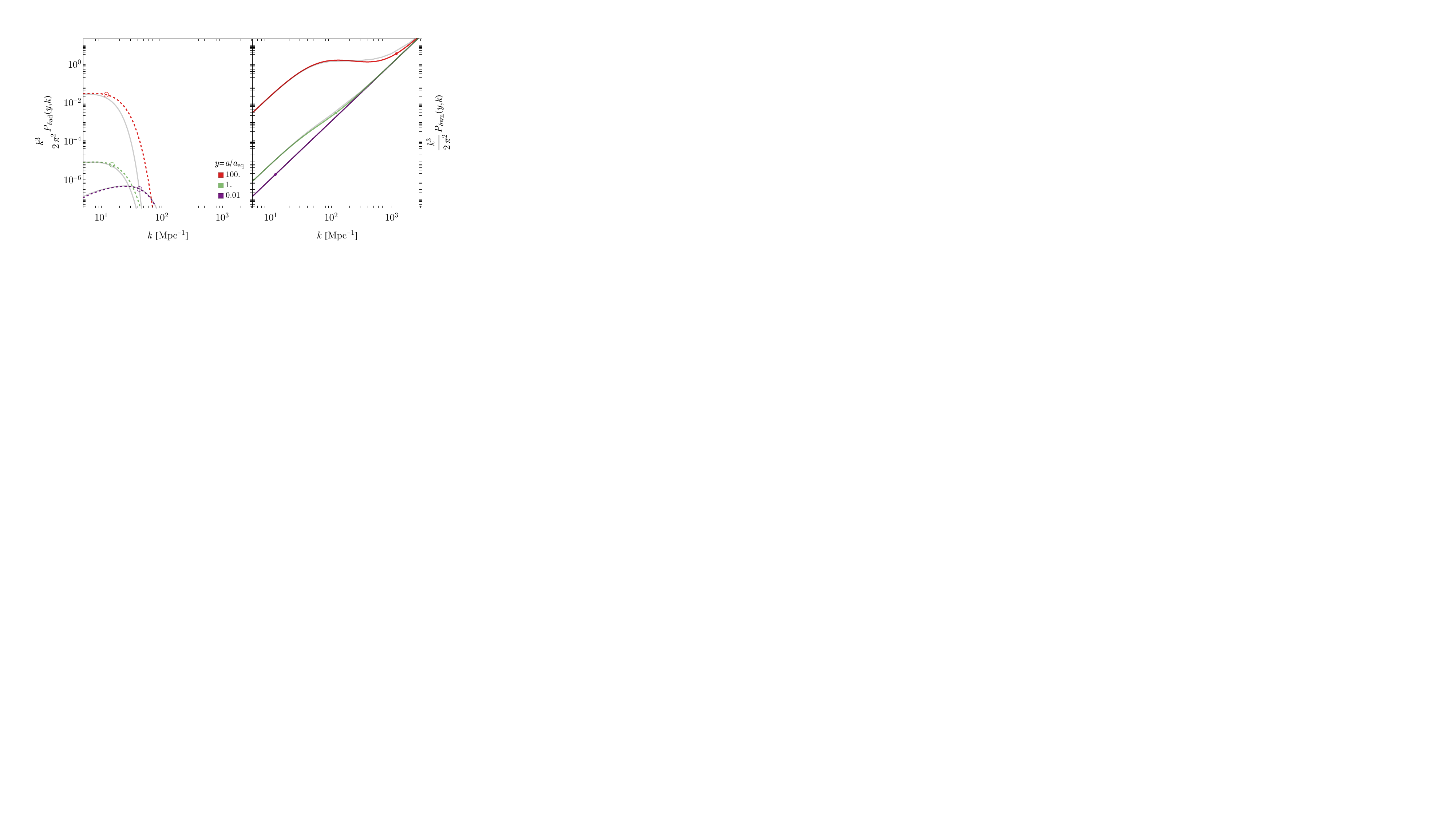}
    \caption{The colored lines are the power spectra from solving the Volterra equations, while the grey lines are based on the simplified ansatz provided in Appendix \ref{sec:Approx}. For the left panel, note that the ansatz $P_{\delta_{\rm ad}}^{\rm CDM}T_{\rm fs}^2$ works well in radiation domination (as expected). Deep in the matter era, the ansatz leads to a $10$-$30\%$ error regarding the characteristic length scale where the power spectrum is suppressed. The error is smaller for colder dark matter. For both panels we adopt $\sigma_{\rm eq}\approx 22\,\textrm{km} \,\textrm{s}^{-1}$ and $\bar{n}\approx 5\times 10^7/\Mpc^3$.}
    \label{fig:Ansatz}
\end{figure}

To a good approximation, $\mathcal{T}^{(a,b)}_k(y,y')$ appearing inside the integral in the evolution of the white noise spectrum \eqref{eq:PwEvolution} can be replaced by (this is empirical, rather than a controlled expansion)
% \beq
% \label{eq:PwnAnsatz}
% \mathcal{T}^{(a,b)}_k(y,y')&\to T_0^{(a,b)}(y,y')\sqrt{T_{\rm fs}(y,y',k)}\,,
% \eeq
\beq
\label{eq:PwnAnsatz}
\mathcal{T}^{(a)}_k(y,y')\mathcal{T}^{(b)}_k(y,y')&\to T_0^{(a)}(y,y')T_0^{(b)}(y,y')T_{\rm fs}(y,y',k)\,,
\eeq
where
\beq
T^{(a)}_0(y,y')&=\left(1+\frac{3}{2} y\right) \left(1+3y'\right)-\frac{3}{2} y' \sqrt{1+y'} \left[3
   \sqrt{1+y}-\left(1+\frac{3}{2} y\right) \ln (\frac{x}{x'})\right],\\
T^{(b)}_0(y,y')&=3 \left(1+\frac{3}{2} y\right) \left(1+\frac{3}{2} y'\right) \left[\frac{\sqrt{1+y}}{1+\frac{3}{2} y}-\frac{\sqrt{1+y'}}{1+\frac{3}{2} y'}-\frac{1}{3}\ln(x/x')\right],
\eeq
with $x\equiv\frac{\sqrt{1+y}+1}{\sqrt{1+y}-1}$ and $x'\equiv\frac{\sqrt{1+y'}+1}{\sqrt{1+y'}-1}$. This ansatz eliminates the need for solving Volterra equations for $T^{(a)}_k$ and $T^{(b)}_k$. Given a $f_0(p)$, we can get an expression for $T_{\rm fs}$, which then yields $P_{\delta_{\rm wn}}$. The right-hand panel of figure~\ref{fig:Ansatz} shows how this approximation compares to the result of the full calculation.

The adiabatic part of the power spectrum can be approximated as
\beq
\label{eq:PadApprox}
P_{\delta_{\rm ad}}(y,k)
&\approx 
P_{\delta_{\rm ad}}^{\rm CDM}(y,k)
T^2_{\rm fs}(y,y_0,k),
\eeq
where $P_{\delta_{\rm ad}}^{\rm CDM}(y,k)$ is the power spectrum for cold dark matter (CDM). This expression is exact during radiation domination but loses some accuracy during the matter era. For $y\gg 1$, it overestimates the length scale at which the power spectrum is suppressed by around 10-30\% (with less error for colder dark matter). The CDM power spectrum $P_{\delta_{\rm ad}}^{\rm CDM}$ may be computed using standard methods (e.g.~\cite{Blas:2011rf}), but on scales that are sufficiently subhorizon at matter-radiation equality ($k\gtrsim\Mpc^{-1}$), we may write it as
\beq\label{P_ad_CDM}
P_{\delta_{\rm ad}}(y,k)
&\approx 36P_\mathcal{R}(k)\left[\left(1+\frac{3}{2}y\right) \ln \left(0.15 \frac{k}{k_{\rm eq}}\right)+3\sqrt{1+y}-\left(1+\frac{3}{2}y\right)\ln\left(\frac{\sqrt{1+y}+1}{\sqrt{1+y}-1}\right)\right]^2,
\eeq
where $P_\mathcal{R}(k)$ is the primordial curvature power spectrum.
The left-hand panel of figure~\ref{fig:Ansatz} compares the result from equations \eqref{eq:PadApprox} and~\eqref{P_ad_CDM} to the result of the full calculation. Note that equation~\eqref{P_ad_CDM} is derived assuming that dark matter is all of the matter. To account for baryons, at late times $z\lesssim 100$, equation~\eqref{P_ad_CDM} should be scaled by a factor of about $1/2$.\footnote{Baryons remain coupled to the cosmic microwave background until $a\approx 1/125$ (e.g.~\cite{Bertschinger:2006nq}), which leads to slower growth of dark matter perturbations for $\aeq\lesssim a\lesssim 1/125$.}

In order to set initial conditions for evolution of adiabatic perturbations, it is also useful to note that at $y=y_0\ll 1$,
\beq
P_{\delta_{\rm ad}}(y_0,k)\approx 36P_{\mathcal{R}}(k)\left[3+\ln\left(0.15\frac{k}{k_{\rm eq}}\right)-\ln\left(\frac{4}{y_0}\right)\right]^2.
\eeq

%~~~~~~~~~~~~~~~~~~~~~~~~~~~~~~~~~~~~~~~~~
\section{Example: the Uniform-Sphere Momentum Distribution}
\label{sec:Flat}
%~~~~~~~~~~~~~~~~~~~~~~~~~~~~~~~~~~~~~~~~~

%~~~~~~~~~~~~~~~~~~~
\begin{figure}
    \centering
\includegraphics[width=0.9\linewidth]{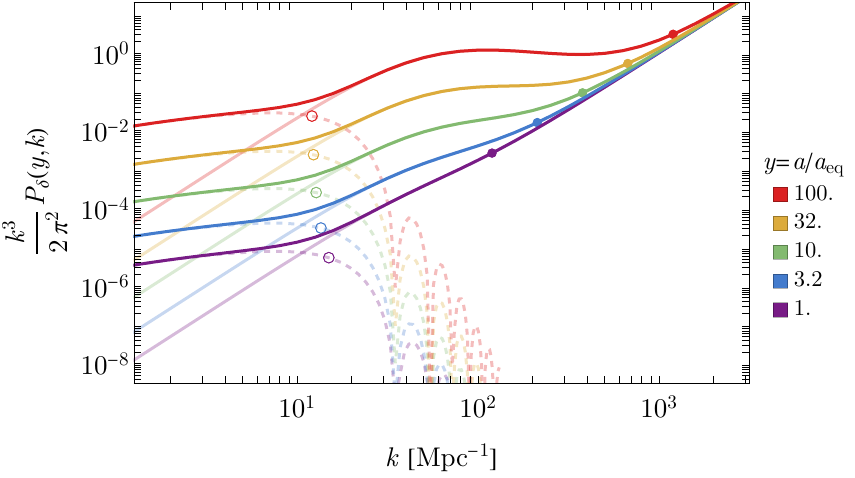}
    \caption{Evolution of the matter power spectrum (solid lines) for the case of a uniform-sphere initial velocity distribution. The lighter curves are the separate adiabatic (dashed) and white-noise (solid) contributions. The open circles on the adiabatic part indicate the free-streaming wavenumber $k_{\rm fs}(y)$, and solid circles on the white noise part are the Jeans wavenumber $k_{\rm J}(y)$. Here we set $\sigma_{\rm eq}\approx 22\,\textrm{km} \textrm{s}^{-1}$ and $\bar{n}\approx 5\times 10^7/\Mpc^3$, and the adiabatic part has an amplitude consistent with Planck 2018 observations.
    Compare Fig.~\ref{fig:PbothEvolutionLateApprox} for the Maxwellian distribution.}
    \label{fig:plotBothLateApproxFlat}
\end{figure}
%~~~~~~~~~~~~~~~

%~~~~~~~~~~~~~~~
\begin{figure}
    \centering
\includegraphics[width=0.875\linewidth]{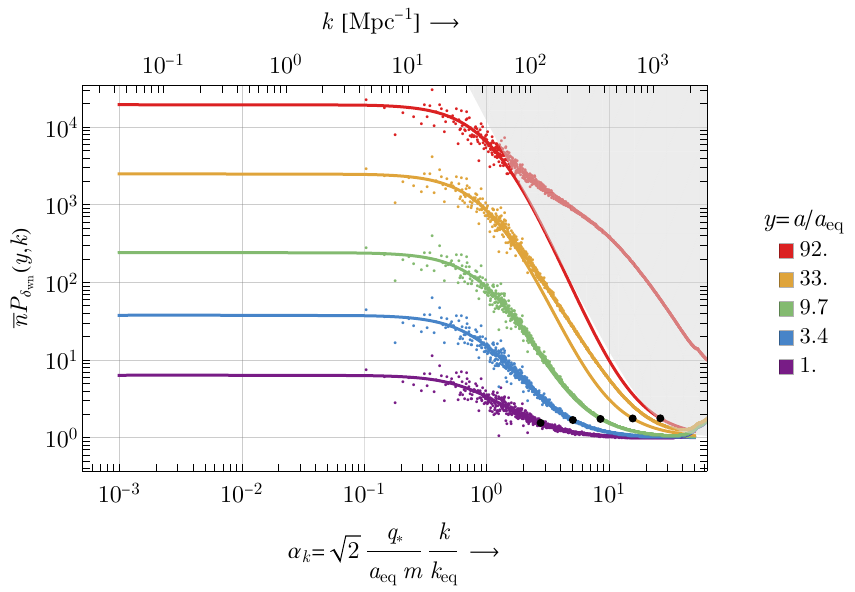}
    \caption{Comparing analytically predicted power spectra (solid curves) with results from an $N$-body simulation (colored points) for the case of the uniform-sphere momentum distribution. The gray region marks the nonlinear regime, while the black dots mark the Jeans wavenumber $k_{\rm J}$. Compare Fig.~\ref{fig:plotNumCompare} for the Maxwellian distribution.}
    \label{fig:flatplotNumCompare}
\end{figure}
%~~~~~~~~~~~~~~~~~~~~~~~~~~

%~~~~~~~~~~~~~~~~~~~
\begin{figure}
    \centering
\includegraphics[width=0.875\linewidth]{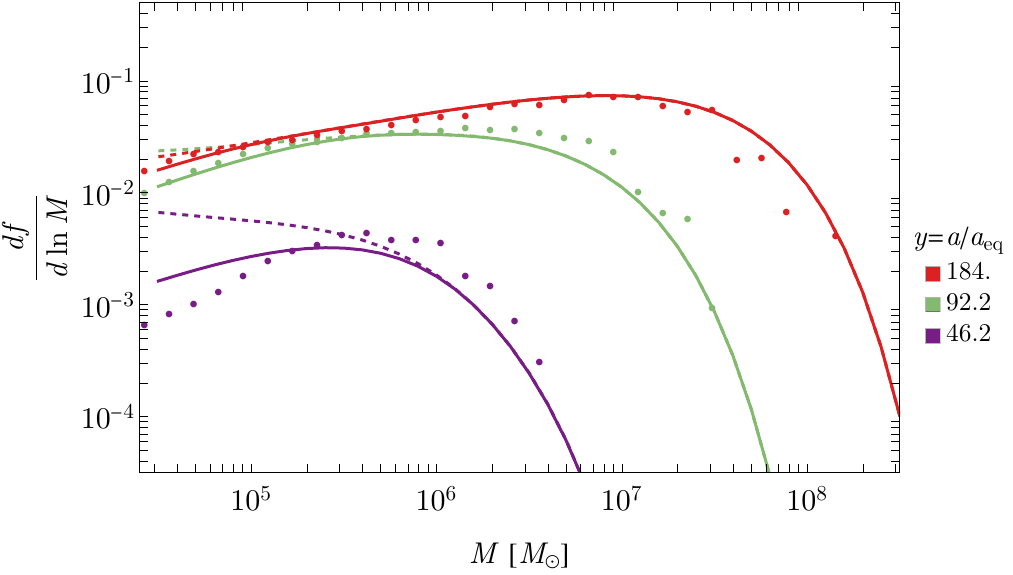}
    \caption{Comparing analytically predicted halo mass functions (solid curves) with results from an $N$-body simulation (points) for the uniform-sphere momentum distribution. We show the differential fraction of mass (per logarithmic mass interval) in halos of mass $M$. We evaluate the analytic predictions with the $k_{\rm max}=k_{\rm J}/4$ ansatz; the dashed lines show the predictions without that ansatz. Compare Fig.~\ref{fig:plotMassFuncCompare} for the Maxwellian distribution.}
    \label{fig:plotMassFuncCompareFlat}
\end{figure}
%~~~~~~~~~~~~~~~~~

The calculations in this work are valid for any isotropic initial momentum distribution $f_0(q)$, but in the main text, we focused on the Maxwell-Boltzmann distribution, $f_0(q)\propto e^{-(q/q_*)^2/2}$. Here we show results for a uniform-sphere momentum distribution, $f_0(q)\propto\Theta(q_*-q)$.
Figure~\ref{fig:plotBothLateApproxFlat} shows the predicted matter power spectrum evolution at late times for a model with this momentum distribution, along with the separate adiabatic and white noise contributions. Compared with the Maxwell-Boltzmann case (Fig.~\ref{fig:PbothEvolutionLateApprox}), here the momentum distribution gives rise to oscillations in the adiabatic spectrum (dashed lines), although these do not significantly contribute to the total power spectrum.

In Fig.~\ref{fig:flatplotNumCompare}, we compare the analytically predicted white noise power spectrum to the results from an $N$-body simulation. This simulation is initialized with a uniform-sphere momentum distribution and no initial (adiabatic) perturbations. As in section~\ref{sec:N-Compare}, the particles have mass $780\,M_{\odot}$, and the 1D velocity dispersion at matter-radiation equality is $\sigma_{\rm eq}\approx 22\,{\rm km}\, {\rm s}^{-1}$.
%(but note that $q_*/m$ is $\sqrt{5}$ times larger).
The predicted power spectrum (solid curves) generally matches the simulation results (dots). Deviations only arise at low $k$ due to cosmic variance in the 1.38-Mpc simulation box and at late times, when the density perturbations are approaching the nonlinear regime (gray shaded area).

Figure~\ref{fig:plotMassFuncCompareFlat} compares the halo mass function in the simulation with the mass function predicted analytically as in section~\ref{sec:SimWNnl}. With the mass variance $\sigma_M$ evaluated from equation~\eqref{eq:sigmaM} with $k_{\rm max}=k_{\rm J}/4$ (the same ansatz that worked well for the Maxwell-Boltzmann momentum distribution), the prediction matches the simulation result generally well, although the abundance of low-mass halos is somewhat overpredicted at early times. Without that ansatz (so $k_{\rm max}\to\infty$), the abundance of low-mass halos is overpredicted to a much greater degree.

Finally, we note that the approximation in appendix~\ref{sec:Approx} does not work as well for the uniform-sphere momentum distribution as it does for the Maxwell-Boltzmann case. For the uniform sphere, it predicts significantly less power at $\alpha_k\gtrsim 1$ than does the exact calculation, leading to a lower predicted halo abundance.

\end{document}